\documentclass[12pt]{article}
\pdfoutput=1
\usepackage{color}
\usepackage{amsmath, amssymb}
\usepackage{epsfig, palatino}
\usepackage{pstricks,pst-node,pst-tree}
\usepackage{epic}

\usepackage{mathrsfs}

\usepackage{ae} 
\usepackage[T1]{fontenc}
\usepackage[ansinew]{inputenc}
\usepackage{amsmath}
\usepackage{amssymb}
\usepackage{graphicx}
\usepackage{color}
\definecolor{darkblue}{cmyk}{0.9,0.9,0,0}
\usepackage[pdftex,colorlinks=true,linkcolor=darkblue,citecolor=darkblue,urlcolor=darkblue]{hyperref}
\usepackage{simplewick}
\usepackage{cite}

 \newcommand{\Blue}[1]{{\color{blue}#1\color{black}}}
\newcommand{\Red}[1]{{\color{red}#1\color{black}}}

\newcommand{\beq}{\begin{equation}}
\newcommand{\eeq}{\end{equation}}
\newcommand\beqa{\begin{eqnarray}}
\newcommand\eeqa{\end{eqnarray}}
\newcommand\bea{\begin{array}}
\newcommand\eea{\end{array}}

\def\XXint#1#2#3{{\setbox0=\hbox{$#1{#2#3}{\int}$}
\vcenter{\hbox{$#2#3$}}\kern-.5\wd0}}

\newcommand{\nn}{\nonumber}

\newcommand{\COMMENT}[1]{}

\newcommand{\neqa}{\nonumber\end{eqnarray}}
\newcommand{\la}[1]{\label{#1}}

\renewcommand{\d}{\partial}

\newcommand{\<}{{\langle}}
\renewcommand{\>}{{\rangle}}

\newcommand{\cA}{{\cal A}}

\newcommand{\re}{\relax{\rm I\kern-.18em R}}

\def\su2{{SU(2)}}

\def\[{\left[}
\def\]{\right]}
\def\Tr{\text{Tr\,}}

\def\({\left(}
\def\){\right)}
\def\[{\left[}
\def\]{\right]}

\def\<{\langle}
\def\>{\rangle}

\def\i2{\frac{i}{2}}

\def\cO{{\cal O}}

\newcommand{\rrangle}{\rangle \hspace{-.15em} \rangle}
\newcommand{\llangle}{\langle\hspace{-.15em}\langle}

\usepackage{varioref}
\usepackage{makeidx}
\makeindex
\usepackage[english]{babel}

\begin{document}

\thispagestyle{empty}

\renewcommand{\thefootnote}{\fnsymbol{footnote}}
\setcounter{footnote}{0}
\setcounter{figure}{0}
\begin{center}
$$$$
{\Large\textbf{\mathversion{bold}
From Polygon Wilson Loops to Spin Chains and Back
}\par}
\vspace{1.0cm}

\textrm{Amit Sever$^{a,b}$, Pedro Vieira$^{a}$, Tianheng Wang$^{a,c}$}
\\ \vspace{0.5cm}

\textit{$^{a}$  Perimeter Institute for Theoretical Physics\\ Waterloo,
Ontario N2J 2W9, Canada} \\
\texttt{} \\
\vspace{.7mm}
\textit{$^{b}$ School of Natural Sciences,\\Institute for Advanced Study, Princeton, NJ 08540, USA.} \\
\texttt{} \\
\vspace{.7mm}
\textit{$^{c}$ Department of Physics and Astronomy \& Guelph-Waterloo Physics Institute,\\
University of Waterloo, Waterloo, Ontario N2L 3G1, Canada} \\
\texttt{} 
\vspace{0mm}

\par\vspace{.5cm}

\textbf{Abstract}
\end{center} 
\vspace{-.3cm}

Null Polygon Wilson Loops in ${\cal N}=4$ SYM can be computed using the Operator Product Expansion in terms of a transition amplitude on top of a color Flux tube. That picture is valid at any value of the 't Hooft coupling and is studied here in the planar limit. So far it has been efficiently used at weak coupling in cases where only a single particle is flowing. At any finite value of the coupling however, an infinite number of particles are flowing on top of the color flux tube. A major open problem in this approach was how to deal with generic multi-particle states at weak coupling. In this paper we study the propagation of any number of flux tube excitations at weak coupling. We do this by first mapping the Wilson loop expectation value into a sum of two point functions of local operators. That map allows us to translate the integrability techniques developed for the spectrum problem back to the Wilson loop. In particular, we find that the flux tube Hamiltonian can be represented as a simple kernel acting on the loop. Having an explicit representation for the flux tube Hamiltonian allows us to treat any number of particles on an equal footing. We use it to bootstrap some simple cases where two particles are flowing, dual to $\text{N}^2$MHV amplitudes. The flux tube is integrable and therefore has other (infinite set of) conserved charges. The generating function of all of these charges is constructed from the monodromy matrix between sides of the polygon. We compute it for some simple examples at leading order in perturbation theory. At strong coupling, these monodromies were the main ingredients of the Y-system solution. To connect the weak and strong coupling computations, we study a case where an infinite number of particles are propagating already at leading order in perturbation theory. We obtain a precise match between the weak and strong coupling monodromies. That match is the Wilson loop analog of the well known Frolov-Tseytlin limit where the strong and weak coupling descriptions become identical. Hopefully, putting the weak and strong coupling descriptions on the same footing is the first step in understanding the all loop structure.

\vspace*{\fill}

\setcounter{page}{1}
\renewcommand{\thefootnote}{\arabic{footnote}}
\setcounter{footnote}{0}

\newpage

\tableofcontents

\section{Introduction}

\begin{figure}[h]
\centering
\def\svgwidth{10cm}
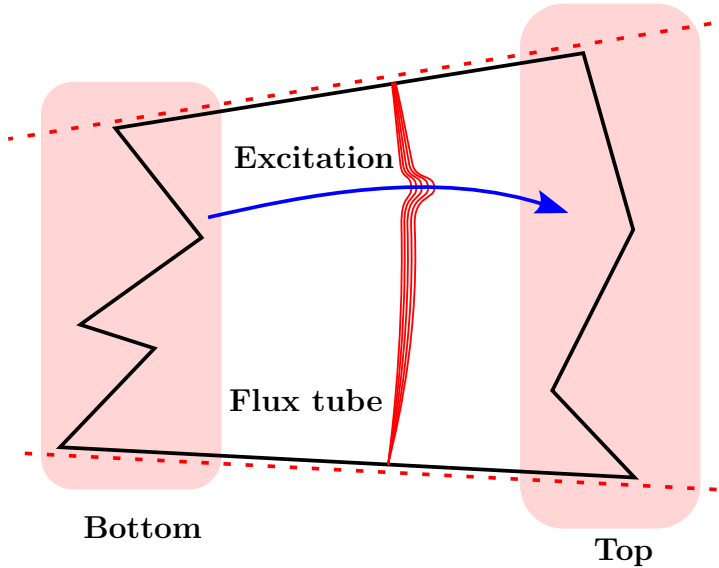
\caption{\small The OPE picture. \textit{Flux tube  excitations} are created at the \textit{bottom} and absorbed in the \textit{top}. The flux is thick in spacetime but is 1+1 dimensional in AdS. Its excitations are integrable.}\label{OPEpicture}
\end{figure}

The Operator Product Expansion (OPE) for Null Polygon Wilson Loops translates the Wilson loop computation into a transition amplitude on top of a color Flux tube \cite{OPEpaper}, see figure \ref{OPEpicture}. The OPE does not rely on perturbation theory. Indeed, it was observed both at weak and at strong coupling in $\mathcal{N}=4$ SYM. The main hope of this program is to fill up the gap and find the same description at any intermediate coupling for this particular theory. 

The flux preserves some symmetries. We can therefore classify the states by the way they transform under these symmetries. In particular the states  have {\it energies} and continuous {\it momenta}.\footnote{Also known as twist and conformal spin.} The states are gapped at any value of the coupling \cite{AMcomments,Benjamin} and therefore we can further classify them according to the \textit{number of particles}. 

One main difference between weak and strong coupling is in the number of particles involved. 
At a given order in the weak coupling perturbative expansion, we have a maximum number of particles that can be excited and flow in the flux tube. Sometimes we can have at most a single excitation. This is the simplest case and is the one that is currently well understood from the OPE point of view.\footnote{Despite the simplicity of the single particle exchange case, it turns
out to be a very powerful tool for computing or constraining
scattering amplitudes at the first few loop orders. For example, under simple assumptions, it fixes completely the two loops MHV amplitudes and the one loop NMHV amplitudes \cite{bootstraping,Hexagonpaper,superOPE}. At higher loops it does not
fix the full result but provides very helpful tight constraints on the
result. They were recently used in constraining the hexagon NMHV amplitude
at two loops \cite{Dixon:2011nj}, the hexagon MHV amplitude at three loops \cite{Dixon:2011pw}
and for the three loop 2D MHV octagon \cite{Heslop:2011hv}. The single particle OPE was also used in deriving the $\bar Q$ equation \cite{SimonHe}, see also \cite{davesametime}.}
On the other hand, at strong coupling, we have a continuous distribution of infinitely many flux tube excitations. 

Hence, to further connect the weak and strong coupling we should go beyond the single particle OPE and get a better handle on the flux tube states. 
That is the main motivation for this paper. 

The reason why a single particle is considerably simpler is that single particle wave functions are always trivial to classify. They are just plane waves with definite momenta. On the other hand, many particles can interact. As a result, what the two or more particle wave functions are becomes a dynamical question that depends on the details of the interaction.  
To handle these states, we need to understand the map to the flux tube wave functions more concretely, as we will do in this paper. 

The flux tube states can be characterized by their charges. The most obvious of all is the energy. By reading the energy of these states we can easily make an infinite number of predictions at any loop order \cite{OPEpaper}. In sections \ref{sec2}\,-\,\ref{sec4} we focus on the Hamiltonian which measures this energy. We start by considering a simple set of examples which illustrate the general method in section \ref{sec2}. In section \ref{sec3} we consider generalizations and further conceptual discussions of the general method. In section \ref{sec4} we consider further examples. Section \ref{sec4} can be read independently of section \ref{sec3}. 

\begin{figure}[t]
\centering
\def\svgwidth{10cm}
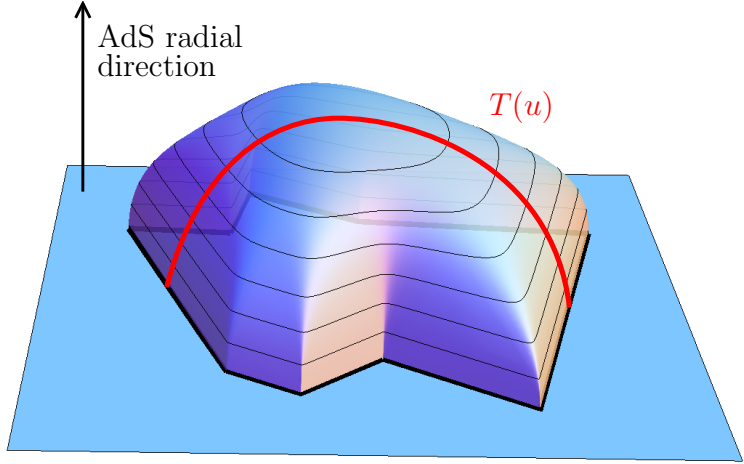
\caption{At strong coupling the polygon Wilson loop expectation value is computed by a minimal surface area. The surface ends on the polygon at the boundary of AdS and is stretched in the AdS radial direction. The holonomy $T(u)$ of the flat connection between two edges of the polygon is drawn in red. It is a measure of all the higher conserved charges of what is flowing through it. Such holonomies are the building blocks in the Y-system strong coupling solution \cite{Alday:2010vh}.}\label{TinAdS}
\end{figure}

The energy is only one out of the infinitely many charges that characterize the flux tube states. The other charges can be encoded in the generating functions of conserved charges constructed from the so called monodromy matrices. At strong coupling these monodromies are the main ingredient of the Y-system solution \cite{Alday:2010vh}. More precisely, at strong coupling one computes the holonomies of the flat connection between edges of the polygon, see figure \ref{TinAdS}. In order to connect the weak and strong coupling descriptions, we study these objects at weak coupling in sections \ref{MonoSec} and \ref{StrongWeakmatch}. In section \ref{MonoSec} we identify these holonomies between edges at weak coupling and compute them for the simplest Wilson loops. The result we find is strikingly different from the strong coupling analogous holonomies!  This is not surprising. As mentioned above, at weak coupling we have a small finite number of excitations flowing from the bottom to the top whereas at strong coupling we have densities describing the propagation of infinitely many particles. To bring the weak and strong coupling descriptions closer -- and hence provide further support for our identifications of section \ref{MonoSec} -- we consider in section \ref{StrongWeakmatch} a scaling limit where infinitely many particles flow already at weak coupling. In this limit we insert, by hand, a large number of insertions in the loop. This would be the kind of situation one would encounter when studying N$^k$MHV amplitudes with $k \gg 1$ and is the analogue of the Frolov-Tseytlin limit \cite{FT} in the spectrum problem. In section \ref{StrongWeakmatch} we compute the weak and strong coupling holonomies  in this limit and find a match between the two.

In a very nice paper \cite{Belitsky}, Belitsky explored the connection of Null Polygon Wilson loops to the correlation function of two Wilson lines from the OPE point of view (we will also use this). He focused on MHV amplitudes with single particle excitations whereas we will consider mainly multi-particle excitations for N$^k$MHV amplitudes. He borrowed the technology of renormalization of Wilson lines while we will mostly explore the connection to Integrability for local operators. Both approaches are obviously tightly related and it is definitely worth exploring this connection further. 

\section{From Wilson loops to two point functions}\la{sec2}

\begin{figure}[t]
\centering
\def\svgwidth{14cm}
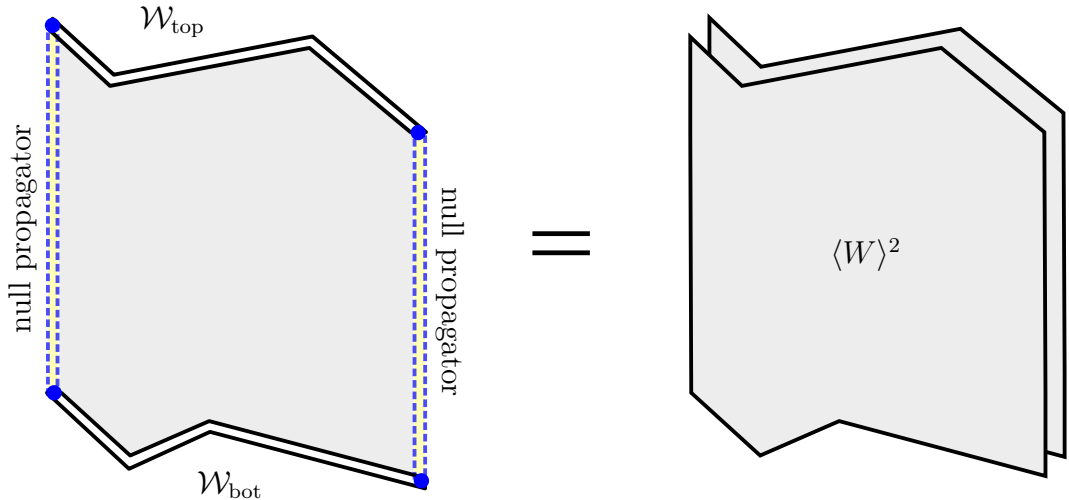
\caption{A relation between a correlation function of two Wilson lines with insertions and a polygonal Wilson loop. On the left we have a correlation function of two Wilson lines with insertions ${\cal W}_\text{bot}$ and ${\cal W}_\text{top}$. We choose these to be  composed of a sequence of null lines. At their tips we have scalar insertions indicated by blue dots in the figure. We take a limit where each one of the two scalars in ${\cal W}_\text{bot}$ becomes null separated from a conjugate scalar in ${\cal W}_\text{top}$. In that null limit the correlator develops a double pole due to two fast particles going along the left and right edges in the figure, indicated by the dashed lines. The residue of the double pole singularity is computed by two copies of a polygon Wilson loop. The polygon loop is composed of the two open curves in ${\cal W}_\text{bot}$ and ${\cal W}_\text{top}$, closed by the right and left edges into a closed loop.}\label{setup}
\end{figure}

We start with a correlation function between two Wilson loops in the fundamental representation with two adjoint field insertions\footnote{Alternatively, one can add massless probe quarks and consider the operators $\bar q(x)W[x,y]q(y)$.}
\beq
{\cal W}_\text{bot}=\Tr\,\cO(x_1) W[x_1,x_2]\cO(x_2)W[x_1,x_2]^\dagger\quad\text{and}\quad{\cal W}_\text{top}=\Tr\,\cO^\dagger(y_1) W[y_1,y_2]\cO^\dagger(y_2)W[y_1,y_2]^\dagger\nn
\eeq
The Wilson line $W[x_1,x_2]$ connects points $x_1$ and $x_2$ along some path. The conjugate Wilson line  $W[x_1,x_2]^\dagger$ connects $x_2$ and $x_1$ along the same (but reversed) path. 
In the limit where the points $x_i$ and $y_i$ become light-like separated,  the correlation function $\<{\cal W}_\text{bot}{\cal W}_\text{top}\>$ develops two poles corresponding to two fast particles going between $\cO(x_i)$ and $\cO^\dagger (y_i)$. Each particle is charged under the gauge group and its interaction with the gauge field is approximated by a pair of Wilson lines in the fundamental and anti-fundamental between $x_i$ and $y_i$ \cite{Alday:2010zy}. \footnote{We do not distinguish between a Wilson line in the adjoint and a pair of lines in the fundamental/anti-fundamental representations.} This joins $W[x_1,x_2]$ and $W[y_1,y_2]$ into a single closed loop $W$ and $W[x_1,x_2]^\dagger$ and $W[y_1,y_2]^\dagger$ into another identical closed loop $W^{\dagger}$. In the planar limit $\<W W^{\dagger}\>=\<W\>^2$.
In other words, we start from a configuration where $x_1-y_1$ and $x_2-y_2$ are space-like. Then, the residue of the double pole when $(x_i-y_i)^2\to0$ is computed by the expectation value of one closed Wilson loop as (see figure \ref{setup}) 
\beq\la{corrtoWilson}
\lim_{(x_i-y_i)^2\to0}\frac{\<{\cal W}_\text{top}{\cal W}_\text{bot}\>}{\<{\cal W}_\text{top}{\cal W}_\text{bot}\>_\text{tree}}=\<W\>^2 \,.
\eeq
Here, $\<{\cal W}_\text{top}{\cal W}_\text{bot}\>_\text{tree}$ stands for the tree level correlator. 
We take the Wilson lines $W[x_i,y_i]$ to be composed of a sequence of null edges. Then the resulting closed Wilson loop $W$ is a null polygon. The relation (\ref{corrtoWilson}) is a simple generalization of the Correlation function/Wilson loop correspondence of \cite{Alday:2010zy}, see also \cite{Belitsky}. 

The expectation value of null polygon Wilson loops needs to be UV regularized. As explained in \cite{Alday:2010zy}, depending on the choice of regularization, there may be some issues in taking the light-like limit. These come about due to a possible recoil of the fast particles. In this note we will always consider ratios of polygons such that these issues cancel out. 

\begin{figure}[t]
\centering
\def\svgwidth{17cm}
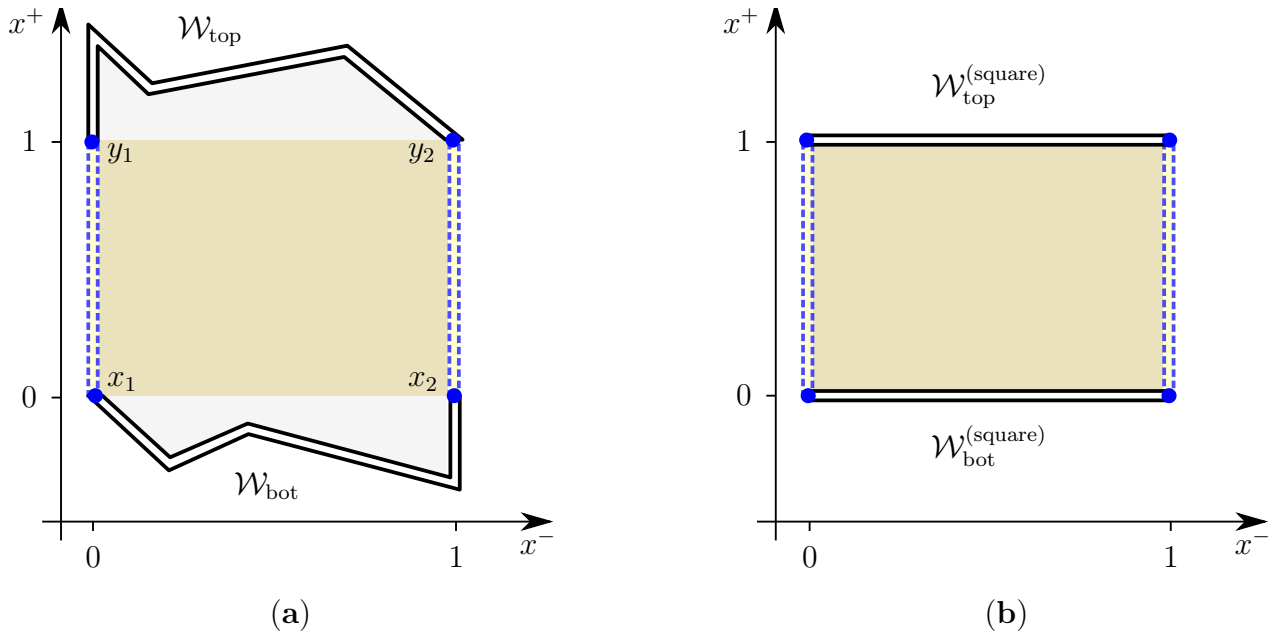
\caption{The setup of Wilson correlators considered in this paper. (b) On the right we have a square representing the OPE vacuum. In that case, the bottom and top operators, ${\cal W}_\text{top}^{(\text{square})}$ and ${\cal W}_\text{bot}^{(\text{square})}$, are composed of a Wilson line along a single null line connecting two scalars. (a) On the left  we replace that single null line by a sequence of null lines connecting the two cusps in the bottom or top of the square.}\label{squaref}
\end{figure}

We will now introduce the OPE vacuum. 
For that, we take the point $x_1$  to be null separated from $x_2$ and the same for $y_1$ and $y_2$, see figure \ref{squaref}. Then, the four points $\{x_1,x_2,y_1,y_2\}$ define a null square called the \textit{reference square}. By a conformal transformation, we fix these points to
\beq\la{squarepoints}
\{x_1,x_2,y_1,y_2\}=\{(0,0,0,0),(0,1,0,0),(1,1,0,0),(1,0,0,0)\}
\eeq
where the first two components are lightcone directions denoted by $+$ and $-$ respectively. The Wilson loop along the reference square, denoted as $W^\text{(square)} $, represents the vacuum state of the flux tube\cite{AMcomments}.

\begin{figure}[t]
\centering
\def\svgwidth{16cm}
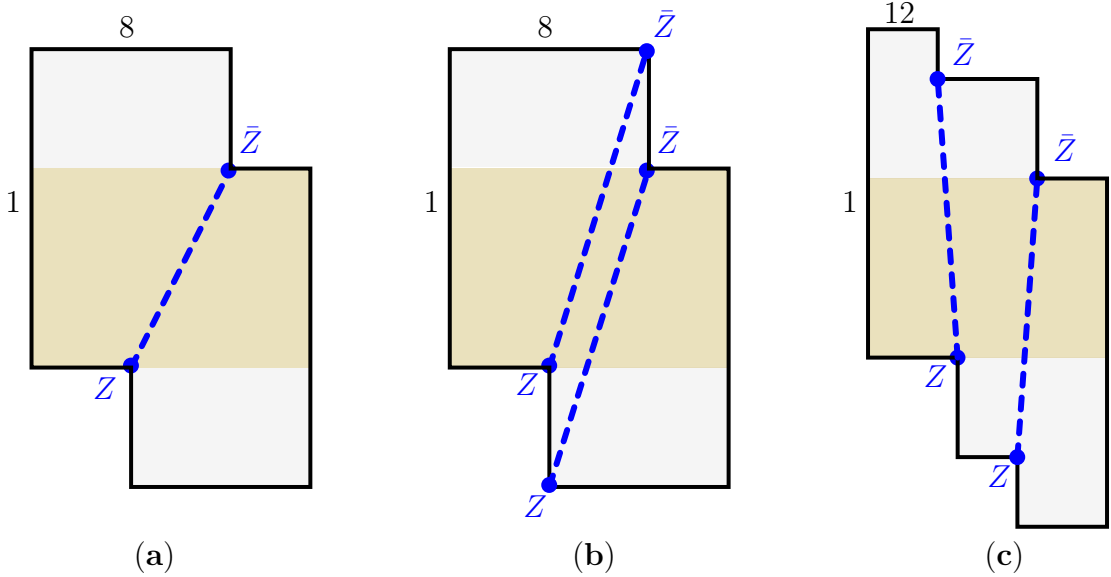
\caption{There is a simple class of N$^k$MHV amplitudes which we use as a laboratory for exploring the OPE physics. We
denoted such components as $Z \to \bar Z$ components, see
(\ref{equation5}). At tree level they are given by $k$ insertions of
scalars $Z$ at the bottom cusps of the polygon and another $k$ insertions
of the complex conjugate scalars $\bar Z$ at the top cusps. At tree level they are given by \textit{exactly} $k$ free scalars flux tube excitations going from the
bottom to the top part of the polygon. At one loop these particles
start interacting and feeling the flux. Such components are pure examples of multiparticles in the OPE.
Figure (a): NMHV octagon, (b): N$^2$MHV octagon, (c): N$^2$MHV
dodecagon. The last two correspond to two particle examples. These
three examples will be discussed in detail in this paper and
illustrate the general method.}\label{Examples}
\end{figure}

The bosonic polygon Wilson loop is dual to the MHV scattering amplitudes \cite{AmplitudeWilson,AmplitudeWilson2}. The other non-MHV amplitudes are captured by a supersymmetric generalization of this loop \cite{Skinner,Simon}. Formula (\ref{corrtoWilson}) can be easily generalized for the super loop case. The new ingredient is that we will also have insertions of various fields at the edges and cusps of the top and bottom Wilson lines as we now describe.\footnote{Insertions on the two null lines of the two fast particle can, in principle, be taken into account by adjusting the operators $\cO$ at the ends of ${\cal W}_\text{top/bottom}$\cite{Alday:2010zy,DaveAndOthers}. We will not consider these components in this note.} The super loop has \textit{components} which are the dual of the several scattering amplitude components. That is, $W$ is a polynomial in dual grassmannian variables $\eta_i^{A} $
where $i=1,\dots,n$ is a particle label and $A=1,2,3,4$ is an R-charge index. We should always have R-charge singlets so that the total power of $\eta$'s in each monomial is a multiple of four. Then it is convenient to factor out the purely bosonic loop which is obtained when setting to zero all grassmannian variables. That is,  
\beq
W/ W_\text{MHV} =  1+ \eta_i^1 \eta_j^2 \eta_k^3 \eta_l^4\,\, \mathcal{R}^{(ijkl)} + \eta_i^1 \eta_j^2 \eta_k^3 \eta_l^4  \eta_{m}^1 \eta_{n}^2 \eta_{o}^3 \eta_{p}^4\,\, \mathcal{R}^{(ijkl)(mnop)}  + \dots   \la{Rdef0}
\eeq
The $\mathcal{R}$'s are the so called ratio functions \cite{RatioCite}. The first one is the NMHV ratio function, the second is the N$^2$MHV one etc. 
These are finite conformal invariant quantities.\footnote{Alternatively, one could consider the BDS stripped amplitude \cite{SimonHe} which is the supersymmetric generalization of the Reminder function \cite{AmplitudeWilson2}. This is also a finite conformal invariant quantity.} If we furthermore divide by their tree level expressions
\beq\la{rratio}
r^{(ijkl)}=\frac{ \mathcal{R}^{(ijkl)}}{\mathcal{R}_\text{tree}^{(ijkl)}} \qquad \text{and similarly for N$^2$MHV and so on} 
\eeq
then the resulting functions $r$ only depend on the conformal cross ratios.\footnote{Without dividing by the tree level quantity we would need to deal with Helicity factors as in \cite{superOPE}. In particular, the conformal transformation (\ref{Tdef}) introduced below should also act on these factors and so on, see \cite{superOPE} for details.  }
 These are the finite objects that we will consider to illustrate the OPE method. More precisely, we will consider a very particular set of components (i.e. choice of $ijkl\dots$) which have a particularly clean OPE interpretation. Namely we consider the components multiplying the monomial
\beq
\prod_{a=1}^{k} \eta_{i_a}^1 \eta_{{i_a}+1}^2  \eta_{j_a}^3 \eta_{{j_a}+1}^4 \qquad \text{where} \qquad \begin{array}{l}
1<i_1< \dots < i_k \le m \\
m+1< j_k< \dots < j_1 \le n \label{equation5}
\end{array}
\eeq
The indices $i_a$ are associated to the bottom and the indices $j_a$ belong to the top part of the polygon. To leading order, a pair $ \eta_{i_a}^1 \eta_{{i_a}+1}^2$ corresponds to inserting a complex scalar $\phi^{12} \equiv Z$ at the bottom cusp $P_{i_a}
$ while the pair $ \eta_{j_a}^3 \eta_{{j_a}+1}^4$ corresponds to inserting the conjugate complex scalar $\phi^{34} \equiv \bar Z$ at the cusp $P_{j_a}
$ in the top. The grassmannian variables associated to the flux tube edges are $\eta_1^A$ and $\eta_{m+1}^A$ and they are never used. To leading order (at tree level), the scalars just propagate from the bottom to the top cusps respecting planarity, see figure \ref{Examples}.

For obvious reasons we denote such components as $Z\to \bar Z$ components. Below, we will consider the three $Z\to \bar Z$ components depicted in figure \ref{Examples} as our examples.
The generalization to other components poses no conceptual obstacle.\footnote{However, for the OPE purpose we do not consider components involving $\eta_1^A$ and $\eta_{m+1}^A$. That is we only consider components that do not involve insertions along the two null edges of the flux tube.}

To render the discussion simpler while retaining all the physics, we will from now on restrict our polygons to lie in an $\mathbb{R}^{1,1}$ subspace. The generalization to $\mathbb{R}^{1,3}$ poses no conceptual obstacle. The null polygons in $\mathbb{R}^{1,1}$ have their edges along two null directions: $x^+$ and $x^-$.
We use (\ref{squarepoints}) so that the points $x_1$ and $x_2$ are separated in the $x^-$ direction, $x_2$ and $y_2$ are separated in the $x^+$ direction and so on.

As mentioned above, by a conformal transformation we set the first cusp of the bottom of the polygon to be at position $P_1=(0,0)$ and the first cusp of the top to be located at $P_{m+1}=(1,1)$. These coincide with two of the cusps of a reference square, see figure \ref{squaref}.
Then we generate a family of polygons by acting on the bottom cusps (only!) with the \textit{time  translation} symmetry of the reference square. This generates a family of polygons parametrized by $T=e^{-2\tau}$ where the $x^+$ coordinates of the bottom cusps are given by
\beq
x_i^{+} \to x_i^+(T)= \frac{T \,x_i^+}{1-x_i^++T \,x_i^+} \,. \la{Tdef}
\eeq
This is a conformal transformation that leaves the points $0$ and $1$ invariant. We now have a family of Wilson loops $W(T)$. We will now study the ratio functions $r(T)$. It admits an OPE decomposition. At leading order
\beq\la{OPEdiscontinuity}
r(T) = 1 + g^2\[ \log(T) D(T)+\tilde D(T) \] + {O}(g^4)
\eeq
The function $D(T)$ is the one loop OPE discontinuity and arises due to the one loop correction to the energy of the flux tube excitations that propagate. To measure the energy of the excitation we should work out what the flux tube Hamiltonian is. 
This operator will act on the bottom or top Wilson lines which we now introduce. Since we are dealing with a ratio function there are two Wilson loops relevant for the ratio (\ref{Rdef0}):
(a) the supersymmetric Wilson loop and (b) the bosonic one or MHV loop. Accordingly there should also be two types of Wilson lines in both the top and the bottom. To leading order in the coupling these Wilson loops read
\begin{eqnarray}
\mathcal W_\text{bot} &=& \text{Tr}\[Z(0,0) \star  Z(x_{i_1}^+(T),x_{i_1}^-) \dots Z(x_{i_k}^+(T),x_{i_k}^-) \star Z(0,1)\,\star\,\]  \la{botS} \\
\mathcal W_\text{bot}^\text{MHV} &=& \text{Tr}\[ Z(0,0) \star Z(0,1)\,\star \,\]\la{botMHV}
\end{eqnarray} 
and similarly for the lines in the top. The $\star$ in these expressions indicates that the fields are connected by the fundamental Wilson lines along the bottom and top contours as depicted in the figures. 

At one loop, the Hamiltonian $\mathbb{H}$ will act on each pair of neighboring $Z$ insertions in these Wilson lines as $\mathbb{H}=g^2 \sum_{a=1}^L \mathcal{H}_{a,a+1}$ where $L$ is the number of scalars $Z$ on the bottom (or top) Wilson line. Hence we can focus on a single pair of neighboring insertions inside the trace. These can be either null separated or not. Lets first consider the case where they are null separated in the $x^+$ direction and omit the direction $x^-$ that plays no role,\footnote{Of course, the same discussion holds true with $x^+$ and $x^-$ interchanged.}
\begin{figure}[t]
\centering
\def\svgwidth{16cm}
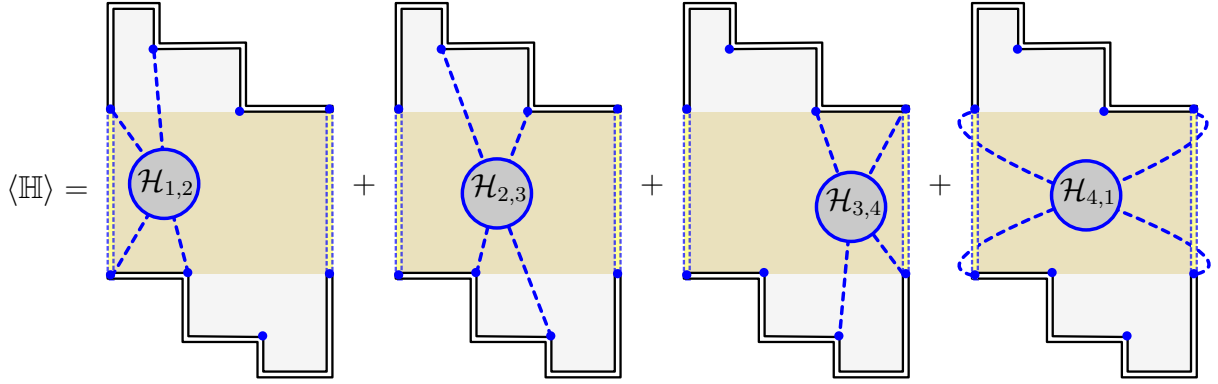
\caption{To compute any one loop OPE discontinuity one computes the expectation
value of the Hamiltonian $\<\mathbb{H}\>=\sum_{i=1}^{L}
\<\mathcal{H}_{i,i+1}\>$. For N$^k$MHV amplitudes of the type
described in the main text the length is $L=k+2$. In this figure the component $r^{(2,3,10,11)(4,5,8,9)}$ of the N$^2$MHV amplitudes for 12 gluons is represented.}\label{Hcircdodecagon}
\end{figure}
\beq\la{SL2kernel0}
\Tr\[\dots Z(0)\star Z(x)\dots\]=\Tr \[\dots\sum{x^k\over k!}Z(0) {D}_+^k Z(0)\dots\]+O(g)
\eeq
where $D_+$ is in the fundamental representation. These operators are very well studied. They can be mapped to the so called SL(2) subsector of the full $\mathcal{N}=4$ PSU(2,2|4) spin chain. The SL(2) chain is a non-compact chain where at each site we can have an arbitrary integer which is identified with the number of covariant derivatives. The corresponding Hamiltonian is discussed in appendix \ref{Spinchainapp}.\footnote{The energy we are interested in is the Twist which differs from the Dimensions by an integer, the Spin. Hence, since we are interested in the anomalous part, we can use the Dilatation operator of $\mathcal{N}=4$ SYM \cite{Beisert} to derive the flux tube Hamltonian.} Once we act with $\mathcal{H}$ on these two neighboring sites we can re-assemble  the result back into a Wilson line easily. This is also reviewed in this appendix. At the end of the day we end up with
\beq\la{SL2kernel}
\mathcal{H}\circ Z(0)\star Z(x) = \int\limits_0^1{dt\over t(1-t)}\[Z(0)\star Z(x)-tZ(0)\star Z(tx)-(1-t)Z(tx)\star Z(x)\]
\eeq
This kernel coincide with the renormalization of a null Wilson line with insertions and is a well known result \cite{Belitsky:2005qn,Belitsky}.

If the two insertions are not null separated then instead of (\ref{SL2kernel}) we will have now both kind of derivatives, $D_+$ and $D_-$. The action of the Hamiltonian is much less studied but it can be read from the Harmonic action of Beisert \cite{Beisert} as discussed in the appendix \ref{Spinchainapp}. 
In this case, we find
\beqa\la{SL2xSL2kernel}
&&\mathcal H\circ Z(0,0)\star Z(x,y)= \int\limits_0^1\!\!{dt\over t}\oint\!\!{dz\over2\pi iz}\oint\!\!{dw\over2\pi iw}\\
&&\qquad\,\,\,\,\, \Big[Z(0)\star Z(x,y) -Z(xt[1+z],yt[1+w])\star Z(x(1-t)[1+\frac{1}{z}],y(1-t)[1+\frac{1}{w}])\Big]\nn\\
&&\qquad\,\,\,\,\, +\[\dots\]\nn
\eeqa
Here, the $\star$ stands for a Wilson line connecting the two points. As we explain in the next section, the one loop result (\ref{SL2xSL2kernel}) does not depend on the path and one can drop the $\star$'s from the right hand side of (\ref{SL2xSL2kernel}).  The dots in the last line, $\[\dots\]$, involve insertions of the $Z$'s out of the $\mathbb{R}^{1,1}$ plane. It can be read easily from the Harmonic action \cite{Beisert} and do \textit{not} contribute when we Wick contract the result with two  $\bar Z$'s in ${\mathbb R}^{1,1}$, for more details see appendix \ref{Spinchainapp}. As a result, it will not contribute in the examples we will consider. 

After acting with the Hamiltonian as in (\ref{SL2xSL2kernel}) we can Wick contract the result with two $\bar Z$ fields. We get
\begin{eqnarray}
\frac{\left\<\right.
\contraction[2ex]{... \star}{\bar Z_d}{\star \bar Z_c\star \dots  \star (\mathcal H\circ Z_a \star}{ Z_b }
\contraction{... \star\bar Z_d\star}{ \bar Z_c}{\star \dots  \star (\mathcal H\circ}{ Z_a }
... \star\bar Z_d \star \bar Z_c\star \dots  \star (\Blue{\mathcal H}\circ Z_a \star Z_b 
)\star ...\left.\right\>}
{
\bcontraction[2ex]{... \star}{\bar Z_d\star}{ \bar Z_c\star \dots  \star  Z_a \star}{ Z_b\star }
\bcontraction{... \star\bar Z_d\star}{ \bar Z_c\star}{ \dots  \star }{ Z_a\star }
\left\< ... \star\bar Z_d \star \bar Z_c\star \dots \star Z_a \star  Z_b\star ...  \right\>} = \frac{(\chi^+ -1)\log\chi^+ -(\chi^- -1)\log\chi^-}{\chi^+ -\chi^-} +O(g^2)\la{actionofH}
\end{eqnarray}
where $Z_a= Z(x_a^+,x_a^-)$ and so on. Finally  $\chi^+$ is the cross ratio  
\beq\la{ccr}
\chi^+ =\chi^+_{abcd}= \frac{(x^+_a - x^+_d) (x^+_b -x^+_c)}{(x^+_a -x^+_c) (x^+_b - x^+_d)}.
\eeq
and similar for $\chi^-$. The result (\ref{actionofH}) will be our fundamental building block. The Hamiltonian is Hermitian so we could have acted on the pair $\bar Z_d\star \bar Z_c$ instead. If the two points in the bottom (or in the top) are null separated the result (\ref{actionofH}) simplifies considerably. For example, if $x_a^+=x_b^+$ we have $\chi^+=1$ and (\ref{actionofH}) reduces to the single logarithm $\log \chi^-$. This last result can of course be derived directly from the simpler representation (\ref{SL2kernel}). 

The OPE discontinuity in (\ref{OPEdiscontinuity}) comes from the correction to the energy of the bottom (or top) operators. We are computing a ratio function which means we should compute the average of $\mathbb{H}=\sum_{i=1}^L \mathcal{H}_{i,i+1}$ in the super loop and subtract the same quantity in the bosonic loop. 
This leads to 
\beq
D(T)=\frac{\langle \text{top}|\mathbb H|\text{bot}\rangle}{\langle \text{top}|\text{bot}\rangle}-\frac{\langle \text{top}|\mathbb H|\text{bot}\rangle_\text{MHV}}{\langle \text{top}| \text{bot}\rangle_\text{MHV}}\,. \la{finD}
\eeq
To compute the first average we  sum over the action of $\mathcal{H}$ on all nearest neighbor scalars in the bottom of the super Wilson loop as depicted in figure \ref{Hcircdodecagon}. The second average is computed similarly using the bosonic loop.
For the action on each such pair we use (\ref{actionofH}). 

It is now very simple to compute the OPE discontinuity of any component with an arbitrary number of $Z$ fields in the bottom. The Hamiltonian density ${\mathbb H}$ acts locally on a pair of insertions at a time and the total number of insertions is irrelevant. This means we now have control over multiparticle states. The first interesting multiparticle states appear for the N$^2$MHV amplitudes at tree level. We can now easily promote them to loop level. 

\begin{figure}[t]
\centering
\def\svgwidth{11cm}
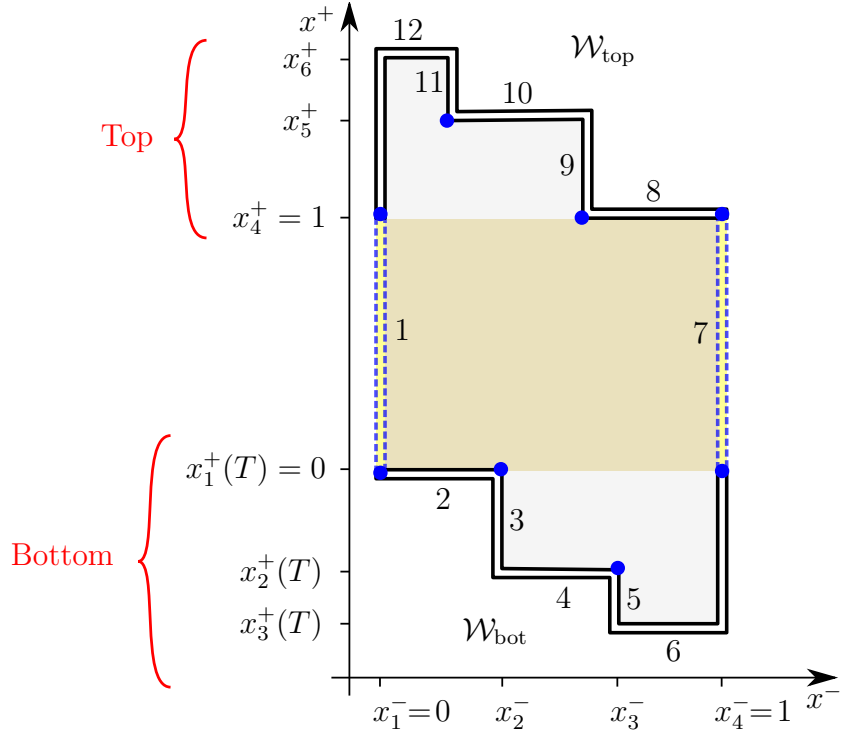
\caption{The Dodecagon Wilson Loop. We generate a family of polygons with
twelve edges $W(T)$ by acting with a conformal transformation (twist) on the bottom
part of the polygon (\ref{Tdef}). That conformal transformation leaves the square invariant and correspond to the ``time" translation of the OPE. By a global conformal transformation we
always set the reference square to be given by the cusps
$(0,0)$,$(0,1)$,$(1,1)$ and $(1,0)$. In particular this fixes the
position of two of the cusps of the polygon, in this figure cusps $12$
and $78$. In the main text we considered the component
$r^{(2,3,10,11)(4,5,8,9)}$. At leading order (tree level) this
N$^2$MHV Wilson loop is given by the propagation of the two scalars from
the two cusps in the bottom to those two in the top. These scalars are denoted by the blue dots in this figure.}\label{Dodecagon}
\end{figure}

For example, for the N$^2$MHV component $r^{(2,3,10,11)(4,5,8,9)}$ of the 12 gluon amplitude depicted in figure \ref{Dodecagon}, we get the one loop discontinuity
\beq
D(T) = \frac{(\chi^-_{2365}-1)\log\chi^-_{2365}-(\chi^+_{1254}-1)\log\chi^+_{1254}}{\chi^-_{2365}-\chi^+_{1254}}+\log\(\chi^-_{1346}\chi^-_{2361}\chi^-_{1543}\), \la{discN2MHV}
\eeq
This is worked out more carefully in section \ref{dodecagondetail}. The result (\ref{discN2MHV}) agrees precisely with the unpublished results by Simon Caron-Huot and Song He!

Following \cite{superOPE}, it should be simple to fully bootstrap all N$^2$MHV amplitudes at one loop by considering the OPE expansion  in several channels and making use of the SUSY Ward identities \cite{Elvang:2009wd}.  We can also work out the $\log(T)^2$ discontinuity at two loop level (and so on at higher loops) by acting twice (and more times) with the one loop Hamiltonian on the bottom of the loop. Later we will consider one such two loop example for an octagon N$^2$MHV component. These will be new predictions.
 
\section{Generalizations and Comments}\la{sec3}

In the previous section we have used the known result for the PSU(2,2|4) spin chain Hamiltonian acting on the single trace operators to read the flux tube Hamiltonian acting on the bottom or top Wilson loops. In this section we will elaborate on the details of the map between the Wilson loop and the single trace operators. We will start from the vacuum $W^\text{(square)}$. We will then move to the full polygon loop $W$, expressed as a sum of excitations on top of the vacuum. The reader who is most interested in further applications and examples can skip this section and move directly to section \ref{sec4} in a first reading. 

The reference square representing the flux tube vacuum is parametrized by the four points (\ref{squarepoints}).  As in (\ref{corrtoWilson}) we have
\beq
\lim_{(x_i-y_i)^2\to0}\frac{\<{\cal W}_\text{top}^\text{(square)}{\cal W}^\text{(square)}_\text{bot}\>}{\<{\cal W}^\text{(square)}_\text{top}{\cal W}^\text{(square)}_\text{bot}\>_\text{tree}}=\<W^\text{(square)}\>^2 \la{squareW}
\eeq
The square bottom and top operators are simple null Wilson lines given by
\beqa\la{square}
{\cal W}_\text{bot}^\text{(square)}&=&\Tr\[Z e^{(1-\epsilon) D_-}Z\]=\sum_{m=0}^\infty\left.{(1-\epsilon)^m\over m!}\,\Tr\!\!\[ZD_-^mZ\]\right |_{(0,0)}\\ {\cal W}_\text{top}^\text{(square)}&=&\Tr\[\bar Z e^{-(1-\epsilon)D_-}\bar Z\]=\sum_{m=0}^\infty \left.{(\epsilon-1)^m\over m!}\,\Tr\!\!\[\bar Z D_-^m\bar Z\]\right|_{(1,1)}\nn
\eeqa
where we have Taylor expanded the Wilson loop operators in terms of single trace operators inserted at $x_1=(0,0)$ and $y_2=(1,1)$. That is, in (\ref{square}), we have mapped the expectation value of the Wilson loop $W^\text{(square)}$ into a sum of two point functions of local operators. 

Here, $\epsilon=(x_i-y_i)^2$ is the null regulator. That is, $\<{\cal W}^\text{(square)}_\text{top}{\cal W}^\text{(square)}_\text{bot}\>_\text{tree}={1\over\epsilon^2}+O(1)$. In (\ref{square}) we have an arbitrary number of $D_-$ derivatives acting on the second operator. However, the pole in the correlator (\ref{corrtoWilson}) is dominated by the terms with infinite number of derivatives, displacing $\cO$ from $x_1$ to $x_2$. These are exactly the large spin operators with finite twist that are dual to the flux tube.\footnote{The map (\ref{square}) also give us an alternative definition of the null correlator (\ref{squareW}) (and (\ref{corrtoWilson})). Instead of taking the null limit $\epsilon=(x_i-y_i)^2\to0$, we can regulate the null correlators by putting a cutoff on the spin $S$ of the operators in these sums. Then, the double pole divergence ${1\over\epsilon^2}$ is replaced by an $S^2$ divergence. When constructing UV safe ratios all these divergences cancel out leaving behind regulator independent quantities.\la{footnoteCut}}

\begin{figure}[t]
\centering
\def\svgwidth{8cm}
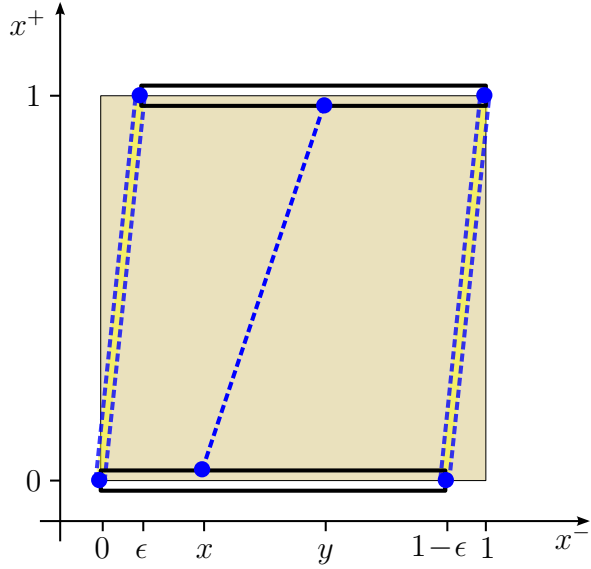
\caption{The square loop obtained as a correlation function of ${\cal W}_\text{top}^{(\text{square})}$ and ${\cal W}_\text{bot}^{(\text{square})}$. We add a scalar $\Blue{Z}(x)$ on the bottom line and a conjugate scalar $\Blue{\bar{Z}}(y)$ on the top. At tree level, these two scalar insertions are connected by a free propagator $1/(y-x)$. 
The null correlator $\<{\cal W}_\text{top}^{(\text{square})}{\cal W}_\text{bot}^{(\text{square})}\>$ is regulated by placing the bottom right scalar at $x^-=1-\epsilon$ and the top left scalar at $x^-=\epsilon$.}\label{OneZ}
\end{figure}

Suppose we now add one scalar $Z$ insertion at $x^-=x$ on the bottom of the square and one $\bar Z$ insertion at $x^-=y$ on the top as prescribe in figure \ref{OneZ}. That is, we now consider the bottom and top operators
\beqa\la{oneZ}
{\cal W}_\text{bot}&=&\sum_{m=0}^\infty\left.{(1-\epsilon)^m\over m!}\sum_{k=0}^\infty{x^k\over k!}\,\Tr\!\!\[ZD_-^kZD_-^mZ\]\right |_{(0,0)}\\
{\cal W}_\text{top}&=&\sum_{m=0}^\infty \left.{(\epsilon-1)^m\over m!}\sum_{k=0}^\infty{(1-y)^k\over k!}\,\Tr\!\!\[\bar Z D_-^kZD_-^m\bar Z\]\right|_{(1,1)}\nn
\eeqa
We now have
\beq\la{Zprop}
\<W\>_\text{tree}=\lim\limits_{\epsilon\to0}\frac{\<{\cal W}_\text{top}{\cal W}_\text{bot}\>_\text{tree}}{\<{\cal W}^\text{(square)}_\text{top}{\cal W}^\text{(square)}_\text{bot}\>_\text{tree}}={1\over y-x}
\eeq
That scalar propagator (\ref{Zprop}) represents the free propagation of the scalar $Z$ excitation from the bottom to the top. The flux tube energy of that excitation is the twist of the corresponding operator ${\cal W}_\text{bot}$ or ${\cal W}_\text{top}$, minus that of the vacuum ${\cal W}_\text{square}$. At tree level, it is just equal to $E=\Delta-S=1+O(g^2)$. The spin cannot get any loop corrections. Therefore, loop corrections to the energy of the flux tube excitations are the same as the anomalous dimension of the corresponding operator. 

To compute the one loop correction to $\<W\>$ due to the energy of the flux tube state, it is convenient to represent the operators ${\cal W}_\text{bot}$ and ${\cal W}_\text{top}$ as spin chain states. We will denote these by $\<\text{top}|$ and $|\text{bot}\>$. The standard representation for the single trace operators in (\ref{oneZ}) as SL(2) spin chain states simply counts the number of derivatives
\beq\la{SL2}
|n_1,\dots,n_L\>\equiv \frac{1}{n_1! \dots n_L!} {\rm Tr}\Big[D_-^{n_1} Z \dots D_-^{n_L} Z\Big] 
\eeq
So we have, for example
\beq
|\text{bot}\>=\sum_{m,k=0}^\infty (1-\epsilon)^mx^k|0,k,m\>\ ,\qquad\<\text{top}|=\sum_{m,k=0}^\infty (\epsilon-1)^m(1-y)^k\<m,k,0|
\eeq
Then, the Wick contraction of the fields between two operators is represented by the overlap of the corresponding spin chain states, for more details see appendix \ref{Spinchainapp}.\footnote{The states in (\ref{SL2}) would have been an orthonormal basis provided that the two operators were inserted at zero and infinity. However, we have chosen to insert the two operators at zero and one. The two choices are related by a conformal transformation that relates the two frames. The corresponding orthonormal base in our (1,0) frame reads  \label{FOOTNOTE}
\beq
|n\rrangle=e^{L_{+1}}|n\>=\sum^n_{k=0}\(\!\!\begin{array}{c}n\\k\end{array}\!\!\)\mathcal |k\>\ ,\qquad\llangle m|n\rrangle=\delta_{m,n}
\eeq
The Hamiltonian commute with the SL(2) generator $L_{+1}$ and is blind to such a change of basis.} The correction to $\<W\>$ due to the one loop energy of the states  is given by $\<\text{top}|{\mathbb H}\circ\text{bot}\>$ where ${\mathbb H}$ is the one loop spin chain Hamiltonian. In that way, we obtain the kernels (\ref{SL2kernel}) and (\ref{SL2xSL2kernel}).  

The one loop Hamiltonian kernel (\ref{SL2xSL2kernel}) acts on a pair of insertions that are separated both in the $x^+$ and $x^-$ directions. As stated before, it does not depend on the path connecting the insertion (parametrized by the "$\star$" in (\ref{SL2xSL2kernel})). That sounds strange at first since the gauge field in the Wilson line connecting the two points does contribute to the one loop Hamiltonian. To better understand what is going on, consider one such pair of insertions connected by a Wilson line $Z(0,0)W[(0,0),(x,y)] Z(x,y)$. We first Taylor expand the Wilson line in covariant derivatives acting at $(0,0)$. Different paths correspond to different ordering of the covariant derivatives at $(0,0)$. For example, one of the operators in the expansion can be
\beq\la{ordering}
ZD_+D_-Z={1\over2}Z\{D_+,D_-\}Z\ +\ {1\over2}\,g\,ZF_{+-}Z
\eeq
Now the main point is that the two operators on the right hand side cannot be mixed by the one loop Hamiltonian! Even though they carry the same charges, they transform in different representations. The first is represented by two sites of a spin chain while the second is represented by three sites. At one loop, the spin chain Hamiltonian does not change the length of the chain (this does happen at higher loops \cite{Beisert:2003ys}). As such, the second term with the $F_{+-}$ insertion can only contribute at higher loops and can therefore be dropped. The first term, with the derivatives symmetrized, is independent of the order of derivatives we started with on the left hand side. The same argument applies for any number of derivatives and hence the result is independent of the path.   

In the previous section, we concentrated on computing the leading OPE discontinuity for non-MHV amplitudes. These are simple as they do not depend on the shape of the loop as explained above. The shape of the loop is important when considering the sub leading OPE discontinuities at higher loops or already at leading order when considering MHV amplitudes. For describing these we will now consider the bosonic loop dual to MHV amplitudes.

\begin{figure}[t]
\centering
\def\svgwidth{17cm}
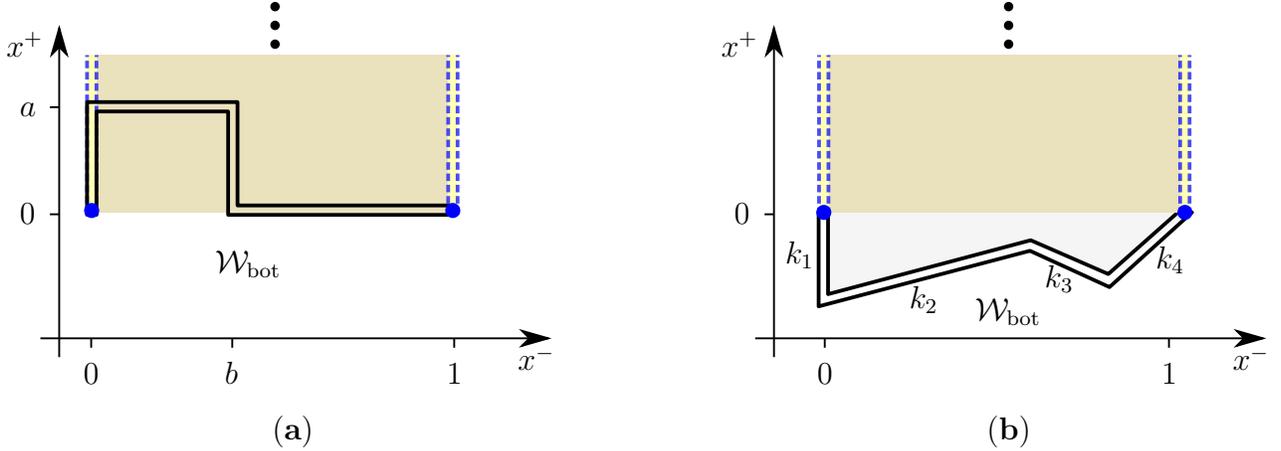
\caption{Bottom Wilson lines. The two endpoints of the bottom and top Wilson
lines are null separated from two other points in the top part of the
polygon. The dashed lines represent two propagators that connect the
two bottom endpoints to the two top endpoints. The two bottom
endpoints (and similar for the top) are also null separated. However
we choose them to be charged in such a way that there is no propagation
from the left to the right. The simplest way to do this is to set the
bottom insertions to be the complex $Z$ scalars and the top insertions to
be their complex conjugate $\bar Z$. Note that the bottom part of the
Wilson loop is made of the edges of the bottom plus a finite part of
one of the two null edges that source the flux tube. That is, $k_1$ in
figure (b) is not the full edge.}\label{bottom}
\end{figure}
Let us concentrate on the operator at the bottom, ${\cal W}_\text{bot}$. Suppose that instead of the single edge on the bottom of the square we consider a bottom with four edges $\{k_1,k_2,k_3,k_4\}=\{(a,0),(0,b),(-a,0),(0,1-b)\}$, see figure \ref{bottom}.a. The corresponding bottom operator is located at $(0,0)$ and reads
\beqa\la{MHVexp}
{\cal W}_\text{bot}-{\cal W}_\text{bot}^\text{(square)}\!\!\!&=&\!\!\!\Tr\[ Z e^{ a D_+}e^{bD_-}e^{ -aD_+}e^{(1-b)D_-}Z\]\ -\ \Tr\[ Z e^{D_-}Z\]\la{label}
\\ \!\!\!
\!\!\!&=&\!\!\!g\sum_{m=0}^\infty {1^m\over m!} \sum_{n=1}^\infty\,{\,b^n \over n!} \,\sum_{k=1}^{\infty} {a^k\over k!}\,\,\Tr\[ Z [(D_{(-}^{n-1}D_{+)}^{k-1}F_{+-}), (D_-^{m}Z)]\]\nn+\dots\nn
\eeqa
where we have subtracted from  ${\cal W}_\text{bot}$ the vacuum ${\cal W}_\text{bot}^\text{(square)}$. In the second line we have expanded the difference to leading order in the {\it number of excitations}. I.e., to leading order in the number of the corresponding spin chain sites. That leading term in (\ref{label}) is basically ``Abelian". It is equal to the integral of $F_{+-}$ on the square of sides $a$ and $b$ as dictated by the Stokes theorem. Once contracted with the top, this term correspond to a single particle insertion on top of the flux. Note that the covariant derivatives in $D_{(-}^{n-1}D_{+)}^{k-1}F_{+-}$ are symmetrized. They correspond to a descendant of the primary $F_{+-}$ represented on a single site of the spin chain. Finally, the two terms in the commutator in (\ref{MHVexp}) correspond to the insertions on the front and back copies of the bottom Wilson line, as indicated by the double line in the figure \ref{bottom}.

At the next order we have operators with two excitations
\beq\la{twoexcitations}
\left.{\cal W}_\text{bot}\right|_\text{Two Excitations}=g^2\sum_{\alpha\beta\gamma\delta} c_{\alpha\beta\gamma\delta}  \,\Tr  Z  \[\mathcal{F}^{\alpha\beta},\[\mathcal{F}^{\gamma\delta}, e^{D_-}Z \]\]  \, ,
\eeq
where $\mathcal{F}^{\alpha\beta} \equiv b^\alpha a^\beta D_{(-}^{\alpha-1}D_{+)}^{\beta-1}F_{+-} $.
The coefficents $ c_{\alpha\beta\gamma\delta}$ are obtained from (\ref{MHVexp}) by symmetrizing the derivatives in the $(+)$ and $(-)$ directions, keeping only the terms with two $F_{+-}$'s. The general expression can be neatly expressed as a sum over paths with integration over a surface bounded by the path, but it does not seem very illuminating.\footnote{
For the reference we quote only the first few terms in the expansion (\ref{twoexcitations}):  
$c_{1111}=3/8 $, $c_{1121}=1/4 $, $c_{2111}=1/9 $, $c_{1112}=1/6 $, $c_{1211}=2/9 $ etc. } 
Once contracted with the top, the operator (\ref{twoexcitations}) correspond to two flux tube excitations. 

Let us finish this section by enumerating the three types of loop corrections in our approach entering the Ratio function (\ref{rratio}) or its bosonic counterpart, the Reminder function. 
\begin{enumerate}
\item {\bf Corrections to the energies of the flux tube excitations}. These are obtained by acting with the flux tube Hamiltonian ${\mathbb H}$ on the bottom or top states. At one loop, ${\mathbb H}$ only acts on the nearest neighbors but at higher loops the range of interaction grows.
\item {\bf Bare form factor}. These are obtained by first Taylor expanding the bottom and top Wilson operators. For example, for the bosonic bottom Wilson operator parametrized by an ordered set of edges\footnote{A small comment: Since we have fixed $x_1$ and $x_2$ to be light-like separated, the first (or the last) edge of the bottom is not the full edge of the original polygon but only part of it, see figure \ref{bottom}; similarly for the top operator.} $k_1,k_2,\dots,k_s$ (see figure \ref{bottom}.b), that expansion takes the form
\beqa\la{maptochain}
{\cal W}_\text{bot} =\Tr\[Ze^{k_1 \cdot D}  e^{k_2 \cdot D}  \dots e^{ik_s \cdot D} Z\] =\sum_{m_1=0}^\infty\dots \sum_{m_s=0}^\infty\Tr\[Z{(k_1\cdot D)^{m_1}\over m_1!}\dots {(k_s\cdot D)^{m_s}\over m_s!}  Z\]
\eeqa
Then, by symmetrizing the covariant derivatives we expand (\ref{maptochain}) in the number of flux tube excitations or spin chain sites. This expansion includes contributions at any loop order.
\item {\bf Renormalization of the bare form factors}. The local single trace operators obtained in the Taylor expansion of the Wilson operators (\ref{label}), (\ref{maptochain}) are not normal ordered. As a result, the decomposition of the bottom and top in flux tube states also gets quantum corrections due to self contractions of the fields in the trace. We hope to come back to this type of corrections in the future. 
\end{enumerate}

\section{Examples in Greater Detail} \la{sec4}
At one loop order, the ratio function can be decomposed as in (\ref{OPEdiscontinuity}). Similarly, at any loop order, we have
 \begin{figure}[t]
\centering
\def\svgwidth{17cm}
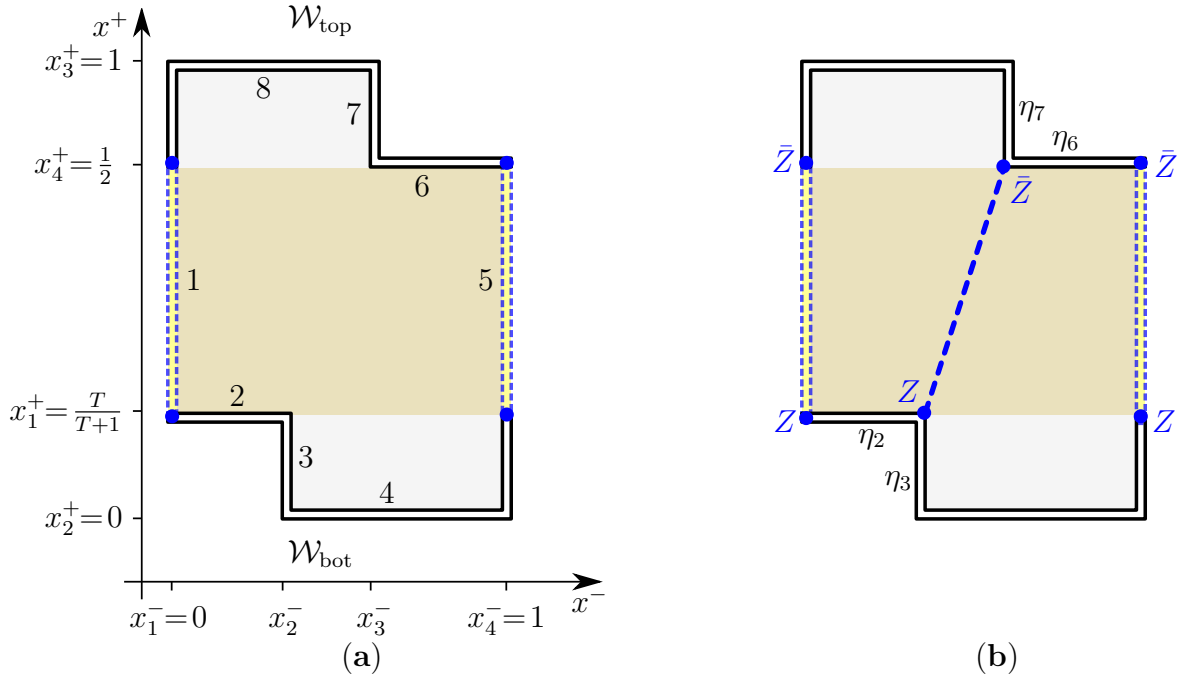
\caption{Left: The octagon in $\mathbb{R}^{1,1}$ kinematics. The shaded region
is the reference square. In the $x^+$ direction we used the $SL(2)_+$
subgroup of the conformal group to fix three of the points. The last
one is parametrized as in (\ref{Tdef}). Right: The $r^{(2367)}$
component at tree level. At tree level is corresponds to a scalar
insertion at the cusp $23$ that propagates from the bottom to the top
cusp $67$. There are also two other scalars at the endpoints of the
bottom and top Wilson lines that generate the left and right fast
particles that complete the two Wilson lines into a closed Wilson
loop; these fast particles source the flux tube on top of which the
middle scalar propagates.}\label{Octagon}
\end{figure}
\beq
r= \sum_{l=0}^\infty \frac{g^{2l}}{l!} \sum_{m=0}^{l} \log(T)^m D_m^{(l)}(T)
\eeq
where $D_m(T)$ admits a regular expansion around $T=0$. This follows from the general arguments of \cite{OPEpaper} generalized to the supersymmetric loops, see e.g. section 5 in \cite{superOPE}.
At each loop order the $m=l$ term is called the \textit{maximal OPE discontinuity} and can be computed easily. It is simply given by 
\beq
  \frac{\< \text{top} |e^{-2g^2\tau \mathbb{H}}|\text{bot}\>}{\< \text{top} |e^{-2g^2\tau \mathbb{H}}|\text{bot}\>_\text{MHV}} /   \frac{\<\text{top}  | \text{bot}\>}{\< \text{top}| \text{bot}\>_\text{MHV}}=
  \sum_{l=0} \frac{\[-2 \tau g^2\]^l}{l!} \Big( D_l^{(l)}(T) + O(g^2) \Big)\la{exp1}
\eeq
 For example,  
\beqa
D_1^{(1)} &=&   \frac{\langle \text{top}|\mathbb H|\text{bot}\rangle}{\langle \text{top}|\text{bot}\rangle}-\frac{\langle \text{top}|\mathbb H|\text{bot}\rangle_\text{MHV}}{\langle \text{top}| \text{bot}\rangle_\text{MHV}}\,, \la{finD1}\\
D_2^{(2)} &=&  \frac{\langle \text{top}|\mathbb H^2|\text{bot}\rangle}{\langle \text{top}|\text{bot}\rangle}-\frac{\langle \text{top}|\mathbb H^2|\text{bot}\rangle_\text{MHV}}{\langle \text{top}|\text{bot}\rangle_\text{MHV}}-2D^{(1)}_1 \frac{\langle \text{top}|\mathbb H|\text{bot}\rangle_\text{MHV}}{\langle \text{top}|\text{bot}\rangle_\text{MHV}} \la{finD2}
\eeqa
These averages can all be easily computed as explained after (\ref{finD}) and elaborated further below. As already emphasized above, a great advantage of writing things as in (\ref{exp1}) is that the single particle and the multiparticle states are treated on the same footing. 
We have already worked out one example of a one loop discontinuity $D_1^{(1)}$ above, see (\ref{discN2MHV}). In this section we will provide more details on that computation and also present a few other examples which will illustrate a few novel structures. 

\subsection{NMHV Octagon - Single Particle.} \la{octsec}
As a first example we consider the octagon NMHV ratio function $r^{(2367)}$. For NMHV amplitudes we have, at leading order, a single scalar propagating from bottom to top, see figure \ref{Octagon}b. In figure \ref{Octagon}a we have used the $SL(2)_+$ conformal symmetry to fix three of the $x^+$ coordinates. The remaining cross-ratio is parametrized by the fourth coordinate which is given in terms of $T$ as discussed around (\ref{Tdef}). For this example, we can choose the reference square as depicted in figure \ref{Octagon}. 
\subsubsection{One loop} \la{oct1}
All the insertions are located at the bottom and top of the reference square. As explained before, in this case, (\ref{actionofH}) reduces to a single logarithm
\begin{eqnarray}
\frac{\left\<\right.
\contraction[2ex]{... \star}{\bar Z_d}{\star \bar Z_c\star \dots  \star (\mathcal H\circ Z_a \star}{ Z_b }
\contraction{... \star\bar Z_d\star}{ \bar Z_c}{\star \dots  \star (\mathcal H\circ}{ Z_a }
... \star\bar Z_d \star \bar Z_c\star \dots  \star (\Blue{\mathcal H}\circ Z_a \star Z_b 
)\star ...\left.\right\>}
{
\bcontraction[2ex]{... \star}{\bar Z_d\star}{ \bar Z_c\star \dots  \star  Z_a \star}{ Z_b\star }
\bcontraction{... \star\bar Z_d\star}{ \bar Z_c\star}{ \dots  \star }{ Z_a\star }
\left\< ... \star\bar Z_d \star \bar Z_c\star \dots \star Z_a \star  Z_b\star ...  \right\>} =  \log \frac{(x^-_a - x^-_d) (x^-_b -x^-_c)}{(x^-_a -x^-_c) (x^-_b - x^-_d)}  \,.\la{actionofHsl2}
\end{eqnarray}
The two insertions at the endpoints of the bottom Wilson line are at $x^-=0$ and $x^-=1$. Similarly, the two insertions at the top are also at $x^-=0$ and $x^-=1$; they are null separated from the bottom insertions. Hence, when acting with the Hamiltonian as in (\ref{actionofHsl2}) we will get UV divergent results. These are physical divergences due to a divergent anomalous energy of the flux tube vacuum. When we construct the ratio function we subtract off that energy and should end up with a finite conformal invariant quantity. To see explicitly how this comes about we regulate these divergences by shifting slightly the position of the top and bottom insertions. We put the two insertions at the endpoints of the bottom Wilson line at $x^-=0$ and $x^-=1-\epsilon$ and the insertions in the top at $x^-=1$ and $x^-=\epsilon$, see figure \ref{OneZ}.

From (\ref{finD1}) and (\ref{actionofHsl2}), one can immediately read off the two-point functions of the top and bottom Wilson lines in the presence of the Hamiltonian insertions,
\beqa
\<\mathbb H \>\!\!&=&\!\!  \< \mathcal{H}_{1,2} \>\qquad\qquad +\< \mathcal{H}_{2,3} \>\qquad \qquad \qquad\qquad\,\,\, +\< \mathcal{H}_{3,1} \> \nn\\
 \!\!&=&\!\! \log\frac{x^-_4 (x^-_2-\epsilon)}{(x^-_2-x^-_4) \epsilon}\,+ \log\frac{(1-x^-_2)(1-x^-_4-\epsilon)}{(x^-_2-x^-_4)\epsilon}\,\,+ \log\frac{2\epsilon-1}{\epsilon^2},\la{Havarage}\\
\<\mathbb H \>_\text{vacuum} \!\!&=&\!\!\<\mathbb H \>_\text{MHV}=  \< \mathcal{H}_{1,3} \>_\text{MHV}+\< \mathcal{H}_{3,1} \>_\text{MHV}=2\< \mathcal{H}_{1,3} \>_\text{MHV}=2\log\frac{2\epsilon-1}{\epsilon^2}\la{Hvacuum}
\eeqa
where 
\beq
\<\mathbb{H}\>_\text{MHV}\equiv \frac{\<\text{top}|\mathbb{H}|\text{bot}\>_\text{MHV}}{\<\text{top}|\text{bot}\>_\text{MHV}}\,,\qquad \<\mathcal{H}_{1,2}\>\equiv \frac{\<\text{top}|\mathcal{H}_{1,2}|\text{bot}\>}{\<\text{top}|\text{bot}\>} \,,
\eeq
and so on. As anticipated, we see that all epsilon divergences neatly cancel out when we subtract (\ref{Hvacuum}) from (\ref{Havarage}). At the end of the day, the one loop OPE discontinuity (\ref{finD1}) for this component is given by
\begin{eqnarray}
D^{(1)}_1 
=\log\Big[\frac{\chi^-_{1243}-1}{(\chi^-_{1243})^2}\Big]. \la{done1}
\end{eqnarray}
That is the coefficient of $\log T=\log\chi^+_{1432}$ in the one loop result. This agrees perfectly with the well known results for the NMHV ratio functions, see e.g. \cite{addpapers1}.

\subsubsection{Two Loops} 
\begin{figure}[t]
\centering
\def\svgwidth{10cm}
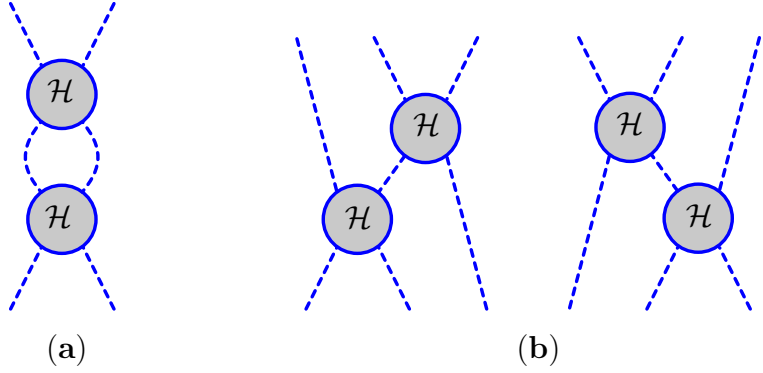
\caption{At two loops one needs to compute the expectation value of a product
of two Hamiltonian insertions. The two nontrivial cases are depicted
in this figure. In figure (a) we depict the case
$\<\mathcal{H}_{i,i+1}^2\>$ where the two insertions sit on top of each
other. In figure (b) we have the case
$\<\mathcal{H}_{i,i+1}\mathcal{H}_{i+1,i+2}\>+\<\mathcal{H}_{i+1,i+2}\mathcal{H}_{i,i+1}\>$
where they are slightly misaligned.}\label{Hcases}
\end{figure}

\la{oct2}
The $l$ loop maximal OPE discontinuity is obtained by simply acting with the Hamiltonian $l$ times, see (\ref{exp1}). For example, at two loops we need to deal with $\<\mathbb{H}^2\>=\sum_{i,j} \<\mathcal{H}_{i,i+1} \mathcal{H}_{j,j+1} \>$. If $i$ and $j$ are very separated then the two Hamiltonian densities do not communicate with each other and we can simply use the one loop results twice. If $i$ and $j$ are close there are two different cases: (a) the two Hamiltonians are on top of each other (for $i=j$) or  (b) the two are not aligned (for $i=j\pm1$), see figure \ref{Hcases}. 

For the NMHV component  $r^{(2367)}$ there is one further simplification: All bottom insertions are located at the same $x^+$, see figure \ref{Octagon}. The same is true for the top insertions. In this case we can use (\ref{SL2kernel}) and (\ref{actionofHsl2}) to readily obtain
the new building blocks mentioned in the previous paragraph, (see appendix \ref{Hav} for more details) 
\begin{eqnarray}
\frac{\left\<\right.
\contraction[2ex]{... \star}{\bar Z_d}{\star \bar Z_c\star \dots  \star (\mathcal H_{a,b}\circ Z_a \star}{ Z_b }
\contraction{... \star\bar Z_d\star}{ \bar Z_c}{\star \dots  \star (\mathcal H_{a,b}\circ}{ Z_a }
... \star\bar Z_d \star \bar Z_c\star \dots  \star (\Blue{\mathcal H_{a,b}^2}\circ Z_a \star Z_b 
)\star ...\left.\right\>}
{
\bcontraction[2ex]{... \star}{\bar Z_d\star}{ \bar Z_c\star \dots  \star  Z_a \star}{ Z_b\star }
\bcontraction{... \star\bar Z_d\star}{ \bar Z_c\star}{ \dots  \star }{ Z_a\star }
\left\< ... \star\bar Z_d \star \bar Z_c\star \dots \star Z_a \star  Z_b\star ...  \right\>}= -2\,\text{Li}_2(1-\chi_{abcd}^-),\la{SL2Li2}
\end{eqnarray}
and likewise,
\begin{eqnarray}
\frac{\left\<\right.
\contraction[2ex]{... }{\bar Z_f}{\star \bar Z_e\star\bar Z_d \dots   (\mathcal H_{b,c}\circ\mathcal H_{a,b}\circ Z_a \star Z_b\star}{ Z_c}
\contraction[1.5ex]{... \bar Z_f\star}{ \bar Z_e}{\star\bar Z_d \dots   (\mathcal H_{b,c}\circ\mathcal H_{a,b}\circ Z_a\star}{Z_b}
\contraction{... \bar Z_f\star\bar Z_e \star}{\bar Z_d}{ \dots   (\Blue{\mathcal H_{b,c}\circ\mathcal H_{a,b}} \circ}{Z_a}
... \bar Z_f\star\bar Z_e \star \bar Z_d \dots   (\Blue{\mathcal H_{b,c}\circ\mathcal H_{a,b}} \circ Z_a\star Z_b \star Z_c
) ...\left.\right\>}
{
\bcontraction[2ex]{... }{\bar Z_f\star}{ \bar Z_e\star\bar Z_d \dots    Z_a \star Z_b\star}{ Z_c\star }
\bcontraction[1.5ex]{... \bar Z_f\star}{ \bar Z_e\star}{\bar Z_d \dots   Z_a\star }{Z_b\star}
\bcontraction{... \bar Z_f\star\bar Z_e\star}{\bar Z_d}{ \dots}{Z_a\star}
\left\< ... \bar Z_f\star\bar Z_e \star \bar Z_d \dots  Z_a \star  Z_b\star Z_c ...  \right\>} = \log\chi_{abde}^-\log\chi_{bcef}^-+\text{Li}_2(1-\chi_{abef}^-)\la{SL2Li2Log}
\end{eqnarray}
where the cross ratios are defined as in (\ref{ccr}).
Now, to compute the OPE discontinuities at higher loops we need to sum over the average of all the possible combinations of the Hamiltonian insertions. At two loops, the following terms involved in (\ref{finD2}) are
\beqa
\langle \mathbb{H}^2 \rangle\!\! &=&\!\! \[\langle \mathcal{H}^2_{1,2}\rangle+\langle \mathcal{H}_{2,3}\mathcal{H}_{1,2}\rangle+\langle \mathcal{H}_{3,1}\mathcal{H}_{1,2}\rangle\]+\[\mathcal{H}_{i,i+1} \to \mathcal{H}_{i+1,i+2}\]+\[\mathcal{H}_{i,i+1} \to \mathcal{H}_{i+2,i+3}\]\,\nn\\
\< \mathbb H^2 \>_\text{MHV}\!\!&=&\!\! 4 \<\mathcal{H}^2_{1,3}\>_\text{MHV}
\eeqa
Putting together all the ingredients, one obtains the following expression for the maximal two loop OPE discontinuity (\ref{finD2})
\begin{eqnarray}
D^{(2)}_2 
= \log^2\Big[\frac{1-\chi^-_{1243}}{(\chi^-_{1243})^2}\Big]-\frac{1}{2}\log^2(1-\chi_{1243}^-)+\frac{2}{3}\pi^2, \la{done2}
\end{eqnarray}
In appendix \ref{usualOPE} we cross check (\ref{done1}) and (\ref{done2}) against the usual OPE promotion. It would be interesting to cross-check this prediction against (more) direct computations.

\subsection{N$^2$MHV Octagon. Two particles.}
We now move on to the 2-particle states. We consider the component $(\eta_2\eta_3\eta_7\eta_8)(\eta_3\eta_4\eta_6\eta_7)$ of the N$^2$MHV 8-point amplitudes, see figure \ref{Examples}. 

\subsubsection{One Loop}
The logic of section \ref{oct1} goes through without any modification for this case. The fact that we are now dealing with two particles instead of one makes no difference since all we have to do is consider the local insertions of the Hamiltonian densities; they do not care about the total number of insertions. The 1-loop OPE discontinuity obtained in this way reads
\begin{eqnarray}
D^{(1)}_1 = \log\Big[\frac{(\chi^-_{1243}-1)\chi^+_{1243}}{(\chi^-_{1243})^2}\Big].
\end{eqnarray}

\subsubsection{Two Loops}
\begin{figure}[t]
\centering
\def\svgwidth{12cm}
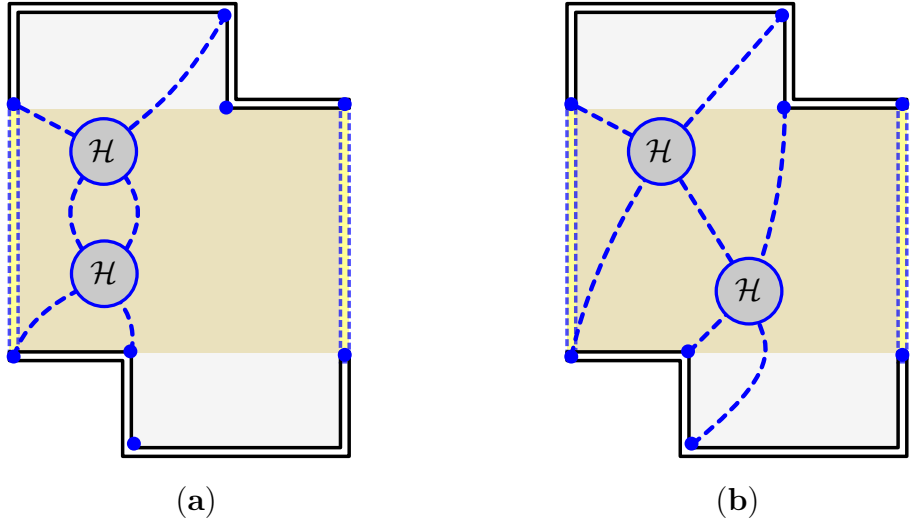
\caption{Examples of Hamiltonian insertions for the octagon N$^2$MHV Wilson
Loop at two loops. To compute the two loop OPE discontinuity we should
sum over all possible pairs of insertions.
For the octagon all cases exhibit an important simplification: at
least one of the two Hamiltonian insertions acts on a pair of null
separated points. For example, in figure (a) the bottom insertion acts
on two points null separated along $x^-$ while in figure (b) the
bottom insertion acts on two points null separated along $x^+$.
Because of this nice feature we can always use (\ref{SL2kernel})
followed by (\ref{actionofH}) and easily compute all building blocks,
see appendix \ref{Hav} for more details.}\label{examplesHH}
\end{figure}

We now move to two loops. At two loops we also need to compute $\<\mathcal{H}_{i,i+1}\mathcal{H}_{j,j+1}\>$. Two such examples are depicted in figure \ref{examplesHH}. For the octagon example under consideration, it turns out that in all cases that one needs to consider, one of the Hamiltonian insertions always acts on two insertions which are null separated. For example, in figure \ref{examplesHH}a the bottom Hamiltonian acts on the two insertions located at the same $x^+$ while in figure \ref{examplesHH}b it acts on the two at the same $x^-$. Hence to compute $\<\mathcal{H}_{i,i+1}\mathcal{H}_{j,j+1}\>$  we replace the action of one of the Hamiltonian insertions on the two null separated points by (\ref{SL2kernel}). This leads to an integration over $t$. Then we use (\ref{actionofH}) to get rid of the second Hamiltonian and finally we do the integral over $t$.\footnote{Note that if neither the top nor the bottom insertions were null separated this computation would be a bit more complicated. More precisely, after acting with the first Hamiltonian we would get two insertions which might leave the $\mathbb{R}^{1,1}$ plane. Then, the dots in the last line of (\ref{SL2xSL2kernel}) will also contribute. For all examples we will consider we can avoid dealing with them as explained in the text, see also section \ref{projections}. It would certainly be interesting to consider an example where these terms need to be taken into account.
}
 For example, 
\begin{eqnarray}
\frac{\<\text{top}|\mathcal H_{1,2}\mathcal H_{2,3}|\text{bot}\>}{\<\text{top}|\text{bot}\>} = \text{Li}_2\Big[1-\frac{x_4^- (x_2^- - \epsilon)}{(x^-_2 - x^-_4) \epsilon}\Big] - \text{Li}_2\Big[1-\chi^+_{1243}\frac{x_4^- (x_2^- - \epsilon)}{(x^-_2 - x^-_4) \epsilon}\Big] - \frac{1}{2}\log^2\chi^+_{1243}
\end{eqnarray}
where $\epsilon$ is a regulator, see section \ref{oct1}. When we add up all contributions and compute (\ref{finD2}) it drops out. 
All other contributions can be easily computed, see  Appendix \ref{Hav} for some useful building blocks. 
Summing up all the contributions as in (\ref{finD2}), the 2-loop OPE discontinuity for this component is predicted to take the following form,
\begin{eqnarray}\la{N2MHVOctD2}
D^{(2)}_2 =  \frac{1}{2}\log^2(1-\chi^-_{1243})-2\log\frac{1-\chi^-_{1243}}{\chi^-_{1243}}\log\chi^-_{1243}+\log\chi^+_{1243}\log\frac{1-\chi^-_{1243}}{(\chi^-_{1243})^2}-\text{Li}_2(1-\chi^+_{1243}).
\end{eqnarray}
In principle, there is no conceptual obstacle to generate in this way infinitely many higher loop predictions. 

\subsection{N$^2$MHV Dodecagon}\la{dodecagondetail}
Let us close this section by giving one final example, the component $r^{( 2 ,3 ,{10} ,{11})( 4 ,5 ,8 ,9)}$ of the N$^2$MHV amplitude for 12 particles at one loop, see figure \ref{Dodecagon}. The computation is basically the same as the previous two examples. But since now the scalar insertions are separated diagonally in both the top and bottom, the rational functions multiplying the logarithms in (\ref{actionofH}) survive and the OPE discontinuity at one loop leads to (\ref{discN2MHV}).

\section{The Monodromy Matrix} \la{MonoSec}
We have seen that it can be quite useful to think of the Wilson loop expectation value as a correlation function of two Wilson lines, a bottom and a top one. These Wilson lines can be expanded in local operators. These local operators can be thought of as non-compact spin chains and a great deal is known about these from Integrability \cite{Beisert:2010jr}. For example, above we used the knowledge of the action of the Dilatation operator on these spin chains to learn about higher loop predictions for Wilson loops. In particular we saw how to tame multi-particle corrections in the OPE. The Dilation operator acts as a spin chain Hamiltonian on the local operators and measures the energy of the state produced at the bottom and flowing to the top. This is one of the conserved charges of the spin chain. The $\mathcal{N}=4$ spin chains are integrable and there are many other interesting charges that are well understood and that one might envisage measuring and making use of. In Integrable models, the generating functions of such higher charges are the so called monodromy matrices and transfer matrices. In this section we shall initiate the study of these objects for Null Polygonal Wilson loop at weak coupling (to leading order in the 't Hooft coupling). 

For guidance and further motivation, it is very instructive to recall how the story goes at strong coupling where the full artillery of Integrability was put to use \cite{Alday:2010vh,AMapril,Hexagonpaper}. 
For large 't Hooft coupling, the Wilson loop is governed by a minimal surface area in $AdS$ \cite{AmplitudeWilson}. The main ingredients entering the solution to the strong coupling problem are holonomies of the flat connection on the worldsheet \cite{Alday:2010vh}. 
The world sheet has a disk topology and therefore the holonomies are not closed. Instead, they can end on an edge of the polygon, see figure \ref{TinAdS}. When restricting to ${\mathbb R}^{1,1,}$ kinematics, these holonomies go between two edges of the polygons in the same null direction. At each of the edges, they are contracted with a flat section associated to that edge. That section is called ``small solution" and is the unique flat section that is decreasing as we approach that edge. To summarize, schematically, the fundamental building blocks are then the holonomies $\<i,j\>$ connecting edges $i$ and $j$
\begin{eqnarray}
\< i,j \> \equiv s_L\cdot \varepsilon\cdot\Big[\mathcal{P}\exp\int\limits^{\infty_i}_{\infty_j}\mathcal A(u)\Big]\cdot s_R \,, \qquad s_{L}\equiv  s_i(\infty_i) \,, \,\,\, s_R\equiv s_j(\infty_j) \la{holonomy} \,.
\end{eqnarray}
Here $\mathcal A(u)$ is the flat connection used to transport the sections from one edge to the other. The connection depends on an arbitrary complex number $u$ called the spectral parameter. $s_i$ is the small section associated with edge $i$; $\infty_i$ is a point at edge $i$ and $\varepsilon$ is the $SL(2)$ metric $\varepsilon_{\alpha\beta}=\beta-\alpha$ with $\alpha,\beta=1,2$. 
Finally, the minimal area is some functional of these holonomies, see \cite{Alday:2010vh} for more details.

We will now try to identify the analogue of the holonomies and small solutions at leading order at weak coupling. 
The hope is that once the strong and weak coupling result are expressed in terms of the same ingredients, we may understand how to extend them to any value of the coupling.  

To present the main ideas we will focus on some particularly simple components of  the super Wilson loop  where the states created at the bottom (and absorbed at the top) admit a very transparent spin chain description. At the end we comment on how to generalize these considerations to any possible components and even present some results for the bosonic Wilson loop. 

\begin{figure}[t]
\centering
\def\svgwidth{7cm}
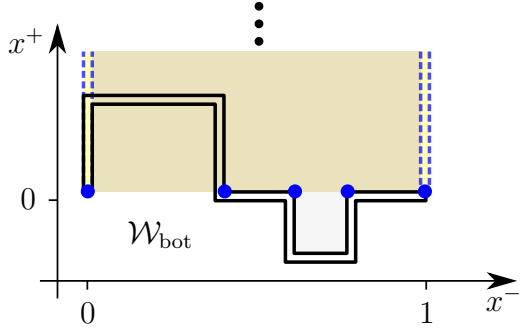
\caption{The bottom Wilson line along which $k$ scalars $Z$ are inserted at $x^+=0$.}\label{ZbarZsameXp}
\end{figure}
For simplicity, we start by considering the $Z \to \bar Z$ components introduced in (\ref{equation5}). Furthermore we will consider the case where all the insertions at the bottom are located at the same $x^+=0$ and the same for all the top insertions, inserted at $x^+=1$, see figure \ref{ZbarZsameXp}. In this case, to leading order at weak coupling, the bottom and top Wilson lines are simply\footnote{To be more precise, all the scalars $Z$ are dressed by extra pre-factors that carry helicity weights \cite{Simon, Skinner}. However in practice, we will be considering ratios constructed from these Wilson operators, for example (\ref{Rauxdef}), and these multiplicative factors drop out.}
\beqa\la{WLexpansion}
W_{bot} &=& {\rm Tr}\[Z(x_0,0) \star Z(x_1,0) \star \dots \star Z(x_k,0)\star Z(x_{k+1},0)\, \star\,\] \,, 
\eeqa
where $x_0=0$ and $x_{k+1}=1-\epsilon$ with $\epsilon$ being a regulator that will drop out in any physical quantity.  Equivalently, 
\beqa
W_{bot} = \sum_{n_j=0}^\infty \frac{(x_1)^{n_1}}{n_1!} \dots \frac{(x_k)^{n_k}}{n_k!}\frac{(1-\epsilon)^{n_{k+1}}}{n_{k+1}!}  {\rm Tr}\[ Z D_-^{n_1}  Z \dots D_-^{n_k} Z D_-^{n_{k+1}} Z \](0,0) \nn
\eeqa
and similarly for $W_{top}$. For N$^k$MHV components, we are dealing with spin chains of length $L=k+2$. In the OPE language, they correspond to $k$ particle states in the OPE flux tube. In the spin chain notation, the bottom operator simply reads
\beqa
|\text{bot}\> = \sum_{n_j=0}^\infty\Big[  (x_1)^{n_1} \dots (x_{k+1})^{n_{k+1}}  \Big]\left| 0,n_1,\dots,n_{k+1}\right\> \,, \la{spinchainexpansion}
\eeqa
see appendix \ref{Spinchainapp} for more details. Such spin chains are the so called $SL(2)$ spin chains where at each site we have a spin transforming in the spin $s=-1/2$ representation of $SL(2)$. 
The $SL(2)$ generators act as $L_a = \sum_{n=1}^L L_a^{(n)}$ where $L_a^{(n)}$ acts on site $n$. Acting on kets we have
\beqa\la{actspin}
\begin{array}{lcl}
L_{-1} |n\>&=&(n+1)|n+1\> \\
L_0\,\,\, |n\>&=&(n+1/2)|n\>  \\
L_{+1} |n\>&=& n |n-1\>
\end{array}
\qquad\text{or}\qquad
\begin{array}{lcl}
L_{-1} \circ Z(x,y)&=& \partial_{x} \qquad\qquad\, Z(x,y)  \\ 
L_{0}\,\,\, \circ Z(x,y)&=& (x\partial_{x}+1/2)\, Z(x,y) \\
L_{+1} \circ Z(x,y)&=& (x^2\partial_{x}+x) \,\,\,\,Z(x,y) 
\end{array}
\eeqa
when acting directly on fields. 
Of course, it is straightforward to go between these two representations. 
The action on kets 
is the convenient one if we want to use (\ref{spinchainexpansion}) while the action on the operators 
can be used to act directly on each of the fields in (\ref{WLexpansion}).

The monodromy matrix can be thought of as the scattering of an auxiliary ghost particle with the physical excitations. It is given by 
\beq
\hat{L} = \mathcal{R}_{1}(u) \dots \mathcal{R}_{L}(u)  \,, \qquad \mathcal{R}_j(u)\equiv R_j(u-i/2) \la{monoDef}
\eeq
where $u$ is the so called spectral parameter and where the $R$-matrix $R_{j}$ acts on a tensor product of two spaces: The physical space at the chain site $j$ and an auxiliary space $V_0$ associated to the auxiliary particle. The length $L=2k+2$, see (\ref{WLexpansion}). The most important property of the $R$-matrix is that it obeys the triangular relation known as Yang-Baxter. One can think about the $R$-matrix as a discretized/quantum version of the strong coupling holonomy. Yang-Baxter is the quantum analogue of the flatness condition at strong coupling. 

The monodromy matrix acts on the physical Hilbert space tensored with the space associated to an extra auxiliary particle. 
That is, denoting the indices of this auxiliary space by capital letters $A,B,\dots$ and  the indices of the physical sites by $i_a$, $j_a$, we have
\beq
\(\hat{L}_{i_1\dots i_L}^{j_1\dots j_L}\)_{A}^{B} = \sum_{A_1,\dots,A_L} \( \mathcal{R}_{1}(u) \)_{Ai_1}^{A_1j_1}\( \mathcal{R}_{2}(u) \)_{A_1i_2}^{A_2j_2} \dots \(\mathcal{R}_{L}(u) \)_{A_{L-1}i_L}^{Bj_L}
\eeq
The ghost particle transforms in some representation of $SL(2)$. Different representations correspond to different auxiliary spaces $V_0$ and define different monodromy matrices. Which one is the relevant one for our purposes? There are two obvious candidates that deserve to be analyzed. 

One natural choice is to take the representation of the auxiliary space to be the same as in the physical space. That is, the non-compact, unitary, spin $-1/2$ representation. In this case, $A,B,\dots$ are of the same nature as the physical indices $i_a$ and $j_a$ and take values $A,B=0,1,2,\dots$. The monodromy matrix in this representation is certainly a very important object. For example, the spin chain Hamiltonian and the next local higher charges can be obtained by expanding the (trace of the logarithm of the) monodromy matrix at a particular value of the spectral parameter. The $R$-matrix for this choice of auxiliary space is given in \cite{Beisert}. It would be very interesting to study this case in greater detail. In this paper we focus on the other natural choice, which is mostly motivated by the story at strong coupling. 

The second natural choice is to let the auxiliary space $V_0$ be the simplest/smallest it can be. Namely, we take the auxiliary particle to transform in the spin $1/2$ compact (and hence non-unitary) representation of $SL(2)$. In this case $V_0=\mathbb{C}^2$ and the auxiliary particle can either be spin up or down, that is $A,B=1,2$. At strong coupling, for polygons in $\mathbb{R}^{1,1}$ the small solutions were $2$-component spinors which were transported between edges using the flat connection in this same compact representation. This is the main motivation for this choice.\footnote{For the spectrum problem the connection between the monodromy and transfer matrices in compact representations at weak and strong coupling was beautifully studied in \cite{BKSZ1,BKSZ2}.} The spin $1/2$ SL(2) R-matrix is very simple and is given by \cite{FaddeevReview}
\beq\la{Rdef}
\qquad\def\svgwidth{14cm} 
\begingroup%
  \makeatletter%
  \providecommand\color[2][]{%
    \errmessage{(Inkscape) Color is used for the text in Inkscape, but the package 'color.sty' is not loaded}%
    \renewcommand\color[2][]{}%
  }%
  \providecommand\transparent[1]{%
    \errmessage{(Inkscape) Transparency is used (non-zero) for the text in Inkscape, but the package 'transparent.sty' is not loaded}%
    \renewcommand\transparent[1]{}%
  }%
  \providecommand\rotatebox[2]{#2}%
  \ifx\svgwidth\undefined%
    \setlength{\unitlength}{799.99516602bp}%
    \ifx\svgscale\undefined%
      \relax%
    \else%
      \setlength{\unitlength}{\unitlength * \real{\svgscale}}%
    \fi%
  \else%
    \setlength{\unitlength}{\svgwidth}%
  \fi%
  \global\let\svgwidth\undefined%
  \global\let\svgscale\undefined%
  \makeatother%
  \begin{picture}(1,0.1330008)%
    \put(0,0){\includegraphics[width=\unitlength]{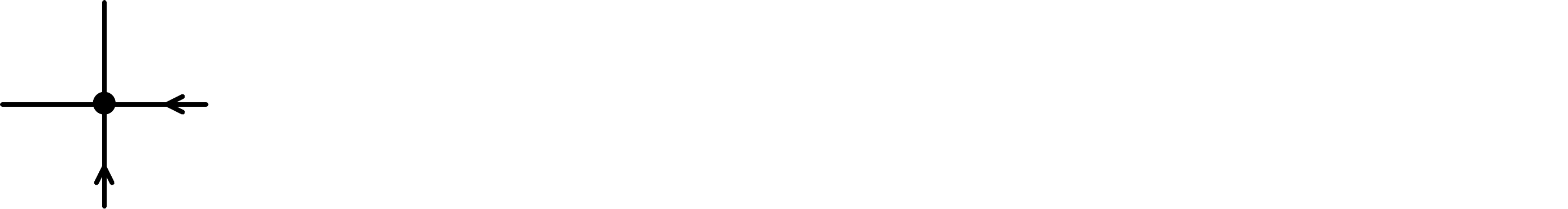}}%
    \put(0.1165007,0.07650046){\color[rgb]{0,0,0}\makebox(0,0)[lb]{\smash{$u$}}}%
    \put(0.07650046,0.00650004){\color[rgb]{0,0,0}\makebox(0,0)[lb]{\smash{$0$}}}%
    \put(0.15892943,0.05981054){\color[rgb]{0,0,0}\makebox(0,0)[lb]{\smash{$\displaystyle{=\mathcal{R}(u)={1\over{\sqrt{u^2+{1}/{4}}}}\[u\,\mathbb I^{(\text{aux})}\otimes \mathbb I^{(\text{phy})}+{i\over2}\sum_a {L}_a^{(\text{aux})} \otimes {L}_a^{(\text{phy})}\] }$}}}%
  \end{picture}%
\endgroup%

\eeq
where ${L}^{(\text{phys})}_a$ are the conformal generators acting on the physical space as in (\ref{actspin}). 
${L}^{(\text{aux})}_a$ are the conformal generators acting on the auxiliary space in the compact spin $1/2$ representation. That is, they are simple Pauli matrices 
\begin{eqnarray}
L^{(\text{aux})}_{-1} = \left(\!
\begin{array}{cc}
0 & 1\\
0 & 0
\end{array}
\!\right),\qquad
L^{(\text{aux})}_0 =\left(\!
\begin{array}{cc}
\frac{1}{2} & 0\\
0 & -\frac{1}{2}
\end{array}
\!\right),\qquad
L^{(\text{aux})}_{+1} =\left(\!
\begin{array}{cc}
0 & 0\\
-1 & 0
\end{array}
\!\right).
\end{eqnarray}
Using (\ref{actspin}) it is easy to read off the ``expectation value'' of the R-matrix after acting with it on a single scalar $Z(x)$ in the bottom spin chain and contracting the result with the corresponding scalar $\bar Z(y)$ in the top spin chain. This way we get rid of the physical space and what is left over is simply a $2\times 2$ matrix that acts on the auxiliary space only,\footnote{We omitted the dependence on the $x^+$ direction which plays no role and drops out.}
\begin{eqnarray}
\mathbb R_{x,y}(u) \equiv   \frac{\< \text{top}|R(u)|\text{bot}\>}{\<\text{top}|\text{bot}\>} = \frac{\contraction{}{\bar Z}{\mathcal{R}(u)\circ Z(x)}{} \bar Z(y) \(\mathcal{R}(u)\circ Z(x)\)}
{\bcontraction{}{\bar Z}{(y)}{ Z}{} \bar Z(y)  Z(x) } \la{Rauxdef}
\end{eqnarray}
which yields
\begin{eqnarray}\la{Raux}
\mathbb R_{x,y}(u) =\frac{1}{\sqrt{u^2+1/4}}
\left(\!
\begin{array}{cc}
u-\frac{i}{2}\frac{x+y}{x-y} & \frac{ixy}{x-y}\\
\frac{-i}{x-y} & u+\frac{i}{2}\frac{x+y}{x-y} 
\end{array}
\!\right) .
\end{eqnarray}

We can now construct the analogue of the most important building block in (\ref{holonomy}), namely the path ordered exponential integral of the flat connection. It corresponds to contracting the monodromy matrix (\ref{monoDef}) with the top and bottom operators.\footnote{At strong coupling the background is classical and hence we never have to discuss the contraction with the top and bottom operators, they are automatically done. Only the auxiliary space is visible. Hence the reduced R-matrix $\mathbb{R}(u)$, which only acts on the auxiliary space, is the analogue of the discretization of the strong coupling path ordered exponential. } These operators are made out of the several insertions at points $x_i$ and $y_i$. That is, we simply need to multiply the several $\mathbb R_{x_i,y_i}(u)$ operators as
\beq\la{holoWeak}
\mathbb R_{x_{0},y_{0}}(u)\cdot\mathbb R_{x_{1},y_{1}}(u)\cdot \ldots \cdot \mathbb R_{x_{k},y_{k}}(u)\cdot \mathbb R_{x_{k+1},y_{k+1}}(u) \equiv \Omega(u) \,. 
\eeq
Each of these (reduced) R-matrices propagates an auxiliary vector from one site to the next. In total this object carries a vector from the rightmost site to the first one. It can be thought as the analogue of the path ordered exponential in (\ref{holonomy}) used to carry the small solution at edge $j$ all the way to edge $i$. 

Next we move to the next ingredient in (\ref{holonomy}), the left and right small solutions $s_L$ and $s_R$. At strong coupling when we are close to an edge and propagate a spinor all the way to that edge there are two possibilities. The spinor can be tuned in such a way that, once propagated all the way to the boundary, it vanishes. If it is not tuned it will explode instead. The directions which lead to the vanishing results close to the left and right edges are $s_L$ and $s_R$ in (\ref{holonomy}). We can now repeat this analysis at weak coupling. The weak coupling analogue of the propagation from a point close to the edge all the way to the edge is given by the action of the left-most and right-most $\mathbb{R}$-matrices in (\ref{holoWeak}). 
So, for example, we define $s_L$ ($s_R$) as the spinor that is annihilated by the left-most (right-most) $\mathbb{R}$-matrix once we propagate it from the right (left), 
\beq
 \mathbb R_{x_0,y_0}(u)\cdot s_L = \mathcal{O}(\epsilon) \,, \qquad s_R\cdot\varepsilon\cdot \mathbb R_{x_{k+1},y_{k+1}}(u)=  \mathcal{O}(\epsilon)
\eeq
Of course, these spinors are defined up to an arbitrary normalization. 
Such spinors exist because these matrices are singular as $y_0-x_0=x_{k+1}-y_{k+1}=\epsilon \to 0$. (Also at strong coupling, once we are propagating all the way towards the null edge we approach the boundary of AdS and the warp factor leads to a divergent contribution.) 

Using (\ref{Raux}), we obtain \begin{eqnarray}
s_L = \left(\!\!\begin{array}{c}0\\1\end{array}\!\!\right) +\epsilon\left(\!\!\begin{array}{c}iu+\frac{1}{2}\\0\end{array}\!\!\right)\,, \qquad s_R=\left(\!\!\begin{array}{c}1\\1\end{array}\!\!\right)-\epsilon\left(\!\!\begin{array}{c}iu-\frac{1}{2}\\0\end{array}\!\!\right) \,.
 \la{sLsR}
\end{eqnarray}These are the small solutions when the edges are located at $x_0=0$ and $x_{k+1}=1$. We could repeat the exactly same computation for edges at arbitrary positions $x$. We find in that case 
\begin{eqnarray}
s_x =  \left(\!\!\begin{array}{c}x\\1\end{array}\!\!\right)+\mathcal{O}(\epsilon)\la{generics},
\end{eqnarray}
This result is a local property of the edge and is independent of whether we define it as acting from the right or from the left. The results for different $x$'s are related by conformal transformations that relate the different $x$'s.\footnote{The conformal transformation relating the points reads: $x\to \frac{ax+b}{cx+d}$. This corresponds to the conformal transformation relating the small solutions $s\to g\circ s$ where $g=\left(\!\!\begin{array}{cc}a &b\\ c& d\end{array}\!\!\right)$. This specific representation of $g$ follows from the fact that the physical and auxiliary space are entangled as in (\ref{Rdef}). For a related discussion see section \ref{gaugetrans}.} To study different OPE channels amounts to choosing different pairs of null edges. It would then be convenient to work with these more general small solutions. 

\begin{figure}[t]
\centering
\def\svgwidth{5cm}
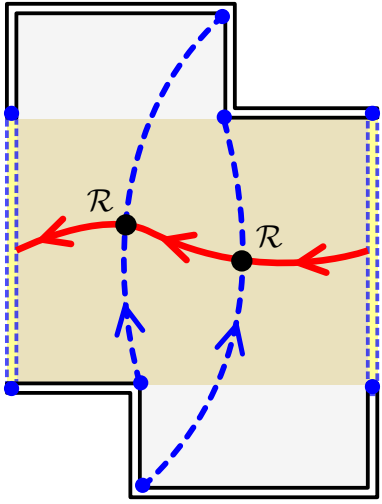
\caption{Weak coupling transfer matrix (\ref{voila}) represented by the solid red line. The dashed (blue) line are the physical space lines while the solid (red) line represents the auxiliary space. This is the analogue of the strong coupling holonomies represented in figure \ref{TinAdS}. }\label{TransferMatrix}
\end{figure}

We now have all the ingredients to write down the analogue of (\ref{holonomy}) at weak coupling. The first and last $\mathbb{R}$-matrices in (\ref{holoWeak}) are used to define the directions $s_L$ and $s_R$. The remaining $\mathbb{R}$-matrices in (\ref{holoWeak}) are the analogue of $\mathcal{P}e^{\int\mathcal A(u)}$  in (\ref{holonomy}). That is, we have 
\beq
\<L,R\>_\text{weak}=s_L\cdot\varepsilon\cdot \mathbb R_{x_{1},y_{1}}(u)\cdot\mathbb R_{x_{2},y_{2}}(u)\cdots\mathbb R_{x_{k},y_{k}}(u)\cdot s_R  \la{voilaPre}
\eeq
Of course, this depends on the overall normalization of the small solutions. We can construct a normalization independent quantity by dividing by their scalar product $s_L\cdot\varepsilon\cdot s_R$. This scalar product can be thought of as the overlap of $s_L$ and $s_R$ in the vacuum. That is, we end up with 
\begin{eqnarray}\la{voila}
 T(u)=  \frac{s_L\cdot\varepsilon\cdot \mathbb R_{x_{1},y_{1}}(u)\cdot\mathbb R_{x_{2},y_{2}}(u)\cdots\mathbb R_{x_{k},y_{k}}(u)\cdot s_R}{s_L\cdot \varepsilon\cdot s_R},
\end{eqnarray}
which we represented in figure \ref{TransferMatrix}.
This is the object that has a nice strong coupling analogue. 
It would be wonderful if we could now extend this object from weak and strong coupling to any intermediate coupling. Also, we would need to know how to extract the Wilson loop expectation value from such holonomies, like we do at strong coupling, see discussion at the end. 

To get some better intuition/familiarity with this object we now present a few more examples and comments. In particular, in section \ref{StrongWeakmatch}, we will explore an example with many $R$-matrices where we will be able to make \textit{direct} contact with strong coupling. 

\subsection{The Simplest Examples}
The simplest possible example is the vacuum. That is quite uninteresting. It corresponds to putting no insertions at all so that in (\ref{voila}) we have no $\mathbb{R}$-matrices and hence $T(u)=1$. The next to simplest example is the case with a single insertion. 
Let us take the component $\eta_2\eta_3\eta_7\eta_8$ of the octagon depicted in figure \ref{Octagon} as an example. There is one scalar insertion between edge 1 and edge 5. Therefore here the monodromy matrix contains only a single $\mathbb R(u)$ corresponding to this scalar. Then we have,
\begin{eqnarray}\la{TNMHVOctagon}
T_{1,5}(u) = \frac{1}{\sqrt{u^2+1/4}}\Big[u-\frac{i}{2}+i\chi^-_{1234}\Big],
\end{eqnarray}
where $\chi^-_{1234}$ is the cross ratio defined previously. 

\subsection{Gauge Transformations and Globally Defined Small Solutions}\la{gaugetrans}
The small solutions of the linear problem are the key ingredients in the computation of Null Polygon Wilson loops at strong coupling. In the previous section we identified the analogous definition of small solutions at weak coupling. More precisely, so far we only needed (and only computed)  the small solution $s_j$  close to the corresponding edge $j$, see (\ref{sLsR}). At strong coupling, the small solution is defined close to the corresponding edge but can then be propagated anywhere in the worldsheet. Similarly, at weak coupling, we can identify the dependence of the small solutions on the position in the spin chain, as we now explain. 

As mentioned above, the analogue of the propagation with the flat connection is given by the reduced $\mathbb{R}$-matrices (\ref{Raux}). Hence we can define the small solution $s_R^{(n)}$ at the $n$th site by propagating it $k-n$ times from its value at the rightmost end of the chain, defined in (\ref{sLsR}). That is
\beq
s_R^{(n)}= \mathbb{R}_{x_n,y_n}(u)\cdot\mathbb{R}_{x_{n+1},y_{n+1}}(u)\cdot \ldots \mathbb{R}_{x_k,y_k}(u) \cdot s_R
\eeq
In particular the constant spinor $s_R$ is the small solution for $n=k+1$. Similarly, for the left small solution we define
$
s_L^{(n)}= s_L\cdot\varepsilon \cdot \mathbb{R}_{x_1,y_1}(u)\cdot\mathbb{R}_{x_{2},y_{2}}(u)\cdot \ldots \mathbb{R}_{x_{n-1},y_{n-1}}(u) \cdot \varepsilon\,.
$
Then the weak coupling analogue of the brackets $\<L,R\>$ as defined in (\ref{voilaPre}) can also be written as 
\beq
\<L,R\>_\text{weak} = s_L^{(n)} \cdot \varepsilon \cdot s_R^{(n)} \la{voilaPos}
\eeq
exactly as at strong coupling.  This quantity is independent of $n$. Similarly, the strong coupling brackets are Wronskians of the linear problem and as such they are independent of the world-sheet coordinates. The story is not yet complete. The $\mathbb{R}$-matrices act on an auxiliary space defined at some given site and map that to a neighboring site. We are free to change our base for any such auxiliary space. In other words, (\ref{voilaPre}) is invariant under \textit{gauge transformations}
\beq
\mathbb{R}_{x_i,y_i} \to g_{i-{1\over2}} \cdot \mathbb{R}_{x_i,y_i} \cdot g_{i+{1\over2}}^{-1} \,, \qquad s_R \to g_{k+{1\over2}} s_R \,, \qquad  s_L \to s_L\cdot g_{1\over2}^T 
\eeq
where $g_{i+{1\over2}}$ is an SL(2) transformation (such that $\varepsilon \cdot g^T_{i+{1\over2}} \cdot \varepsilon= g_{i+{1\over2}}^{-1}$). In particular we get 
\beq
s_R^{(n)} \to g_{n-{1\over2}}\cdot s_R^{(n)} \,, \qquad  s_L^{(n)}\cdot \varepsilon  \to s_L^{(n)} \cdot \varepsilon \cdot g_{n-{1\over2}}^{-1} 
\eeq
which leaves (\ref{voilaPos}) invariant. This  is the direct analogue of the local gauge transformations at strong coupling which do not affect physical quantities such as well defined holonomies of the flat connection.

\subsection{Open vs Folded} \la{OpVsCl}
Above we constructed an holonomy obtained by starting with a small solution at edge $i$, carrying it all the way to edge $j$ and contracting the resulting vector with the small solution at edge $j$. In this procedure the two end-points play an important role. This is expected since we are dealing with open strings. On the other hand, we also know that for most practical purposes, a closed folded string behaves as two decoupled open strings. Basically, each side of the fold has an infinite effective length. Its two sides decouple since excitations cannot propagate from one side to the other in finite time. Therefore, it should be possible to express our main objects (\ref{voila}) in a language closer to the spectrum problem, where the key objects are closed holonomies. That is, traces of the monodromy matrix. This is the analogue of the usual adjoint versus fundamental story which we encountered already in the introduction. In the planar limit, these are trivially related. Indeed, it is possible to define (\ref{voila}) purely in terms of closed string like quantities, without ever introducing any small solutions as
\beq\la{voilaClosed}
T(u) = \lim_{\epsilon\to 0} \frac{{\rm Tr}\[ \mathbb R_{x_{0},y_{0}}\cdot \mathbb R_{x_{1},y_{1}}\cdot\mathbb R_{x_{2},y_{2}}\cdot\ldots\cdot\mathbb R_{x_{k},y_{k}}\cdot \mathbb R_{x_{k+1},y_{k+1}}\]}{{\rm Tr}\[ \mathbb R_{x_{0},y_{0}}\cdot \mathbb R_{x_{k+1},y_{k+1}}\]} \,. 
\eeq
Note that in this expression we included the first and last $\mathbb{R}$-matrices which are absent in (\ref{voila}); they depend on $\epsilon$ through  $y_0-x_0=y_{k+1}-x_{k+1}=\epsilon$. It is straightforward to prove (\ref{voilaClosed}).\footnote{Here is a pedestrian explanation: Using (\ref{sLsR}), we have
$
s_L \cdot \varepsilon \cdot M \cdot s_R = - M_{11} - M_{12} =- {\rm Tr}\[ \mathcal{P} \cdot M \] 
$
where the projector $\mathcal{P}=\(\! \begin{array}{cc} 1 & 0 \\ 1 & 0 \end{array}\!\)$. On the other hand, as $\epsilon \to 0$ we have
$
\mathbb R_{x_{k+1},y_{k+1}}(u)\cdot \mathbb{R}_{x_0,y_0}(u) \to \frac{1}{\epsilon^2} \, \frac{1}{u^2+\frac{1}{4}} \times \mathcal{P}
$.
 Hence (\ref{voilaClosed}) and (\ref{voila}) are equivalent; they can both be written as ${\rm Tr}\[\mathcal{P}\cdot \mathbb R_{x_{1},y_{1}}(u)\cdot\ldots\cdot \mathbb R_{x_{k},y_{k}}(u)\]/{\rm Tr}\[\mathcal{P}\]$. It is easy to write the same proof in a way that does not rely on a specific choice of frame using the basis $\{s, \varepsilon\cdot s\}$ at each edge.} That relation between the closed and open monodromies also makes clear that $T(u)$ generates conserved charges.
 
In the trace in (\ref{voilaClosed}) there are only R-matrices between $R_{x_0,y_0}$ and $R_{x_{k+1},y_{k+1}}$ but none between $R_{x_{k+1},y_{k+1}}$ and $R_{x_0,y_0}$. This is because, so far, we have chosen to put excitations only between $Z(0,0)$ and $Z(1,0)$, leaving the other side of the bottom and top Wilson loops empty, see for example figures \ref{Dodecagon} and \ref{OneZ}.  More generally, one could choose to have excitations on both sides of the bottom and top loops. In such cases, due to the projective property of $\mathbb R_L=\mathbb R_{x_0,y_0}$ and $\mathbb R_R=\mathbb R_{x_{k+1},y_{k+1}}$, the trace of the monodromy matrix factorize as
\beq\la{folded}
 \frac{\Tr\Big[ \mathbb R_L\cdot\overbrace{\,\mathbb R\cdot\ldots\cdot\mathbb R}^\text{front}\ \cdot \ \mathbb R_R\cdot\overbrace{\,\mathbb R\cdot\ldots\cdot\mathbb R}^\text{back} \Big]}{{\rm Tr}\[ \mathbb R_L\cdot \mathbb R_R\]}={\Tr\[\Omega_\text{front}(u)\]\over\Tr\[\mathbb R_L\cdot\mathbb R_R\]}{\Tr\[\Omega_\text{back}(u)\]\over\Tr\[\mathbb R_R\cdot\mathbb R_L\]}
\eeq
A symmetric choice -- which we will make in section \ref{StrongWeakmatch} -- is to have a folded string where the back of the loop is just the mirror of its front. In that case, the front and back monodromy matrices are related as $\Omega_\text{back}(u)=\Omega_\text{front}(-u)^{-1}$.

\subsection{Monodromy for general Wilson loops and MHV amplitudes}

Above we saw how to define the analogue of the strong coupling holonomies at weak coupling. We did it for a particularly simple set of examples, namely we considered $Z\to \bar Z$ components of the super loop and, furthermore, we assumed that all the bottom excitations were created at the same $x^+$ (and similarly for the top). In fact, it is not hard to lift these simplifying assumptions as we now explain.

First, suppose we still consider the $Z \to \bar Z$ components in $\mathbb{R}^{1,1}$ but with insertions at arbitrary positions in the $x^+$ direction. Then the relevant $R$-matrices would be $SL(2)_{-} \times SL(2)_+$ $R$-matrices which can be written as a four dimensional block diagonal matrix with each $SL(2)$ occupying one block. Then the claim is that (\ref{voila}) is simply untouched. Namely the result only depends on the $x^-$ coordinates of the insertions which are the variables appearing in (\ref{voila}). The reason is that the small solutions -- associated to edges along the $x^+$ direction and localized in $x^-$ -- only have non-zero components in the $SL(2)_{-}$ block which is the only block that becomes degenerate as $\epsilon \to 0$. Hence the $SL(2)_{+}$ block is simply never probed. 
This is related to the $AdS_3$ reduction discussed in section 4.9 of \cite{Alday:2010vh}.

Next we could consider different components of the super loop in $\mathbb{R}^{1,1}$ where the excitations are not the scalars $Z$ but some of the other fields such as the fermions, field strengths, etc. The $SL(2)_{\sigma}$ $R$-matrix still takes the same form as in (\ref{Rdef}). The only difference is that the $SL(2)$ generators act on fields in other representations. These representations are parametrized by the conformal spin $s$. When acting on a field inserted at a cusp, instead of (\ref{actspin}), 
they take the form
\beq\la{higher}
\begin{array}{lll}
L_{+1}|n\>_s&  =&\(n+2s-1\)|n-1\>_s\\
L_0|n\>_s&=&(n+s)|n\>_s\\
L_{-1}|n\>_s & =&(n+1)|n+1\>_s
\end{array}\qquad\text{or}\qquad
\begin{array}{lll}
L_{+1}\circ\cO_s&=&\[x^2\d_x+2sx\]\cO_s\\
L_0\ \circ\, \cO_s&=&\[x\d_x+s\]\cO_s\\
L_{-1}\circ\cO_s&=&\ \d_x\cO_s
\end{array}
\eeq
where
\beq\la{states}
|n\>_s\equiv {1\over n!}|D_-^n\cO_s\>
\eeq
Finally, as $R$ is an $SL(2)$ matrix, its overall normalization is modified from $1/\sqrt{1/4+u^2}$ to $1/\sqrt{s(1-s)+u^2}$. Before, for the scalars, we had $s=1/2$ but for the fermions $s$ can be either $1$ or $1/2$, for example. Hence our previous computations can be straightforwardly generalized to these cases as well. When a field is integrated on an edge, the integration measure may shift the conformal spin of the state accordingly. For example, for the one loop MHV octagon we have a $D_+^kF_{+-}$ insertion integrated along the bottom of the square. That field has conformal spin $s=1$, but the integration shifts it back to $s=0$. In that case, using the expansion of the bottom state (\ref{MHVexp}) and similarly for the top, we find
\beq\la{MHVT}
T_\text{MHV}(u)=1-{i\over u}{e^{-2\sigma}\over(1-e^{-2\sigma})\log(1-e^{-2\sigma})}={1\over r_\text{MHV}}\(1-{i\over2u}\d_\sigma\)\circ r_\text{MHV}
\eeq
where $\sigma=-{1\over2}\log\chi^-_{1342}$ is the cross ratio in the $x^-$ direction (see figure \ref{Octagon}) and $r_\text{MHV}=-{g^2\over2}\log(1+e^{-2\tau})\log(1-e^{-2\sigma})$ is defined as the following logarithm of a ratio of polygons, (see \cite{OPEpaper,bootstraping} for more details) 
\beq\la{smallr}
\def\svgwidth{10cm}
\begingroup%
  \makeatletter%
  \providecommand\color[2][]{%
    \errmessage{(Inkscape) Color is used for the text in Inkscape, but the package 'color.sty' is not loaded}%
    \renewcommand\color[2][]{}%
  }%
  \providecommand\transparent[1]{%
    \errmessage{(Inkscape) Transparency is used (non-zero) for the text in Inkscape, but the package 'transparent.sty' is not loaded}%
    \renewcommand\transparent[1]{}%
  }%
  \providecommand\rotatebox[2]{#2}%
  \ifx\svgwidth\undefined%
    \setlength{\unitlength}{1247.63515625bp}%
    \ifx\svgscale\undefined%
      \relax%
    \else%
      \setlength{\unitlength}{\unitlength * \real{\svgscale}}%
    \fi%
  \else%
    \setlength{\unitlength}{\svgwidth}%
  \fi%
  \global\let\svgwidth\undefined%
  \global\let\svgscale\undefined%
  \makeatother%
  \begin{picture}(1,0.13861686)%
    \put(0,0){\includegraphics[width=\unitlength]{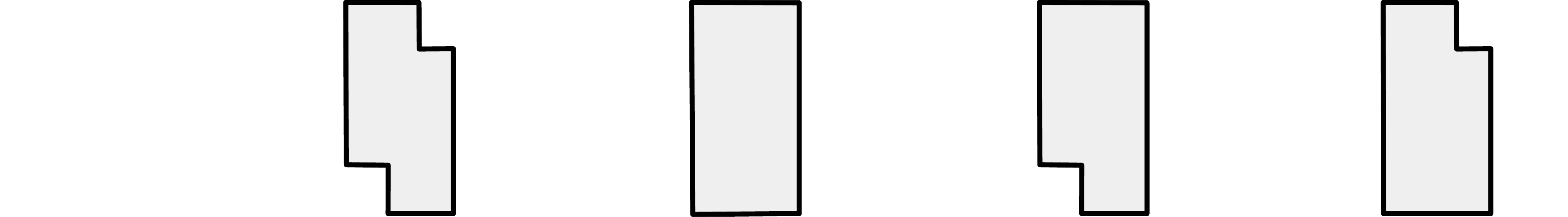}}%
    \put(-0.00164957,0.09502505){\color[rgb]{0,0,0}\makebox(0,0)[lt]{\begin{minipage}{1.33058235\unitlength}\raggedright $ r_\text{MHV} =\log\<\qquad\>+\log\<\qquad\>-\log\<\qquad\>-\log\<\qquad\>$\end{minipage}}}%
  \end{picture}%
\endgroup%

\eeq
It measures what is flowing from the bottom to the top for the case of the bosonic loop.\footnote{Note that the combination $\(u-{i\over2}\d_\sigma\)=(u-p/2)$ is the form of the spin chain transfer matrix for an excitation on top of the GKP vacuum \cite{bootstraping}. There, $p$ is the spin chain momenta of the hole.}

\section{ \textit{Many} Particles and Weak/Strong Coupling Match}\la{StrongWeakmatch}

At this point let us summarize the simplest non-trivial holonomies encountered thus far. At weak coupling the simplest holonomies are those that measure single particle states. These are the tree level NMHV octagon (\ref{TNMHVOctagon}) and the one loop MHV octagon (\ref{MHVT}). These expressions are both very simple. Up to a trivial multiplicative factor, they are given by a simple linear polynomial in $u$ whose coefficients depend on the cross-ratios of the octagon. Let us now compare these weak coupling holonomies with a strong coupling one. At strong coupling, the simplest holonomies are those of the octagon MHV Wilson loop which (in some simple gauge) read \cite{AMapril,Alday:2010vh} 
\beq\la{Tstrong}
T_\text{strong}(u) = \exp\int\limits_{-\infty}^{+\infty} \frac{d\theta'}{2\pi i} \frac{ \log\(1+e^{\cosh(\theta')\log \chi_{1432}^+-i\sinh(\theta')\log \chi_{1342}^-} \)}{\cosh[\theta(u)-\theta']} \,,
\eeq
where\footnote{In the strong coupling classical limit $u$ is large (of order $g$) so that the spectral parameter $\theta$ which appears in the flat connection is of order one.}
\beq
\theta(u)
={1\over4}\log{u+2g\over u-2g}
\eeq
Comparing this result with the simple polynomials obtained at weak coupling (\ref{MHVT}), we conclude that the weak and strong coupling results could hardly look more different!

We should not be surprised however. The simplest weak coupling examples describe the propagation of a single particle whereas at strong coupling we have an infinite number of excitations flowing. They ought to look very different. 

In order to connect these two seemingly very different limits, and hopefully learn about the finite coupling interpolation, we would like to consider a limit where, already at weak coupling, we have an infinite distribution of OPE particles. 
For that purpose, we will now consider a scaling limit with infinitely many scalar insertions in the square Wilson loop.\footnote{Alternatively, we can have many Z insertions by going to very high MHV degree instead of just inserting these excitations by hand on the reference square. For the purpose of understanding the underlying physics both are fine but the latter is simpler and hence that is the example we will follow. }
 In this setup, the weak and strong coupling holonomies will become very similar (even identical in some limit). This limit, which allows us to establish a bridge between weak and strong coupling is the direct analogue of the algebraic curve classical limit \cite{KMMZ,KZ,BKSZ1,BKSZ2} which was of great importance in the study of the spectrum problem.\footnote{Since we are considering mostly polygons in $\mathbb{R}^{1,1}$, described at strong coupling by minimal surfaces in $AdS_3$, the most closely related algebraic curve reference would be \cite{KZ}.}  

We will consider the weak coupling side of this many particle example in subsection \ref{weakSec} and the strong coupling side in section \ref{strongSec}.
As we will consider many complex scalar $Z$ insertions, the dual string will have a non-negligible motion in the sphere. We will describe a simple string solution corresponding to a null polygon in $AdS$ together with point-like motion around an equator in the sphere. We will see that in a particular limit, which is the analogue of the Frolov-Tseytlin limit \cite{FT}, the weak and strong coupling holonomies match perfectly.

\subsection{Weak Coupling} \la{weakSec}

We would like to study a simple case where there are infinitely many particles flowing already at weak coupling. That is, we will insert infinitely many scalars $Z$ on the bottom of the square and corresponding $\bar Z$'s at the top. There is a natural translation symmetry that preserves the two null edges that source the flux tube. Under this translation symmetry, points transform as 
\begin{eqnarray}\la{sigmatrans}
x_i^- \to x_i^-(\sigma) = \frac{e^{2\sigma} x_i^-}{1-x_i^- +e^{2\sigma} x_i^-}\,,
\end{eqnarray}
This is the analogue of (\ref{Tdef}) in the $x^-$ direction.  Note that this conformal transformation indeed preserves $x^-=0$ and $x^-=1$ which are the location of the two null edges. We would like to insert the $Z$'s in such a way that preserves as much symmetry as possible. Therefore we consider insertions which are homogeneously distributed with respect to this $\sigma$ translation symmetry. More precisely, we will insert scalars $Z$ in the bottom of the square at positions
\begin{eqnarray} \la{topX}
\{x_n^-\}_\text{bottom}=\left\{ \frac{e^{2\sigma_n} }{1 +e^{2\sigma_n} } \right\} \qquad \qquad \text{with}\qquad \sigma_n=\log(S) \frac{n}{J}\, ,  \qquad n = -J/2,\dots,J/2\,,
\end{eqnarray}
Here $\log(S)$ is the cut-off in the $\sigma$ direction.\footnote{It is very much related to the spin cut-off discussed in footnote \ref{footnoteCut} but not exactly the same as will hopefully become clear latter. We will come back to this point in a latter footnote. In practice this is a detail that does not play an important role in our discussion.} At the top we consider insertions of scalars $\bar Z$ at positions
\begin{eqnarray} \la{botX}
\{y_n^-\}_\text{top}=\left\{ \frac{-e^{2\sigma_n+\alpha} }{1 -e^{2\sigma_n+\alpha} } \right\} 
\end{eqnarray}
where $\alpha$ is a free parameter, parametrizing how much the propagators from the bottom to the top are tilted. In (\ref{topX}) all points are inserted \textit{inside} the interval $(0,1)$. To make the configuration spacelike we consider the top part of the polygon to go from $0$ to $1$ ``from the outside". That is, in (\ref{botX}) we start close to $x^-=0$ at large negative $\sigma$, go all the way to $|x^-|=\infty$ at $\sigma=0$ and come back to $x^-=1$ at large positive $\sigma$. Note that from a conformal point of view there is no difference between the (\ref{botX}) and (\ref{topX}).  Indeed they are related by a simple analytic continuation, $2\sigma \to 2\sigma + i \pi + \alpha$.

We want to consider the classical limit where the number of insertions is very large and so is $u$. This scaling limit was studied intensively in the spectrum problem, see e.g. \cite{KMMZ} and \cite{someReviewChapter}. 
For large $u$ the reduced $\mathbb{R}$-matrices (\ref{Raux}) exponentiates 
\beq\la{exponentiation}
\mathbb{R}_{x,y}(u) \simeq \exp ({i\, \mathcal{A}_\text{weak}(u;x,y)}) \,, \qquad \mathcal{A}_\text{weak}(u;x,y) = \frac{1}{u}
\left(
\begin{array}{cc}
-\frac{1}{2}\frac{x+y}{x-y} & \frac{xy}{x-y}\\
\frac{-1}{x-y} & \frac{}{2}\frac{x+y}{x-y} 
\end{array}
\right) 
\eeq
That is, if we have a smooth sequence of points (like we do) then the product (\ref{holoWeak}) becomes a simple path ordered exponential integral, exactly as in (\ref{holonomy})! The only difference is that the strong coupling connection $\mathcal{A}$ is now replaced by the weak coupling one, $\mathcal{A}_\text{weak}$. This reinforces strongly the identifications of section \ref{MonoSec}. 
In sum, for our case we have simply 
\beq
\Omega(u)=\mathbb R_{x_{-J/2},y_{-J/2}}\cdot\mathbb R_{x_{-J/2+1},y_{-J/2+1}}\cdot \ldots\cdot \mathbb R_{x_{J/2},y_{J/2}}  =\mathcal{P} \exp\[i\,j\!\!\int\limits_{-\frac{1}{2}\log (S)}^{\frac{1}{2}\log (S)}\!\!\!  d \sigma\, \mathcal{A}_\text{weak}(u;\sigma)\] \,. \la{PexpAppears}
\eeq
Here, $j=J/(\log S)$ is the density of insertions. For example, at $\alpha=0$ we have
\beq\la{Aweak}
\mathcal{A}_\text{weak}(u;\sigma) = \frac{1}{2u}\left(
\begin{array}{cc}
e^{2\sigma} & -e^{2\sigma}\\
2\sinh(2\sigma)& -e^{2\sigma}
\end{array}
\right)\qquad \text{for}\qquad \alpha=0\,.
\eeq
For $\alpha\neq 0$ the expression is not more complicated but it is slightly uglier. 
A simple way to compute the path ordered exponential is to convert it into the solution to a linear problem. That linear problem is the weak coupling counterpart of the strong coupling linear problem \cite{AMapril}. That is, the path ordered exponential in (\ref{PexpAppears}) can be obtained from the monodromy matrix $L(\sigma)$ through $\Omega(u)=L\[\log(S)/2\]$ with
\beq
\[ \partial_{\sigma} - i j \mathcal{A}_\text{weak}(u;\sigma) \]L[\sigma] = 0 \,, \qquad L[-\log(S)/2]=\(\begin{array}{cc} 1 & 0 \\ 0 & 1 \end{array} \)\,.
\eeq
The linear problem can be easily solved and leads to some matrix $\Omega(u)$ which we computed.\footnote{Naively it is not obvious that we are allowed to integrate up to $\log(S)/2$ in (\ref{PexpAppears}), since the entries in the weak coupling connection (\ref{Aweak}) explode as $S\to \infty$. They naively invalidate (\ref{exponentiation}). However the matrix (\ref{Aweak}) is \textit{not} large in the sense that the small solutions do not diverge when acted with (\ref{Aweak}) many times repeatedly. An easy way to see this is the following. There exists a discrete gauge transformation $\mathcal A_\text{weak}(u;\sigma) \to \mathcal A^\text{new}_\text{weak}(u;\sigma) = g(\sigma)\,\mathcal A_\text{weak}(u;\sigma)\, g^{-1}(\sigma+1/j)$ such that (see section \ref{gaugetrans} for discussion)
\begin{eqnarray}
\mathcal A^\text{new}_\text{weak}(u;\sigma) = \frac{1}{2uj}\left(
\begin{array}{cc}
-i-2u\tanh{\sigma} & 2u\coth{\sigma}\\
2u\tanh{\sigma} & i-2u\coth{\sigma}
\end{array}\right) + \mathcal{O}(1/j^2)\; , \qquad \text{for } \alpha=0,
\end{eqnarray}
where $g$ is such that $g(\sigma)(1+\cA_\text{weak}(\sigma))g^{-1}(\sigma)$ is diagonal. This connection does not diverge and therefore we can compute the path ordered exponential by solving the linear problem in this gauge. This is equivalent to solving the linear problem in the previous gauge (where the connection appears large) and applying the gauge transformation in the end. Since it is legitimate to integrate the solution to the linear problem up to $\log(S)/2$ in the new gauge and transform the result back to the previous gauge, we can simply do the computation in the original gauge and the result (\ref{Tcontinuum}) is valid. Of course we also confirmed (\ref{Tcontinuum2}) numerically.}
That is, we just computed the right hand side of (\ref{PexpAppears}). This is an open holonomy so it is not a gauge invariant quantity. One possibility to construct a gauge invariant quantity, as discussed in section \ref{MonoSec}, is to contract this holonomy with two small solutions as in (\ref{voila}). Another simple option, presented in section \ref{OpVsCl}, is to consider the folded situation where the back of the bottom and top loops are the mirrors of their fronts. Then instead of (\ref{voila}) we can compute the trace of the {\it closed} holonomy. The two options are related by (\ref{folded}). That second option is more convenient at strong coupling since it will correspond to a folded string where the two folds are symmetrical. That is, we compute 
\beqa\la{Tcontinuum}
T_\text{closed}&\equiv& \Tr\Big[ \mathbb R_L\cdot\overbrace{\,\mathbb R\cdot\ldots\cdot\mathbb R}^\text{front}\ \cdot \ \mathbb R_R\cdot\overbrace{\,\mathbb R\cdot\ldots\cdot\mathbb R}^\text{back} \Big]=\Tr\(\Omega(u)\cdot\Omega(-u)^{-1}\)
\eeqa
To obtain the continuum limit (\ref{PexpAppears}), we need the R-matrices to change slowly, that is for the insertions to be inserted densely, with $j \gg 1$. Since the exponent in (\ref{PexpAppears}) scales as $j/u$ we also need $u\sim j \gg 1$ to have a good scaling. 
In this limit we find
\beqa\la{Tcontinuum2}
T_\text{closed}(u)=\frac{2u^2}{P(u)}+ 2 \[1-\frac{u^2}{P(u)}\] \cosh\frac{2\sqrt{P(u)}\log S }{u}
\eeqa
where
\beq
P(u) = u^2-\frac{j^2}{4}-i\, j\, u\tanh\frac{\alpha}{2}
\eeq
The transfer matrix is the trace of the monodromy matrix or, equivalently, it is the sum of its two eigenvalues $e^{ip(u)}$ and $e^{-ip(u)}$. The functions $p(u)$ are the so called quasi-momenta introduced for closed strings in $AdS_3$ in \cite{KZ}. 
Next we take $\log(S)$ to be large. Then $T_\text{closed}$ is exponentially large and the quasi-momenta reads
\beq \la{quasi}
p(u) \simeq 2 \frac{\sqrt{j^2/4-u^2+i\, j\, u\tanh\frac{\alpha}{2}} }{u} \log S 
\eeq
This is the main result of this section. 

The quasi-momenta encodes the various charges of the state. For example, the energy can be obtained by expanding it at $u=0$ \cite{KZ}
\beq
p'(u) = -\frac{J}{u^2} - \frac{1}{2g^2}E + \mathcal{O}(u)
\eeq
From (\ref{quasi}) we then derive
\beq
E= \frac{4 g^2 \log^2 S}{J \cosh^2(\alpha/2)}\,.  \la{E1}
\eeq
As a consistency check we can recalculate this energy directly using our Hamiltonian representation (\ref{actionofH}). By acting with the Hamiltonian, one can easily read off the one-loop energy,
\begin{eqnarray} \la{E2}
E  =2\, (2g^2)\!\!\! \int\limits^{\frac{1}{2}\log(S)}_{-\frac{1}{2}\log(S)}\!\!\! d\sigma\,j \log\frac{\cosh\alpha+\cosh\frac{2}{j}}{1+\cosh\alpha} = \frac{4 g^2 \log^2 S}{J \cosh^2(\alpha/2)}\[1+\mathcal O(1/j)\]\,,
\end{eqnarray} 
The first $2$ comes from the fact that the Wilson line is folded, $2 g^2$ arises from the normalization of the spin chain Hamiltonian and $j$ appearing in the integral is the density of insertions. As we see, the direct computation (\ref{E2}) and the algebraic curve one (\ref{E1})  agrees perfectly. Next, we move to strong coupling.

\subsection{Strong Coupling} \la{strongSec}

In this section we will analyze the string solution dual to the strong coupling limit of the square Wilson loop with insertions considered in the previous section. 

The four cusp  solution with no scalar insertions is related by an analytic continuation to the GKP string \cite{GKP} which is dual to a twist two single trace operator at large spin \cite{Tseytlinetal}, see also appendix B of \cite{OPEpaper}. Similarly here, in the presence of the scalar insertions and at $\alpha=0$, that solution is related by an analytic continuation to the folded string solution dual to a single trace operator at large twist and spin. That solution was studied before in \cite{Tseytlinetal,Freyhult:2007pz,Gromov:2008en} (see also references therein) in the context of the spectrum. The solution at $\alpha\ne 0$ is included in a family of solutions that were studied in \cite{Drukker} in the context of polygon Wilson loops.

The solution can be embedded in an $AdS_3\times S^1$ subspace. The embedding coordinates parametrizing the $AdS$ and the circle obey
\beq
X_{-1}^2+X_0^2-X_1^2-X_2^2=1\qquad\text{and}\qquad Y_1^2+Y_2^2=1
\eeq
correspondingly. In these coordinates the solution reads
\beq\la{stringsolution}
\(\!\!\begin{array}{cc}X_0-X_1&X_2-X_{-1}\\ X_2+X_{-1}& X_0+X_1\end{array}\!\!\)=e^{t\sigma_3\cot\varphi}e^{i\theta\sigma_2}e^{\sigma\sigma_3}\equiv g\qquad\text{and}\qquad Y_1+iY_2=e^{w_t t+iw_\sigma\sigma }
\eeq
The original Alday-Maldacena solution, without any R-charge, corresponds to $\theta=\varphi=\pi/4$. 
Here we used the notation $\sigma$ for one of the worldsheet coordinates because it coincides with the $\sigma$ direction introduced above (\ref{sigmatrans}). This was one of the two symmetries of the reference square. The other symmetry was translations in the OPE \textit{time} $\tau$. It almost corresponds to the other worldsheet coordinate, namely $t=\tau\cot\varphi$. The range of $\sigma$ is related to the spin $S$ as $\sigma\in[-{1\over2}\log S,{1\over2}\log S]$. The frequency $w_t$ is directly related to the angular momentum in the sphere which corresponds to the total R-charge which is flowing, namely $J$. More precisely,\footnote{Note that the embedding of the Euclidian solution (\ref{stringsolution}) in the sphere is complex. It is related by analytic continuation to a real rotating timelike solution.} 
\beq
w_t = \frac{2\pi J}{\sqrt{\lambda} \log S} \equiv \frac{\mathfrak{j}}{2}\,.
\eeq
Finally, we can relate the parameters appearing in $AdS$ and in the sphere through the Virasoro constraints. We find
\beq
\cot^2 \varphi =\frac{ \mathfrak{j}^2(4+\mathfrak{j}^2)}{4[\mathfrak{j}^2+4\cos^2(2\theta)]} \,, \qquad w_\sigma =  \frac{2\cos(2\theta)\cot(\varphi)}{\mathfrak{j}} \,. \la{resultV}
\eeq
The quasi-momenta are the eigenvalues of the total monodromy, 
\beq
U^{-1}\cdot\Omega(x)\cdot \Omega^{-1}(-x)\cdot U=\text{diag}\(e^{ i p(x) },e^{- i p(x) }\) \la{pis}
\eeq
where \cite{KZ}
\beq
\Omega(x) = \mathcal{P} \exp\[ \int\limits_{-\frac{1}{2}\log S}^{\frac{1}{2}\log S}\!\!\! d\sigma\, \mathcal{A}_{\sigma}(x) \]\ ,\qquad \mathcal{A}_{\sigma} = \frac{g^{-1}\partial_{\sigma} g + i x\, g^{-1}\partial_{t} g }{1-x^2} \,.
\eeq
Here, the Zhukowsky variable $x$ is related to the spectral parameter $u$ as
\beq\la{Zhukowsky}
x(u)= \frac{u+\sqrt{u^2-\lambda/4\pi^2}}{\sqrt{\lambda}/2\pi}
\eeq
It is yet another convenient parameterization of the spectral parameter. 
As usual, the product of the two monodromies in (\ref{pis}) comes about due to the folded nature of the string. In practice they just yield an overall factor of $2$ for $p(x)$. As in the previous section, these holonomies can be easily computed and $p(x)$ can be read off straightforwardly. Let us now discuss the results and, in particular, the comparison with weak coupling. 

First let us consider the simplest case where $\theta=\pi/4$. In this case the string is point like in the sphere since $w_\sigma=0$, see (\ref{resultV}). For this case, the quasimomentum reads
\beq \la{first}
p(x) = \frac{2x}{x^2-1} \sqrt{1+\frac{\mathfrak{j}^2}{4}-x^2} \log(S) \qquad \text{for}\qquad \theta=\pi/4\,.
\eeq
There are two limits of this expression which we would like to comment on. The first is the large $u$ limit where 
\beq
x(u) \simeq \frac{4\pi u}{\sqrt{\lambda}} \gg 1 \la{xu}
\eeq
This limit corresponds to the Frolov-Tseytlin limit \cite{FT,KMMZ}. This is the relevant limit for large charge $\mathfrak{j} \gg 1$. In this limit we can drop the $1$'s in the quasi-momenta, that is 
\beq
p(x)  \to  \frac{2}{u} \sqrt{\frac{{j}^2}{4}-u^2} \log(S) \qquad \text{for}\qquad\left\{\begin{array}{l}\theta=\pi/4\\ \mathfrak{j} \gg 1\\ x/\mathfrak{j}\ \text{fixed}\end{array}\right.\,.
\eeq
This matches precisely with the weak coupling result (\ref{quasi}) for $\alpha=0$.
The second interesting limit of (\ref{first}) is when the charge $\mathfrak{j}\to 0$ is small. In that limit
\beq
p(x)  \to  - \frac{2x}{\sqrt{1-x^2}}   \log(S) \,, \qquad \text{for}\qquad\left\{\begin{array}{l}\theta=\pi/4\\ \mathfrak{j} \ll 1\end{array}\right.\,.
\eeq
This form matches precisely the behavior of the strong coupling Y-functions in the limit where the the conformal cross ration in the $x^-$ direction is large. That is, under the identification $\log \chi_{1342}=-2\log S$ and $\theta'={1\over2}\log{1+x\over1-x}$ in (\ref{Tstrong}) we get that $2\sinh(\theta')\log\chi^-_{1342}=-\frac{2x}{\sqrt{1-x^2}}   \log(S)$.

Finally, let us move to the general case with generic $\theta$. The expression for arbitrary $x$ and $\mathfrak{j}$ is not particularly illuminating\footnote{\la{footnotep}It is given by $p(x)=\frac{\sqrt{2} x \log (S) \csc (\varphi ) \sqrt{\left(x^2+1\right)
   \cos (2 \varphi )-x^2-2 i x \cos (2 \theta ) \sin (2 \varphi
   )+1}}{x^2-1}\,$.} but it simplifies quite a lot in the Frolov-Tseytlin like limit where $\mathfrak{j}$ and $x$ are large (with fixed ratio). Then 
   \beq
   p(x) \simeq 2 \frac{\sqrt{\mathfrak{j}^2/4-x^2 -  i\, \mathfrak{j} \,x \cos(2\theta)}}{x} \log(S)
   \eeq
Using (\ref{xu}), this matches precisely the weak coupling result (\ref{quasi}) provided we identify 
\beq
\cos(2\theta) = -\tanh(\alpha) \,.
\eeq

While all these matches are great, we should of course be very clear and keep our original motivation in mind: the main point of all these sections is \textit{not} the precise match of the weak and strong coupling holonomies. After all, we know that these kinds of matches are accidental and very particular to the near BPS nature of the Frolov-Tseytlin limit. Instead, the main motivation for studying the continuum limit was to make sure we had correctly identified the weak coupling analogue of the strong coupling Integrability structures. That claim merited further justification due to the striking difference between the weak (\ref{MHVT}) and strong (\ref{Tstrong}) coupling results. We claimed that the very different looking structures were not unexpected since in one case we were dealing with a few flux tube excitations (weak coupling) whereas in the other case we had infinitely many particles (strong coupling). For justifying the claim, we showed that the weak and strong coupling structures become much more similar to each other once many excitations are flowing already at weak coupling. Checking this statement was the goal of these sections and indeed, it was clearly fully confirmed, to an even larger extent that we really needed!

\section{Discussion and Future Directions}

In this paper we studied the OPE for null polygon Wilson loops at weak coupling. We mapped the computation of the expectation value of the loop to a sum of two point functions of local operators. A lot is known about the Integrability properties of local operators and this allowed us to translate back this knowledge to the Wilson loops. For example, up to now, it was hard to predict the leading OPE discontinuities for examples where multiparticles states play a role. From the local operator point of view, the number of particles is simply related to the length of the corresponding spin chains. From this point of view, different lengths  are equally easy to deal with and hence we can tame the multiparticle example in the OPE and predict leading OPE discontinuities for any amplitude. This was the subject of the first sections (\ref{sec2}, \ref{sec3} and \ref{sec4}). 

Of course, ultimately, we want to find an alternative, Integrable, approach towards computing the \textit{full} amplitudes at any value of the coupling and not only their leading discontinuities. To progress in this direction we went to strong coupling for inspiration. There, the key objects are holonomies of the flat connections between edges of the null polygons. These holonomies are very natural in Integrable models; they are generating functions of all the conserved charges. They contain information about the energy  of the state -- which was the key ingredient for the OPE discontinuity story -- but they also contain information about all the other higher charges. We identified the analogue of these objects at weak coupling and computed them for a few examples. This was the subject of the last sections (\ref{MonoSec} and \ref{StrongWeakmatch}).

There are now two obvious pressing questions:
\begin{itemize}
\item We have the holonomies at weak and strong coupling. How to compute them at higher loops at weak coupling\footnote{Note that higher loop R-matrices are not known. Can one construct them? We have a lot of data we can use from the studies of amplitudes. Can this data be used to learn about these higher loop objects?} or, rather,  at finite coupling? Can we make an educated guess for their finite coupling form? Note that there is a major difference between perturbative computations and finite coupling. In perturbation theory we have, at each order, a maximum number of particles propagating in the flux tube. On the contrary, at finite coupling, there are always infinitely many excitations flowing. From this point of view, we expect the strong coupling holonomies to be much more representative of the full quantum solution.\footnote{At strong coupling, the holonomies often enter with shifts in the spectral parameter. These shifts are known as crossing shifts under which the Zhukowsky parameter $x(u)$ (\ref{Zhukowsky}) transforms as $x\to1/x$. That transformation does not commute with perturbation theory (see \cite{Janik:2006dc} and \cite{Benjamin,Basso:2011rc} for its study in the context of polygon Wilson loops). Therefore, one may need to go to finite coupling, where crossing is not degenerate, in order to reveal the all loop structure. The anomalous dimension of a Wilson loop with a cusp was recently studied in \cite{Correa:2012hh}. In particular, in \cite{Gromov:2012eu} an effective description of the full quantum system emerged where the fundamental shift in the spectral parameter is the crossing shift. It suggests that there may by a description of polygon Wilson loops where the crossing shift is the fundamental one at any value of the coupling.}
\item Note that computing the holonomies is not the end of the story. After all, what we are interested in is the expectation value of the Wilson loops! At strong coupling, the expectation value of the Wilson loop is the area of a minimal surface. This area is encoded in the asymptotic values of the holonomies. What is the analogue of this last step at weak coupling or at finite coupling?\footnote{Should we first build Y-functions out of ratios of holonomies? And, if so, how should we think of these ratios and products once they involve holonomies that cross? At the operatorial level with quantum R-matrices at the crossing points or at the level of products of classical expectation values? We suspect the former.} As mentioned in the previous point, it might be that the perturbative regime is too degenerate and it might be somehow simpler to guess the full quantum result in one go. 
\end{itemize}

We should have these big questions in mind but they are probably too hard and vague to tackle as stated. Hence, we end with some more pragmatic next steps. At weak coupling:
\begin{itemize}
\item It would be interesting to consider more examples at weak coupling to gain some more experience and intuition. For example, using the techniques explained in sections \ref{sec2}-\ref{sec4}, we can now compute the leading OPE discontinuities of N$^2$MHV amplitudes. It might be interesting to bootstrap them following what was done for MHV and NMHV amplitudes. The main difference will be that for N$^2$MHV amplitudes one is dealing with two-particle states while for MHV and NMHV single particle states were enough. We already computed one example of an OPE discontinuity for one N$^2$MHV example at one and two loops, see (\ref{discN2MHV}) and (\ref{N2MHVOctD2}). 
\item The two main ingredients in the OPE are the energies of the flux tube states and the probability amplitudes for creating and absorbing these states. The latter are denoted as the OPE form factors. As discussed in section \ref{sec3}, they receive contributions from the geometrical expansion of the top and bottom parts of the loop but also from quantum corrections due to interactions between the insertions. Can the latter corrections be taken into account by a spin chain operator like the spin chain Hamiltonian? In principle, at leading order in the coupling, it would be straightforward (and very interesting) to compute these quantum corrections. Technically, this should be very similar to the computations in \cite{Alday:2005nd} and \cite{Plefka:2012rd}.
\item So far, the power of supersymmetry has not been combined with the OPE (apart from a simple use of SUSY Ward identities in \cite{superOPE}). That is, we normally consider particular components of the super Wilson loop. Supersymmetry relates the different components and therefore one should be able to package together different OPE results in a compact, manifestly supersymmetric form. The use of supersymmetry was very important in fixing the key ingredients of the spectrum problem solution, namely the dispersion relation and the world-sheet S-matrix. It might be that it will be equally important in the OPE. Related to this point, a remarkable alternative approach towards computing the all loop S-matrix in $\mathcal{N}=4$ SYM was put forward recently by Caron-Huot and Song He \cite{SimonHe} and by Bullimore and Skinner \cite{davesametime}. In this approach a recursive relation for the Wilson loop was derived based on the Yangian symmetry of the problem and supersymmetry plays an absolutely central role in this construction. Another instance where supersymmetry and the Yangian are manifest is in the Grassmanian formulation of scattering amplitudes \cite{Grass}. It would be wonderful if one could establish further connections between the OPE and these two exciting developments. 
\end{itemize}
At strong coupling:
\begin{itemize}
\item As mentioned above, we expect the strong coupling result to be a very good representative of the full quantum solution. The strong coupling result takes the form of a thermodynamic Bethe ansatz free energy. We interpret this free energy as describing a sum over densities of particles flowing in the OPE flux tubes in the different channels. In other words, the free energy is the full re-summation of the OPE. However, this is just an \textit{interpretation} of the final result, the derivation is a purely geometrical one and did not rely on this physical picture in any way whatsoever! It would be extremely instructive to re-derive the strong coupling result from the OPE point of view. We expect such a derivation to have a rather direct finite coupling generalization.  
\item  In the spectrum problem the R-charge of the local operators was directly related to the length of the corresponding spin chains. Solving the problem for large lengths was the first step towards the full solution. At the end, one can take the zero R-charge limit and also study purely gluonic operators.\footnote{This idea was recently used in the quark--anti-quark potential problem in \cite{Correa:2012hh}.} Furthermore, in a classical large length limit it is possible to make direct connections between weak and strong coupling. In the classical limit both are described by algebraic curves \cite{KMMZ} that become identical in the so called Frolov-Tseytlin limit \cite{FT}. Having this common language (the algebraic curves) was instrumental in guessing the quantum Bethe equations \cite{Arutyunov:2004vx}. It is natural to expect that by adding large amounts of R-charge to Wilson loop, we might obtain a less degenerate situation where the weak and strong coupling descriptions become considerably more similar. This is exactly what we observed in section \ref{StrongWeakmatch}. We saw that the weak and strong coupling holonomies for a four-cusped polygon with R-charge insertion match precisely in an analogous limit. Hence, it might be simpler to first solve this problem by artificially adding a lot of R-charge that is then removed at the very end. As a first step, one could try to generalize the strong coupling Y-system to the case where we also have movement in the sphere.\footnote{In the Euclidean case, the sphere is complexified and it might be possible to think of it as another AdS. The Y-system dispersion relation should now be modified. It should be something like in footnote \ref{footnotep}. The two Y-systems should be coupled.}  
\end{itemize}
We hope to come back to these points in the future. 

\subsection*{Acknowledgments}
We thank N.~Beisert, S.~Caron-Huot, D.~Gaiotto, N.~Gromov, S.~He, J.~Maldacena and D.~Volin for discussions. We thank S.~Caron-Huot and S.~He for sharing their result for the one loop N$^2$MHV Dodecagon. We thank J.~Bourjaily for sharing his notebooks for one loop non-MHV amplitudes. 
Research at the Perimeter Institute is supported in part by the Government of Canada through NSERC and by the Province of Ontario through MRI. The research of A.S. has been supported in part by the Province of Ontario through ERA grant ER 06-02-293 and by the U.S. Department of Energy grant \#DE-FG02-90ER4054. P.V. is partially supported by NSERC and MEDT of Ontario.

\appendix

\section{NMHV Octagon from the Super OPE in Momentum Space} \la{usualOPE}
In the usual OPE approach for super loops \cite{superOPE} we start with an NMHV tree level amplitude and OPE promote it to higher loops using the dispersion relation $\gamma(p)$ of the relevant particle propagating from the bottom to the top of the polygon. In the octagon Wilson Loop studied in section \ref{octsec}, see figure \ref{Octagon}, the particle is a scalar and \cite{bootstraping,Benjamin,Belitsky:2006en} 
\beq\la{gamma}
\gamma(p)=2g^2 \[ \psi(1/2+ip/2)+\psi(1/2-ip/2)-2\psi(1)\] 
\eeq
where $\psi(x)=\Gamma'(x)/\Gamma(x)$. Tree level amplitudes are most conveniently written using dual momentum twistors. In $\mathbb{R}^{1,1}$ these twistors can be parametrized as 
\beq
Z_{2n-1}\propto\(0,0,1,{x_n^-\over1-x_n^-}\)^T\ ,\qquad Z_{2n}\propto\(1,{x_n^+\over1-x_n^+},0,0\)^T
\eeq
Conformal transformations act linearly as SL(4) transformation of the twistors. We can act with the conformal transformation\footnote{Obviously, in this appendix $S=e^{2\sigma}$ parametrizes the OPE translations, one of the symmetries of the reference square. It should not be confused with the spin $S$ used in most of the main text. There $S$ was a cut-off, the total spin of the local operators.} 
\beq
M=\( \begin{array}{cccc}
T^{-1/2} & 0 & 0 & 0 \\
0 & T^{+1/2} & 0 & 0 \\
0 & 0 & S^{+1/2} & 0 \\
0 & 0 & 0 & S^{-1/2}
\end{array}\) \la{Mdef}
\eeq
on the bottom twistors. These generates a family of twistors where the bottom coordinates are changed according to $x_i^+ \to x_i^+(T)$ and $x_i^-\to x_i^-(S)$ where $x^+_i(T)$ is defined in (\ref{Tdef}) and $x^-_i(S)$ is defined similarly. 

To summarize, we can start with some random numerical twistors $Z_{i}$. For the octagon we will have eight such twistors. Then we act with the transformation (\ref{Mdef}) on the bottom twistors. This generates a family of polygons $W(S,T)$. There are only two independent cross ratios for the octagon so the two parameters $S$ and $T$ are all we need to describe all conformally inequivalent octagons. 

With this prescription all brackets $\<ijkl\>$ which involve a bottom twistor will get $T$ or $S$ dependence. We have
\beq
\mathcal{R}^{(2367)}_{\text{tree}} = \frac{1}{\< 2367\>} =\frac{T^{1/2}}{S^{-1/2}+S^{+1/2}},
\eeq
for a particular choice of the initial twistors.\footnote{The twistors are now chosen to be $Z_1=(0,0,1,0)^T$, $Z_2=M (1,0,0,0)^T$, $Z_3=M (0,0,1,1)^T$, $Z_4=M (1,1,0,0)^T$, $Z_5=(0,0,0,1)^T$, $Z_6=(0,1,0,0)^T$, $Z_7=(0,0,1,-1)^T$ and $Z_8=(1,-1,0,0)^T$. With this parameterization, the cross ratios are $\chi^-_{1243}=1+\frac{1}{S}, \chi^+_{1243}=1+T$.} Other choices of the initial twistors would only differ by translations of $\log(T)$ or $\log(S)$. Next we Fourier transform this result with respect to $\sigma = -\frac{1}{2} \log S$ to get
\beq
\mathcal{R}^{(2367)}_\text{tree} = T^{1/2}\int dp \;\frac{e^{ip\sigma}}{4\cosh(\pi p/2)}.
\eeq
The OPE promotion is now trivial. We simply have to include powers of the anomalous dimension (\ref{gamma}) to get \cite{OPEpaper,superOPE}
\beqa
&&\!\!\!\!\!\!\!\!\!\! 2g^2\,\mathcal{R}^{(2367)}_\text{tree}\,D_1^{(1)}= \int dp \;{\color{red}\gamma(p)}\frac{e^{ip\sigma}}{4\cosh(\pi p/2)}= 2g^2 \mathcal R^{(2367)}_\text{tree}\log\frac{S}{(1+S)^2}, \\ 
&&\!\!\!\!\!\!\!\!\!\! 4g^4\,\mathcal{R}^{(2367)}_\text{tree}\,D_2^{(2)}  = \int dp \;{\color{red}\gamma^2(p)}\frac{e^{ip\sigma}}{4\cosh(\pi p/2)}= 4g^4\mathcal R^{(2367)}_\text{tree}\[\frac{1}{2}\log^2\frac{S}{(S+1)^2}-\frac{1}{3}\log^2(1+S)+\frac{\pi^2}{6}\]\nn
\eeqa
which coincide precisely with the predictions (\ref{done1}) and (\ref{done2}) derived in the main text using the Hamiltonian insertion method!

\section{Hamiltonian Density Averages} \la{Hav}  

In this appendix we quote some results for Hamiltonian averages $\<\mathcal{H}_{ii+1}\mathcal{H}_{jj+1}\>$. As discussed in section \ref{oct2}, there are two type of averages which are non-trivial, see figures \ref{Hcases} and \ref{examplesHH}. We will present the results for the two cases only in the kinematic regimes needed for the two loop examples discussed in the main text. To compute any of these examples we apply (\ref{SL2kernel}) once followed by (\ref{actionofH}), see main text for more details. For $\<\mathcal{H}_{i,i+1}^2\>$ we find 
\beq\la{Hsquaregenera}
\def\svgwidth{12cm}
\begingroup%
  \makeatletter%
  \providecommand\color[2][]{%
    \errmessage{(Inkscape) Color is used for the text in Inkscape, but the package 'color.sty' is not loaded}%
    \renewcommand\color[2][]{}%
  }%
  \providecommand\transparent[1]{%
    \errmessage{(Inkscape) Transparency is used (non-zero) for the text in Inkscape, but the package 'transparent.sty' is not loaded}%
    \renewcommand\transparent[1]{}%
  }%
  \providecommand\rotatebox[2]{#2}%
  \ifx\svgwidth\undefined%
    \setlength{\unitlength}{1788.65097656bp}%
    \ifx\svgscale\undefined%
      \relax%
    \else%
      \setlength{\unitlength}{\unitlength * \real{\svgscale}}%
    \fi%
  \else%
    \setlength{\unitlength}{\svgwidth}%
  \fi%
  \global\let\svgwidth\undefined%
  \global\let\svgscale\undefined%
  \makeatother%
  \begin{picture}(1,0.25315168)%
    \put(0,0){\includegraphics[width=\unitlength]{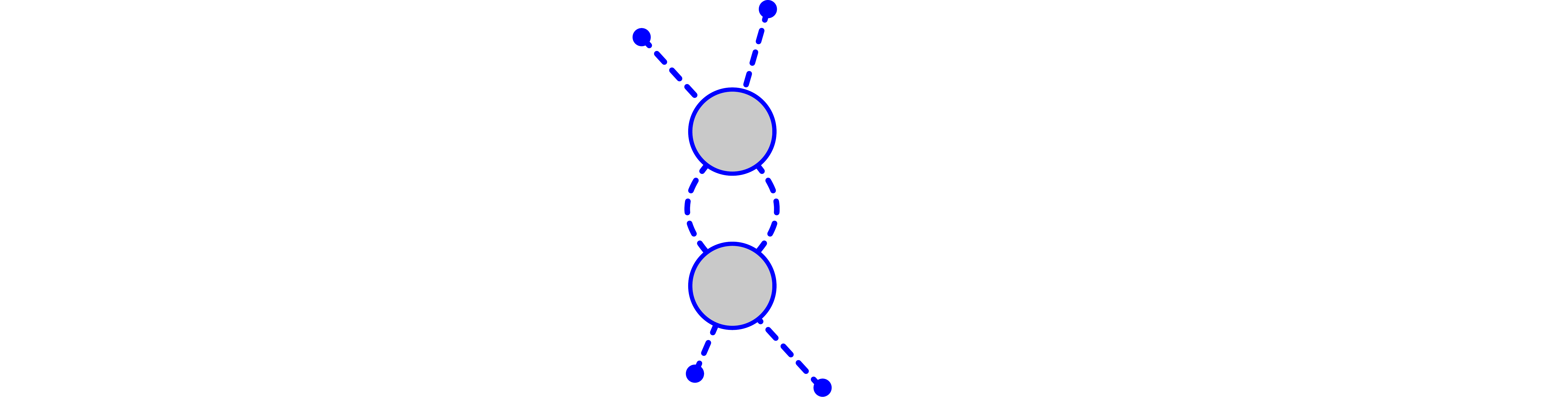}}%
    \put(0.45278046,0.08513564){\color[rgb]{0,0,0}\makebox(0,0)[lt]{\begin{minipage}{0.08318954\unitlength}\raggedright ${\cal H}$\end{minipage}}}%
    \put(0.45278046,0.18532287){\color[rgb]{0,0,0}\makebox(0,0)[lt]{\begin{minipage}{0.08318954\unitlength}\raggedright ${\cal H}$\end{minipage}}}%
    \put(0.52702636,0.15848701){\color[rgb]{0,0,0}\makebox(0,0)[lt]{\begin{minipage}{0.53746302\unitlength}\raggedright $\displaystyle{= -2\frac{(\chi^+ -1){\rm Li}_2\chi^+ -(\chi^- -1){\rm Li}_2\chi^-}{\chi^+ -\chi^-} }\ ,$\end{minipage}}}%
    \put(-0.00074569,0.20231892){\color[rgb]{0,0,0}\makebox(0,0)[lt]{\begin{minipage}{0.27402426\unitlength}\raggedright $\displaystyle{\frac{\<\contraction[2ex]{}{\bar Z_d}{\star\bar Z_c\; (\mathcal H^2\circ Z_a \star}{Z_b}\contraction{\bar Z_d\star}{Z_c}{\; (\mathcal H^2\circ}{Z_a}\bar Z_d\star\bar Z_c\; (\mathcal H^2\circ Z_a \star Z_b)\>}{\<\bcontraction[2ex]{}{\bar Z_d}{\star\bar Z_c\; Z_a\star}{Z_b}\bcontraction{\bar Z_d\star}{\bar Z_c}{\;}{Z_a}\bar Z_d\star\bar Z_c\; Z_a\star Z_b\>}=}$\end{minipage}}}%
    \put(0.42057742,0.03414749){\color[rgb]{0,0,0}\makebox(0,0)[lt]{\begin{minipage}{0.08318954\unitlength}\raggedright $\,_a$\end{minipage}}}%
    \put(0.52613183,0.02967484){\color[rgb]{0,0,0}\makebox(0,0)[lt]{\begin{minipage}{0.03294873\unitlength}\raggedright $\,_b$\end{minipage}}}%
    \put(0.37853456,0.23631102){\color[rgb]{0,0,0}\makebox(0,0)[lt]{\begin{minipage}{0.08318954\unitlength}\raggedright $\,_c$\end{minipage}}}%
    \put(0.49571785,0.2542016){\color[rgb]{0,0,0}\makebox(0,0)[lt]{\begin{minipage}{0.08318954\unitlength}\raggedright $\,_d$\end{minipage}}}%
  \end{picture}%
\endgroup%

\eeq
where $\chi^\pm$ is given in (\ref{ccr}). It is quite remarkable that the difference from the one loop average (\ref{actionofH}) is a simple replacement of a logarithm by a dilogarithm! Probably a similar replacement will give us the higher loop results $\<\mathcal{H}_{i,i+1}^n\>$. 

The averages of $\<\mathcal{H}_{i+2,i+1}\mathcal{H}_{i,i+1}\>$ will depend on the six cross-ratios (3 in the $+$ direction and 3 in the $-$ direction)
constructed out of the three points in the bottom and the three points on the top which these Hamiltonians probe. There is no conceptual difficulty in computing $\<\mathcal{H}_{i,i+1}\mathcal{H}_{i+1,i+2}\>$ in general ${\mathbb R}^{1,1}$ kinematics. Instead, we will only present the cases needed for examples in the main text. For these cases some of the points are null separated. 

For the Octagon N$^2$MHV example (\ref{N2MHVOctD2}), the $\<{\cal H}_{i,i+1}^2\>$ result (\ref{Hsquaregenera}) reduce to 
\beq\la{H2sl22}
\def\svgwidth{16cm}
\begingroup%
  \makeatletter%
  \providecommand\color[2][]{%
    \errmessage{(Inkscape) Color is used for the text in Inkscape, but the package 'color.sty' is not loaded}%
    \renewcommand\color[2][]{}%
  }%
  \providecommand\transparent[1]{%
    \errmessage{(Inkscape) Transparency is used (non-zero) for the text in Inkscape, but the package 'transparent.sty' is not loaded}%
    \renewcommand\transparent[1]{}%
  }%
  \providecommand\rotatebox[2]{#2}%
  \ifx\svgwidth\undefined%
    \setlength{\unitlength}{2555.01425781bp}%
    \ifx\svgscale\undefined%
      \relax%
    \else%
      \setlength{\unitlength}{\unitlength * \real{\svgscale}}%
    \fi%
  \else%
    \setlength{\unitlength}{\svgwidth}%
  \fi%
  \global\let\svgwidth\undefined%
  \global\let\svgscale\undefined%
  \makeatother%
  \begin{picture}(1,0.1798816)%
    \put(0,0){\includegraphics[width=\unitlength]{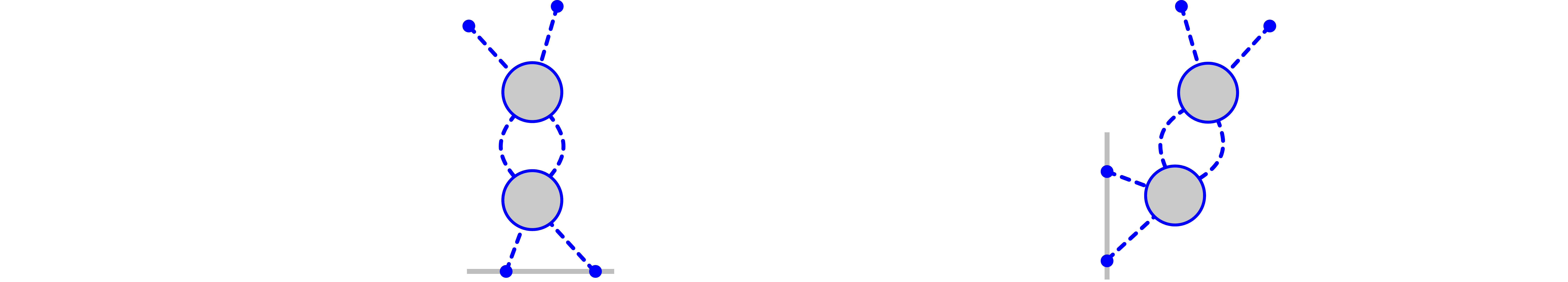}}%
    \put(0.3294957,0.0622611){\color[rgb]{0,0,0}\makebox(0,0)[lt]{\begin{minipage}{0.05823727\unitlength}\raggedright ${\cal H}$\end{minipage}}}%
    \put(0.3294957,0.13239769){\color[rgb]{0,0,0}\makebox(0,0)[lt]{\begin{minipage}{0.05823727\unitlength}\raggedright ${\cal H}$\end{minipage}}}%
    \put(0.38147193,0.09795561){\color[rgb]{0,0,0}\makebox(0,0)[lt]{\begin{minipage}{0.1918321\unitlength}\raggedright $\displaystyle{=-2\text{Li}_2(1-\chi^-_{abcd})}\ ,$\end{minipage}}}%
    \put(-0.00052203,0.14429586){\color[rgb]{0,0,0}\makebox(0,0)[lt]{\begin{minipage}{0.1918321\unitlength}\raggedright $\displaystyle{\frac{\<\contraction[2ex]{}{\bar Z_d}{\star\bar Z_c\; (\mathcal H^2\circ Z_a \star}{Z_b}\contraction{\bar Z_d\star}{Z_c}{\; (\mathcal H^2\circ}{Z_a}\bar Z_d\star\bar Z_c\; (\mathcal H^2\circ Z_a \star Z_b)\>}{\<\bcontraction[2ex]{}{\bar Z_d}{\star\bar Z_c\; Z_a\star}{Z_b}\bcontraction{\bar Z_d\star}{\bar Z_c}{\;}{Z_a}\bar Z_d\star\bar Z_c\; Z_a\star Z_b\>}=}$\end{minipage}}}%
    \put(0.3069518,0.02030438){\color[rgb]{0,0,0}\makebox(0,0)[lt]{\begin{minipage}{0.05823727\unitlength}\raggedright $\,_a$\end{minipage}}}%
    \put(0.38084571,0.02343548){\color[rgb]{0,0,0}\makebox(0,0)[lt]{\begin{minipage}{0.02306593\unitlength}\raggedright $\,_b$\end{minipage}}}%
    \put(0.27751947,0.16809221){\color[rgb]{0,0,0}\makebox(0,0)[lt]{\begin{minipage}{0.05823727\unitlength}\raggedright $\,_c$\end{minipage}}}%
    \put(0.35955424,0.1806166){\color[rgb]{0,0,0}\makebox(0,0)[lt]{\begin{minipage}{0.05823727\unitlength}\raggedright $\,_d$\end{minipage}}}%
    \put(0.80103906,0.09795561){\color[rgb]{0,0,0}\makebox(0,0)[lt]{\begin{minipage}{0.1918321\unitlength}\raggedright $\displaystyle{=-2\text{Li}_2(1-\chi^+_{abcd})}$\end{minipage}}}%
    \put(0.70762929,0.08409022){\color[rgb]{0,0,0}\makebox(0,0)[lt]{\begin{minipage}{0.05823727\unitlength}\raggedright $\,_a$\end{minipage}}}%
    \put(0.70762929,0.01207497){\color[rgb]{0,0,0}\makebox(0,0)[lt]{\begin{minipage}{0.02306593\unitlength}\raggedright $\,_b$\end{minipage}}}%
    \put(0.7333043,0.18115426){\color[rgb]{0,0,0}\makebox(0,0)[lt]{\begin{minipage}{0.05823727\unitlength}\raggedright $\,_c$\end{minipage}}}%
    \put(0.81408663,0.16612499){\color[rgb]{0,0,0}\makebox(0,0)[lt]{\begin{minipage}{0.05823727\unitlength}\raggedright $\,_d$\end{minipage}}}%
    \put(0.73894027,0.06530364){\color[rgb]{0,0,0}\makebox(0,0)[lt]{\begin{minipage}{0.05823727\unitlength}\raggedright ${\cal H}$\end{minipage}}}%
    \put(0.75960552,0.13230913){\color[rgb]{0,0,0}\makebox(0,0)[lt]{\begin{minipage}{0.05823727\unitlength}\raggedright ${\cal H}$\end{minipage}}}%
  \end{picture}%
\endgroup%

\eeq
where the cross ratios are defined as in (\ref{ccr}). 
The other cases  need for the example (\ref{N2MHVOctD2}) are
\beq\la{HijHjkForOctagonN2MHV}
\def\svgwidth{16cm}
\begingroup%
  \makeatletter%
  \providecommand\color[2][]{%
    \errmessage{(Inkscape) Color is used for the text in Inkscape, but the package 'color.sty' is not loaded}%
    \renewcommand\color[2][]{}%
  }%
  \providecommand\transparent[1]{%
    \errmessage{(Inkscape) Transparency is used (non-zero) for the text in Inkscape, but the package 'transparent.sty' is not loaded}%
    \renewcommand\transparent[1]{}%
  }%
  \providecommand\rotatebox[2]{#2}%
  \ifx\svgwidth\undefined%
    \setlength{\unitlength}{2431.88808594bp}%
    \ifx\svgscale\undefined%
      \relax%
    \else%
      \setlength{\unitlength}{\unitlength * \real{\svgscale}}%
    \fi%
  \else%
    \setlength{\unitlength}{\svgwidth}%
  \fi%
  \global\let\svgwidth\undefined%
  \global\let\svgscale\undefined%
  \makeatother%
  \begin{picture}(1,0.44607316)%
    \put(0,0){\includegraphics[width=\unitlength]{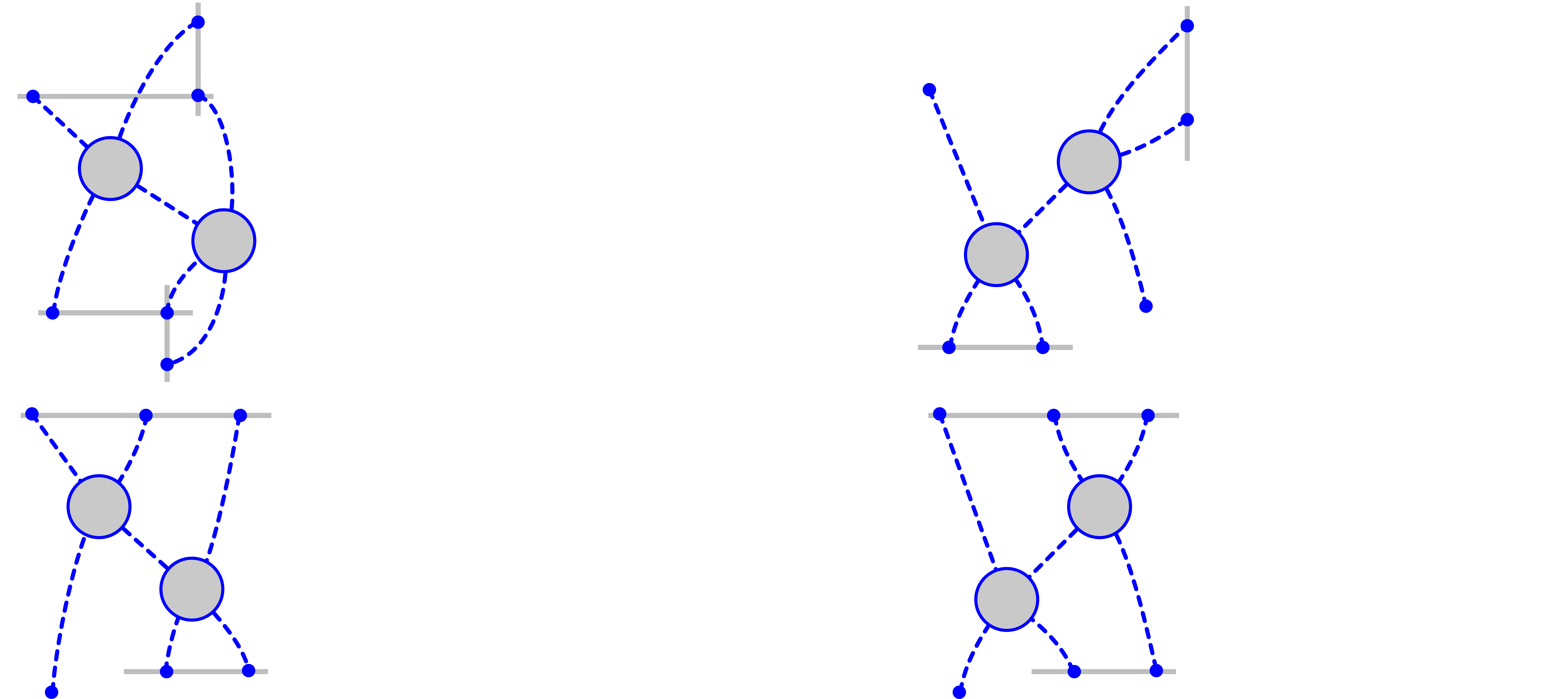}}%
    \put(0.79345876,0.32306245){\color[rgb]{0,0,0}\makebox(0,0)[lt]{\begin{minipage}{0.20154453\unitlength}\raggedright $\displaystyle{=\log\chi^+_{acef}\log\chi^-_{abde}}$\end{minipage}}}%
    \put(0.58303103,0.23844767){\color[rgb]{0,0,0}\makebox(0,0)[lt]{\begin{minipage}{0.06118582\unitlength}\raggedright $\,_a$\end{minipage}}}%
    \put(0.66856127,0.24239522){\color[rgb]{0,0,0}\makebox(0,0)[lt]{\begin{minipage}{0.06118582\unitlength}\raggedright $\,_b$\end{minipage}}}%
    \put(0.73698547,0.25818542){\color[rgb]{0,0,0}\makebox(0,0)[lt]{\begin{minipage}{0.06118582\unitlength}\raggedright $\,_c$\end{minipage}}}%
    \put(0.76264455,0.37661192){\color[rgb]{0,0,0}\makebox(0,0)[lt]{\begin{minipage}{0.06118582\unitlength}\raggedright $\,_f$\end{minipage}}}%
    \put(0.76264455,0.43714102){\color[rgb]{0,0,0}\makebox(0,0)[lt]{\begin{minipage}{0.06118582\unitlength}\raggedright $\,_e$\end{minipage}}}%
    \put(0.59882122,0.39634967){\color[rgb]{0,0,0}\makebox(0,0)[lt]{\begin{minipage}{0.06118582\unitlength}\raggedright $\,_d$\end{minipage}}}%
    \put(0.62239826,0.2954296){\color[rgb]{0,0,0}\makebox(0,0)[lt]{\begin{minipage}{0.06118582\unitlength}\raggedright ${\cal H}$\end{minipage}}}%
    \put(0.68226944,0.35464286){\color[rgb]{0,0,0}\makebox(0,0)[lt]{\begin{minipage}{0.06118582\unitlength}\raggedright ${\cal H}$\end{minipage}}}%
    \put(0.80003801,0.14904129){\color[rgb]{0,0,0}\makebox(0,0)[lt]{\begin{minipage}{0.20132386\unitlength}\raggedright $\displaystyle{=\log\chi^-_{abde}\log\chi^-_{bcef}}$\\$\ $ \\$\displaystyle{+\text{Li}_2(1-\chi^-_{abef})}$\end{minipage}}}%
    \put(0.58961028,0.01837175){\color[rgb]{0,0,0}\makebox(0,0)[lt]{\begin{minipage}{0.06118582\unitlength}\raggedright $\,_a$\end{minipage}}}%
    \put(0.68566732,0.03679365){\color[rgb]{0,0,0}\makebox(0,0)[lt]{\begin{minipage}{0.06118582\unitlength}\raggedright $\,_b$\end{minipage}}}%
    \put(0.74356472,0.03218818){\color[rgb]{0,0,0}\makebox(0,0)[lt]{\begin{minipage}{0.06118582\unitlength}\raggedright $\,_c$\end{minipage}}}%
    \put(0.57842556,0.17495791){\color[rgb]{0,0,0}\makebox(0,0)[lt]{\begin{minipage}{0.06118582\unitlength}\raggedright $\,_d$\end{minipage}}}%
    \put(0.62897751,0.07535368){\color[rgb]{0,0,0}\makebox(0,0)[lt]{\begin{minipage}{0.06118582\unitlength}\raggedright ${\cal H}$\end{minipage}}}%
    \put(0.68884869,0.13456694){\color[rgb]{0,0,0}\makebox(0,0)[lt]{\begin{minipage}{0.06118582\unitlength}\raggedright ${\cal H}$\end{minipage}}}%
    \put(0.58961028,0.01837175){\color[rgb]{0,0,0}\makebox(0,0)[lt]{\begin{minipage}{0.06118582\unitlength}\raggedright $\,_a$\end{minipage}}}%
    \put(0.64882352,0.17561583){\color[rgb]{0,0,0}\makebox(0,0)[lt]{\begin{minipage}{0.06118582\unitlength}\raggedright $\,_e$\end{minipage}}}%
    \put(0.73369585,0.17495791){\color[rgb]{0,0,0}\makebox(0,0)[lt]{\begin{minipage}{0.06118582\unitlength}\raggedright $\,_f$\end{minipage}}}%
    \put(0.22106401,0.14904129){\color[rgb]{0,0,0}\makebox(0,0)[lt]{\begin{minipage}{0.20132386\unitlength}\raggedright $\displaystyle{=\log\chi^-_{abde}\log\chi^-_{bcef}}$\\$\ $ \\$\displaystyle{+\text{Li}_2(1-\chi^-_{bcde})}$\end{minipage}}}%
    \put(0.01063627,0.01837175){\color[rgb]{0,0,0}\makebox(0,0)[lt]{\begin{minipage}{0.06118582\unitlength}\raggedright $\,_a$\end{minipage}}}%
    \put(0.10669332,0.03679365){\color[rgb]{0,0,0}\makebox(0,0)[lt]{\begin{minipage}{0.06118582\unitlength}\raggedright $\,_b$\end{minipage}}}%
    \put(0.16459072,0.03218818){\color[rgb]{0,0,0}\makebox(0,0)[lt]{\begin{minipage}{0.06118582\unitlength}\raggedright $\,_c$\end{minipage}}}%
    \put(-0.00054845,0.17495791){\color[rgb]{0,0,0}\makebox(0,0)[lt]{\begin{minipage}{0.06118582\unitlength}\raggedright $\,_d$\end{minipage}}}%
    \put(0.10921676,0.08193293){\color[rgb]{0,0,0}\makebox(0,0)[lt]{\begin{minipage}{0.06118582\unitlength}\raggedright ${\cal H}$\end{minipage}}}%
    \put(0.05066143,0.13456694){\color[rgb]{0,0,0}\makebox(0,0)[lt]{\begin{minipage}{0.06118582\unitlength}\raggedright ${\cal H}$\end{minipage}}}%
    \put(0.01063627,0.01837175){\color[rgb]{0,0,0}\makebox(0,0)[lt]{\begin{minipage}{0.06118582\unitlength}\raggedright $\,_a$\end{minipage}}}%
    \put(0.06984952,0.17561583){\color[rgb]{0,0,0}\makebox(0,0)[lt]{\begin{minipage}{0.06118582\unitlength}\raggedright $\,_e$\end{minipage}}}%
    \put(0.15472184,0.17495791){\color[rgb]{0,0,0}\makebox(0,0)[lt]{\begin{minipage}{0.06118582\unitlength}\raggedright $\,_f$\end{minipage}}}%
    \put(0.22172193,0.41747469){\color[rgb]{0,0,0}\makebox(0,0)[lt]{\begin{minipage}{0.1763227\unitlength}\raggedright $\displaystyle{= \text{Li}_2 (1-\chi^-_{abde})}$\\$\ $\\$\displaystyle{-\text{Li}_2(1-\chi^-_{abde}\chi^+_{bcef})}$\\$\ $\\$\displaystyle{-\frac{1}{2}\log^2\chi^+_{bcef}}$\end{minipage}}}%
    \put(0.01129419,0.24075041){\color[rgb]{0,0,0}\makebox(0,0)[lt]{\begin{minipage}{0.06118582\unitlength}\raggedright $\,_a$\end{minipage}}}%
    \put(0.08289975,0.26319128){\color[rgb]{0,0,0}\makebox(0,0)[lt]{\begin{minipage}{0.06118582\unitlength}\raggedright $\,_b$\end{minipage}}}%
    \put(0.13103654,0.40786336){\color[rgb]{0,0,0}\makebox(0,0)[lt]{\begin{minipage}{0.06118582\unitlength}\raggedright $\,_f$\end{minipage}}}%
    \put(0.13169446,0.43944376){\color[rgb]{0,0,0}\makebox(0,0)[lt]{\begin{minipage}{0.06118582\unitlength}\raggedright $\,_e$\end{minipage}}}%
    \put(0.01458381,0.40654751){\color[rgb]{0,0,0}\makebox(0,0)[lt]{\begin{minipage}{0.06118582\unitlength}\raggedright $\,_d$\end{minipage}}}%
    \put(0.1315862,0.30299574){\color[rgb]{0,0,0}\makebox(0,0)[lt]{\begin{minipage}{0.06118582\unitlength}\raggedright ${\cal H}$\end{minipage}}}%
    \put(0.0578986,0.35036634){\color[rgb]{0,0,0}\makebox(0,0)[lt]{\begin{minipage}{0.06118582\unitlength}\raggedright ${\cal H}$\end{minipage}}}%
    \put(0.08553145,0.22371577){\color[rgb]{0,0,0}\makebox(0,0)[lt]{\begin{minipage}{0.06118582\unitlength}\raggedright $\,_c$\end{minipage}}}%
  \end{picture}%
\endgroup%

\eeq
The notation should be self-explanatory: a gray line uniting two points indicates that they are null separated (not necessarily adjacent). So for example in the second figure we have $x_a^+ =x_b^+$ and $x_e^-=x_f^-$. 

For computing $D_2^{(2)}$ in (\ref{N2MHVOctD2}) we need $\<\text{top}|\mathbb H^2|\text{bot}\>_\text{MHV}$, $\<\text{top}|\mathbb H|\text{bot}\>_\text{MHV}$, $\<\text{top}|\mathbb H|\text{bot}\>$ and 
\beq \la{H2oct}
 \<\text{top}|\mathbb H^2|\text{bot}\>=\<\text{top}|\(\mathcal H_{4,1}+\mathcal H_{1,2}+\mathcal H_{2,3}+\mathcal H_{3,4}\)\mathcal H_{1,2}|\text{bot}\>+\ \text{cyclic}
\eeq
Using (\ref{H2sl22}) and (\ref{HijHjkForOctagonN2MHV}), we obtain (\ref{N2MHVOctD2}).

\section{Non-compact Spin Chain Hamiltonians}\la{Spinchainapp}

As explained in the main text, in the OPE approach the top and bottom parts of null polygons can be identified with the single trace operators made out of covariant derivatives and all other fields. 

The covariant derivatives are always present. They create the extended Wilson lines. We can also have other fields such as the scalars and the fermions in $\mathcal{N}=4$. Which fields are involved depends on which excitations are being considered in the super Wilson Loop. In other words, it  depends on the helicities of the particles being scattered in the dual Scattering Amplitude picture. 

The single trace operators can be thought of as non-compact spin chains. For the examples considered in this paper two kinds of single trace operators appear:  $SL(2)$ operators
\beqa
|n_1,\dots,n_L\>\equiv \frac{1}{n_1! \dots n_L!} {\rm Tr}\Big[D_+^{n_1} Z \dots D_+^{n_L} Z\Big] 
\eeqa
and $SL(2)\times SL(2)$ operators
\beqa
 |(n_1,m_1),\dots,(n_L,m_L)\>\equiv\frac{1}{n_1! m_1!\dots n_L!m_L!} {\rm Tr}\Big[D_+^{n_1}D_-^{m_1} Z \dots D_+^{n_L} D_{-}^{m_L} Z\Big] \,. \la{sl2sl2}
\eeqa
where $Z$ is a complex scalar.
The integers $n_i,m_j=0,\dots,\infty$ so we are dealing with non-compact spin chains where at each site we have an infinite number of states. 

One small parenthesis on the scalar products of two spin chains: Consider a ket $|n_1,\dots,n_L\>$ representing an operator at $(1,1)$ and a  bra $\<m_1,\dots,m_L|$ representing an operator at $(x^+,x^-)=(0,0)$.\footnote{Recall that in the approach outlined in the main text we always expand the bottom and top Wilson lines around the points $(0,0)$ and $(1,1)$ respectively. } Their scalar product is inherited from the tree level Wick contractions. More precisely we write $\<m_1,\dots,m_L | n_1,\dots,n_L\> $ to indicate that the $a$th field of the first single trace operator is Wick contracted with the $a$th field of the second operator. 
Then we should take $n_1+m_1$ derivatives of the first propagator -- leading to a $(-1)^{m_1}(n_1+m_1)!$ factor -- then $n_2+m_2$ derivatives acting on the second wick contractions and so on. Hence\footnote{We are using $\<Z(x',y')Z(x,y)\>=\frac{1}{(x-x')(y-y')}$. Of course there could be a constant multiplier but in all the physical quantities that we will consider it would drop out. }
\beq
\<m_1,\dots,m_L | n_1,\dots,n_L\> \equiv \prod_{a=1}^L (-1)^{m_a} \frac{(n_a+m_a)!}{n_a! m_a!}  \nn
\eeq
It is sometimes useful to preform a change of basis where the states are orthonormal, see footnote \ref{FOOTNOTE}. 

To OPE promote Null Polygon Wilson loops we need to know how to act on these states with the $\mathcal{N}=4$ planar dilatation operator. By planarity, the one loop dilatation operator $\mathbb{D}$ is a sum of Hamiltonian densities  $\mathbb{H}_{i,i+1} $ which only act on the two sites $i$ and $i+1$, 
\beq
\mathbb{D} = g^2 \sum_{i=1}^{L} \mathbb{H}_{i,i+1} + \mathcal{O}(g^4)
\eeq
The two loop dilatation operator involves interactions between the three sites $i,i+1,i+2$ and so on. For the purpose of the current paper all we need is the one loop dilatation operator. That is, we need the action of the Hamiltonian densities on $SL(2)$ and $SL(2)\times SL(2)$ states,
\beq
\mathbb{H} |n,n'\>  \qquad \text{and} \qquad \mathbb{H} |(n,m),(n',m')\>  \,. \la{forus}
\eeq
This action can be read off from the works of Beisert \cite{Beisert}. In fact, from these papers one can extract the action of the one loop dilation operator on \textit{any} single trace operator.

In this appendix we review what the action (\ref{forus}) is and derive some useful representations that turn out to be very convenient for our OPE purposes. In \ref{sl2first} we discuss the action on $SL(2)$ states while in \ref{sl2sl2next} we consider the more involved action on $SL(2)\times SL(2)$ states. 

\subsection{Action on $SL(2)$ states}\la{sl2first}
Here we discuss the action $\mathbb{H} |n,n'\>$. We will first present the action on states where the first site is empty. As we will review below, this is enough to fix the full Hamiltonian.
\subsubsection*{Action on states with nothing on the first site}
The action of the one loop Dilatation operator can be neatly written using the harmonic numbers as
\beq
\mathbb{H} |0,n\>  = h(n) |0,n\> -  \sum_{d=1}^n \frac{1}{d}|d,n-d\> \,, \qquad h(n) \equiv \sum_{j=1}^n \frac{1}{j} \la{Hon} \,
\eeq
and $h(0)=0$.
There is a very elegant and equivalent way of packaging the relations (\ref{Hon}) using a generating function
\beq
W[x|y]=Z(x)\, e^{(y-x)D_+} Z(x) \,.
\eeq
as\footnote{When checking or deriving this relation is is useful to know about a simple integral representation for the Harmonic numbers which can be derived as
\beq
h(n)= \sum_{j=1}^n \frac{1}{j}=\sum_{j=1}^n\int_0^1\!\! dt  \,t^{j-1} =\int_0^1\!\! dt \frac{1-t^n}{1-t}
\eeq}
\beq
\mathbb{H} \, W[0|y]
= \int_0^1 \frac{dt}{t(t-1)}\Big( W[0|y]-t W[0|ty]-(1-t)W[ty|y] \Big) \la{HonW}
\eeq
In fact, in practice it is often simpler to derive (\ref{HonW}) by a direct Wilson loop computation and from that read off the action (\ref{Hon}) on local operators. 
\subsubsection*{Action on generic states} \la{rec}
From (\ref{Hon}) or (\ref{HonW}) we can easily derive the action on a generic state $|n,n'\>$. Indeed,  the Hamiltonian commutes with the total $D_+$ operator 
\beq
D_+: \qquad D_+ |n,n'\>=(n+1) |n+1,n'\> + (n'+1)|n,n'+1\> \,.
\eeq 
Hence, we have the recursion relation
 \beq
\mathbb{H} |n+1,n'\> =\frac{1}{n+1}\Big( D_+ \mathbb{H} |n,n'\>  -  (n'+1)\mathbb{H} |n,n'+1\> \Big) \,.
\eeq
Combined with the initial condition (\ref{Hon}) this yields the $SL(2)$ Hamiltonian \cite{Beisert}
\beq
\mathbb{H} |n,n'\>  = \[ h(n)+h(n') \] |n,n'\> -  \sum_{d\neq 0} \frac{1}{d}|n-d,n'+d\> \la{fullHSL2}
\eeq
where the sum goes over $d=-n',\dots,n$ except $d=0$. 
An even simpler way of deriving this result is by using the action on Wilson lines (\ref{HonW}). Translating the full expression by $x$ (and relabeling $y+x \to y$) we have
\beqa
\mathbb{H} \, W[x|y]
= \int_0^1\!\! \frac{dt}{t(t-1)}\Big( W[x|y]-t W[x|x+t(y-x)]-(1-t)W[x+t(y-x)|y] \Big) \la{HonW2}
\eeqa
Expanding each Wilson line 
\beq
W[x|y]=\sum_{n=0}^\infty \sum_{n'=0}^\infty x^n y^{n'} |n,n'\>
\eeq
and collecting powers of $x$ and $y$ in each side of (\ref{HonW2}) directly leads to (\ref{fullHSL2}).

\subsection{Action on $SL(2)\times SL(2)$ states}\la{sl2sl2next}
Next we move to the more complicated and considerably  less studied case of $\mathbb{H} |(n,m),(n',m')\>$. As before we start by $\mathbb{H} |(0,0),(n,m)\>$ which is actually all we will use in the main text. 
This sector is not closed, that is 
the action of the Hamiltonian density on $SL(2)\times SL(2)$ states is \textit{not} a linear combination of $SL(2)\times SL(2)$ states (\ref{sl2sl2}) \textit{alone}.\footnote{We can use the equations of motion to relate $D_zD_{\bar z}Z$ to $D_+D_- Z$ which means that we can choose a base of operators where $D_z$ or $D_{\bar z}$ never appear in pairs acting on the same site (if they would we could replace them by $(D_+D_-)^\text{number of pairs}$. With this choice, we can have at most $D_z$'s or $D_{\bar z}$'s in a given site; never both. With this prescription the translation between the oscillator language of \cite{Beisert} and covariant derivatives is unambiguous.}
\beqa
&&\mathbb{H} \,{\color{blue} Z \(D_+^n D_-^m Z \)}= h(n+m) \, {\color{blue}Z \(D_+^n D_-^m Z \)} \la{ouch} \\ \nn
&&\qquad- \sum_{n'=0}^n\sum_{m'=0}^m \delta_{n'+m'>0}  \binom{n}{n'}^2 \binom{m}{m'}^2 B(n'+m',1+n+m-n'-m') \times \\
&&\qquad\qquad\qquad\qquad\qquad\qquad\qquad\times{\color{blue} \(D_+^{n' }D_-^{m'} Z \)  \(D_+^{n-n'} D_-^{m-m'} Z \)} \nn\\
\nn
&&\qquad- \sum_{n'=0}^n\sum_{m'=0}^m\sum_{l=1}^{{\rm min}(n-n',m-m')} \!\!\!\!\delta_{n'+m'>0} \binom{n}{n'} \binom{m}{m'} \binom{m}{m'+l} \binom{n}{n'+l} \times\\
&&\qquad\,\,\,\,\,\,\,\,\, \times\, B(n'+m'+l,1+n+m-n'-m'-l) {\color{red}\(D_+^{n' }D_-^{m'} D_{z}^l Z \)  \(D_+^{n-n'-l} D_-^{m-m'-l} D_{\bar z}^l Z \) }\nn\\
\nn
&&\qquad- \sum_{n'=0}^n\sum_{m'=0}^m\sum_{l=1}^{{\rm min}(n-n',m-m')} \!\!\!\! \delta_{n'+m'>0}\binom{n}{n'} \binom{m}{m'} \binom{m}{m'+l} \binom{n}{n'+l} \times\\
&&\qquad\,\,\,\,\,\,\,\,\, \times\, B(n'+m'+l,1+n+m-n'-m'-l) {\color{red}\(D_+^{n' }D_-^{m'} D_{\bar z}^l Z \)  \(D_+^{n-n'-l} D_-^{m-m'-l} D_{ z}^l Z \) }\nn
\eeqa
where $B(a,b)=\Gamma(a)\Gamma(b)/\Gamma(a+b)$ and $D_z$ and $D_{\bar z}$ are the other two null direction. Schematically, 
\beq
D_{+}= D_{0}-D_{1} \, , \qquad D_{-}= D_{0}+D_{1}\, , \qquad D_{z}= D_{2}+iD_{3} \, , \qquad D_{\bar z}= D_{2}-iD_{3}\,.
\eeq 
The simplest way to obtain (\ref{ouch}) is to read it from the Harmonic action of \cite{Beisert}.\footnote{In this language one first maps the covariant derivatives to bilinears of oscillators. Then the Hamiltonian can hop these oscillators from the first site to the second site and vice versa. When there is no hopping the corresponding weight is an harmonic number of (half) the total number of oscillators $N=2n+2m$. That gives the first line in (\ref{ouch}). The other lines in (\ref{ouch}) correspond to some ($n_{21}$) oscillators jumping from the second site to the first site. There are no jumps in the opposite direction ($n_{12}=0$) since the first site was empty to start with. The corresponding weight is $-B(n_{21}/2,1+N/2-n_{21}/2)$, see (F.5) in the second reference in \cite{Beisert}. This explains all the $B$ functions in (\ref{ouch}). Finally, the binomial coefficients that multiply these $B$ functions come from the number of ways of choosing which oscillators jump from the second to the first site. } The expression (\ref{ouch}) looks scary and hard to work with. Fortunately, we will only need the first two lines of this expression which we will furthermore bring to a more useful form for our OPE applications.

\subsubsection{Projected Hamiltonian action} \la{projections}
First note that we consider polygons in $\mathbb{R}^{1,1}$. The fields in the bottom and top part of the polygon are therefore displaced in the $x_+$ and $x_-$ directions but not in the $x_z$ or $x_{\bar z}$ directions. Hence, suppose we act with a single Hamiltonian on the top. That generates fields which contain derivatives in the $z$ and $\bar z$ directions, the last terms in (\ref{ouch}). When contracting with the bottom we will end up with propagators like 
\beq
\left\< \Big(D_z^l D_{+}^n D_-^m Z \Big) \[x_{+},x_{-},x_{z},x_{\bar z}\]  \Big( D_{+}^{n'} D_-^{m'} Z \Big) \[y_+,y_-,y_z,y_{\bar z}\] \right\>_{x_z=x_{\bar z}=y_z=y_{\bar z}=0}
\eeq
where we used $x\equiv x^{(top)}$ and $y\equiv x^{(bottom)}$. 
To leading order in the coupling this is simply proportional to 
\beq
\(\frac{\partial}{\partial x^{+}}\)^n\(\frac{\partial}{\partial x^{+}}\)^m \(\frac{\partial}{\partial y^{+}}\)^{n'} \(\frac{\partial}{\partial y^{-}}\)^{m'} \(\frac{\partial}{\partial x^z}\)^{l}  \frac{1}{(x_+-y_+)(x_--y_-)-(x_z-y_z)(x_{\bar z}-y_{\bar z}) } \nn
\eeq
After taking the derivatives we end up with positive powers of $(x_{\bar z}-y_{\bar z})$ in the numerator. Since at the end we should set $x_z=x_{\bar z}=y_z=y_{\bar z}=0$, we get zero. Hence we can drop the last two lines in (\ref{ouch}) if we are only interested in leading perturbative computations. 

At higher loops this argument no longer works since we can act on the top \textit{and} on the bottom and easily end up with (dropping the $+$ and $-$ derivatives that play no role in these arguments)
\beq
\(\frac{\partial}{\partial x^z}\frac{\partial}{\partial y^{\bar z}}\)^{l}  \frac{1}{(x_+-y_+)(x_--y_-)-(x_z-y_z)(x_{\bar z}-y_{\bar z}) } \nn
\eeq
In this case the $z$ and $\bar z$ derivatives are \textit{balanced} and we no longer get zero. Sometimes, even for some higher loop examples, the polygon kinematics are such that we can still drop the last terms in (\ref{ouch}).

Given this discussion, from now on we drop the last terms in (\ref{ouch}) and consider the projection of the Hamiltonian action into the $SL(2)\times SL(2)$ sector, 
\beqa
&&\!\!\!\!\left.\mathbb{H} \,{\color{blue} Z \(D_+^n D_-^m Z \)}\right|_{SL(2)\times SL(2)}= h(n+m) \, {\color{blue}Z \(D_+^n D_-^m Z \)} \la{ouch2} \\ \nn
&&\!\!\!\!\quad- \sum_{n'=1}^n\sum_{m'=1}^m \binom{n}{n'}^2 \binom{m}{m'}^2 B(n+m,1+n+m-n'-m') {\color{blue} \(D_+^{n' }D_-^{m'} Z \)  \(D_+^{n-n'} D_-^{m-m'} Z \)} \nn \,.
\eeqa

\subsubsection{Wilson Line Action}
In this section we re-write (\ref{ouch2}) as a Kernel action on a Wilson line connecting two points in $\mathbb{R}^{1,1}$, 
\beq
W[x',y'|x,y] \equiv Z(x',y') \,e^{(x-x')D_-+(y-y')D_+} \, Z(x',y') \,.
\eeq
That is, we want to generalize (\ref{HonW}) to this more involved case and compute  
\beq
\mathbb{H}\,W[0,0|x,y]= \sum_{n,m=0}^\infty \frac{x^n}{n!}\frac{y^n}{m!}\,  \mathbb{H}\,{\color{blue} (Z D_-^n D_+^m Z )} \la{starting}
\eeq
using (\ref{ouch2}). Since the computation is fun, we will do it in detail; the final result is (\ref{hence}). When plugging (\ref{ouch2}) in (\ref{starting}) we have two terms coming from the first and second lines in (\ref{ouch2}). The first one is trivial to deal with following the same steps as in $SL(2)$, section \ref{sl2first}. Hence we consider the contribution from the second line in (\ref{ouch2}):
\beq
- \sum_{n,m=0}^\infty \frac{x^n}{n!}\frac{y^n}{m!}\,   \sum_{n'=0}^n\sum_{m'=0}^m \binom{n}{n'}^2 \binom{m}{m'}^2 B(n'+m',1+n+m-n'-m')  {\color{blue} \(D_+^{n' }D_-^{m'} Z \)  \(D_+^{n-n'} D_-^{m-m'} Z \)} \nn
\eeq
Note that we dropped the constraint $\delta_{n'+m'>0}$ which imposes that there is some hopping. That means that at the end we should take care of subtracting the contribution of this expression with $n'=m'=0$ which is quite simple to do. 
Let us now simplify this expression. First we use
\beq
B(a,b)=\int_0^1\!\!dt\,  t^{a-1} (1-t)^{b-1}
\eeq
and introduce $n=p+n'$ and $m=q+m'$  to get
\beq
-\int_0^1\!\! dt \sum_{p,q,n',m'=0}^\infty   x^{p+n'} y^{m'+q}\,    \binom{n'+p}{n'} \binom{m'+q}{m'} t^{n'+m'-1} (1-t)^{p+q}  {\color{blue}\frac{ \(D_+^{n' }D_-^{m'} Z \)  \(D_+^{p} D_-^{q} Z \)}{n'!m'!p!q!}} \nn
\eeq
Now we use that
\beq
\binom{a+b}{b} = \oint \frac{dz}{2\pi i} \frac{(1+z)^{a+b}}{z^{b+1}} \nn
\eeq
where the contour goes around the origin and $a,b$ are positive integers. 
We end up with a totally factorized integrand/summand
\beq
-\int_0^1 \frac{dt}{t} \oint \frac{dz}{2\pi i z} \oint \frac{dw}{2\pi i w}  \mathcal{I}
\eeq
where
\beqa
\mathcal{I}\!\!&=&\!\!\sum_{p,q,n',m'=0}^\infty \Big[x (1+z) t \Big]^{n'}\Big[y (1+w) t \Big]^{m'}  \Big[x (1+\frac{1}{z}) (1-t) \Big]^{p}\Big[y (1+\frac{1}{w}) (1-t) \Big]^{q} \times  \\
\!\!&&\!\!\qquad\qquad\qquad\qquad\qquad\qquad\qquad\qquad\qquad\qquad\qquad\qquad \times\, {\color{blue}\frac{ \(D_+^{n' }D_-^{m'} Z \)  \(D_+^{p} D_-^{q} Z \)}{n'!m'!p!q!}}\nn\\
\!\!&=&\!\!W[x (1+z) t ,y (1+w) t ][x (1+{1}/{z}) (1-t) ,y (1+{1}/{w}) (1-t) ]  \nn
\eeqa
Now we just need to take care of the remaining contributions. They are the $n'=m'=0$ term that we dropped above plus the first line in (\ref{ouch2}). They are quite trivial to work out to let us simply quote the final result
\beqa
&&\mathbb{H}\,W[0,0|x,y]\Big|_{SL(2)\times SL(2)}=\int_0^1 \frac{dt}{t} \oint \frac{dz}{2\pi i z} \oint \frac{dw}{2\pi i w}\times \la{hence}\\
&&\qquad\quad\times \(W[0,0|x,y]- W[x (1+z) t ,y (1+w) t |x (1+{1}/{z}) (1-t) ,y (1+{1}/{w}) (1-t) ] \) \nn
\eeqa

We could now get the general case for the action on a state containing also derivative in the first site by either translating this expression or by using a recursion relation together with the initial condition (\ref{ouch2}). See section \ref{rec} for a more careful discussion in the $SL(2)$ context. For our purpose (\ref{ouch2}) or (\ref{hence}) is all we need since we can always translate a pair of points so that the origin coincides with one of them.


\begin{thebibliography}{99}


  \bibitem{OPEpaper}
  L.~F.~Alday, D.~Gaiotto, J.~Maldacena, A. Sever and P. Vieira,
  ``An Operator Product Expansion for Polygonal null Wilson Loops,''
 [arXiv:1006.2788 [hep-th]].
 
  



\bibitem{AMcomments}
  L.~F.~Alday and J.~M.~Maldacena,
  ``Comments on operators with large spin,''
  JHEP {\bf 0711}, 019 (2007)
  [arXiv:0708.0672 [hep-th]].
  



  \bibitem{Benjamin}
  B.~Basso,
  ``Exciting the GKP string at any coupling,''
  arXiv:1010.5237 [hep-th]. 
  


\bibitem{bootstraping}
 D.~Gaiotto, J.~Maldacena, A.~Sever and P.~Vieira,
  ``Bootstrapping Null Polygon Wilson Loops,''
  JHEP {\bf 1103}, 092 (2011)
  [arXiv:1010.5009 [hep-th]].

  



\bibitem{Hexagonpaper}
  D.~Gaiotto, J.~Maldacena, A.~Sever and P.~Vieira,
  ``Pulling the straps of polygons,''
  arXiv:1102.0062 [hep-th].





\bibitem{superOPE} 
  A.~Sever, P.~Vieira and T.~Wang,
  ``OPE for Super Loops,''
  JHEP {\bf 1111}, 051 (2011)
  [arXiv:1108.1575 [hep-th]].
  


\bibitem{Dixon:2011nj}
  L.~J.~Dixon, J.~M.~Drummond and J.~M.~Henn,
  ``Analytic result for the two-loop six-point NMHV amplitude in N=4 super Yang-Mills theory,''
  JHEP {\bf 1201} (2012) 024
  [arXiv:1111.1704 [hep-th]].


\bibitem{Dixon:2011pw}
  L.~J.~Dixon, J.~M.~Drummond and J.~M.~Henn,
  ``Bootstrapping the three-loop hexagon,''
  JHEP {\bf 1111} (2011) 023
  [arXiv:1108.4461 [hep-th]].



  \bibitem{Heslop:2011hv}
  P.~Heslop and V.~V.~Khoze,
  ``Wilson Loops @ 3-Loops in Special Kinematics,''
  JHEP {\bf 1111} (2011) 152
  [arXiv:1109.0058 [hep-th]].



\bibitem{SimonHe} 
  S.~Caron-Huot and S.~He,
  ``Jumpstarting the All-Loop S-Matrix of Planar N=4 Super Yang-Mills,''
  arXiv:1112.1060 [hep-th].
  


 \bibitem{davesametime} 
  M.~Bullimore and D.~Skinner,
  ``Descent Equations for Superamplitudes,''
  arXiv:1112.1056 [hep-th].
  



\bibitem{Alday:2010vh} 
  L.~F.~Alday, J.~Maldacena, A.~Sever and P.~Vieira,
  ``Y-system for Scattering Amplitudes,''
  J.\ Phys.\ A A {\bf 43}, 485401 (2010)
  [arXiv:1002.2459 [hep-th]].
   


\bibitem{FT} 
  S.~Frolov and A.~A.~Tseytlin,
  ``Rotating string solutions: AdS / CFT duality in nonsupersymmetric sectors,''
  Phys.\ Lett.\ B {\bf 570}, 96 (2003)
  [hep-th/0306143].



\bibitem{Belitsky} 
  A.~V.~Belitsky,
  ``OPE for null Wilson loops and open spin chains,''
  Phys.\ Lett.\ B {\bf 709}, 280 (2012)
  [arXiv:1110.1063 [hep-th]].
   


\bibitem{Alday:2010zy}
  L.~F.~Alday, B.~Eden, G.~P.~Korchemsky, J.~Maldacena and E.~Sokatchev,
  ``From correlation functions to Wilson loops,''
  arXiv:1007.3243 [hep-th]. $\bullet$ B.~Eden, G.~P.~Korchemsky and E.~Sokatchev,
  ``From correlation functions to scattering amplitudes,''
  arXiv:1007.3246 [hep-th]. $\bullet$ B.~Eden, G.~P.~Korchemsky and E.~Sokatchev,
  ``More on the duality correlators/amplitudes,''
  arXiv:1009.2488 [hep-th].
  

\bibitem{'tHooft:1973jz}
M.~Bullimore and D.~Skinner,
  ``Holomorphic Linking, Loop Equations and Scattering Amplitudes in Twistor Space,''
  arXiv:1101.1329 [hep-th]. $\bullet$   A.~V.~Belitsky, G.~P.~Korchemsky and E.~Sokatchev,
  ``Are scattering amplitudes dual to super Wilson loops?,''
  arXiv:1103.3008 [hep-th].  



\bibitem{AmplitudeWilson}
  L.~F.~Alday, J.~M.~Maldacena,
  ``Gluon scattering amplitudes at strong coupling,''
  JHEP {\bf 0706}, 064 (2007).
  [arXiv:0705.0303 [hep-th]].
  

  \bibitem{AmplitudeWilson2}
 G.~P.~Korchemsky, J.~M.~Drummond, E.~Sokatchev,
  ``Conformal properties of four-gluon planar amplitudes and Wilson loops,''
  Nucl.\ Phys.\  {\bf B795}, 385-408 (2008).
  [arXiv:0707.0243 [hep-th]].
$\bullet$  A.~Brandhuber, P.~Heslop, G.~Travaglini,
  ``MHV amplitudes in N=4 super Yang-Mills and Wilson loops,''
  Nucl.\ Phys.\  {\bf B794}, 231-243 (2008).
  [arXiv:0707.1153 [hep-th]].
  $\bullet$  Z.~Bern, L.~J.~Dixon, D.~A.~Kosower, R.~Roiban, M.~Spradlin, C.~Vergu and A.~Volovich,
  ``The Two-Loop Six-Gluon MHV Amplitude in Maximally Supersymmetric Yang-Mills
  Theory,''
  Phys.\ Rev.\  D {\bf 78}, 045007 (2008)
  [arXiv:0803.1465 [hep-th]].
$\bullet$  J.~M.~Drummond, J.~Henn, G.~P.~Korchemsky and E.~Sokatchev,
  ``Hexagon Wilson loop = six-gluon MHV amplitude,''
  Nucl.\ Phys.\  B {\bf 815} (2009) 142
  [arXiv:0803.1466 [hep-th]].
$\bullet$  N.~Berkovits, J.~Maldacena,
  ``Fermionic T-Duality, Dual Superconformal Symmetry, and the Amplitude/Wilson Loop Connection,''
  JHEP {\bf 0809}, 062 (2008).
  [arXiv:0807.3196 [hep-th]].
  



\bibitem{Skinner}
L.~J.~Mason, D.~Skinner,
  ``The Complete Planar S-matrix of N=4 SYM as a Wilson Loop in Twistor Space,''
  JHEP {\bf 1012}, 018 (2010).
  [arXiv:1009.2225 [hep-th]]. 
  



\bibitem{Simon}
S.~Caron-Huot,
  ``Notes on the scattering amplitude / Wilson loop duality,''
 [arXiv:1010.1167 [hep-th]]. 
  
 



\bibitem{DaveAndOthers}
  T.~Adamo, M.~Bullimore, L.~Mason and D.~Skinner,
  ``A Proof of the Supersymmetric Correlation Function / Wilson Loop Correspondence,''
  JHEP {\bf 1108} (2011) 076
  [arXiv:1103.4119 [hep-th]].



\bibitem{RatioCite}
  J.~M.~Drummond, J.~Henn, G.~P.~Korchemsky and E.~Sokatchev,
  ``Dual superconformal symmetry of scattering amplitudes in N=4 super-Yang-Mills theory,''
  Nucl.\ Phys.\ B {\bf 828} (2010) 317
  [arXiv:0807.1095 [hep-th]].


  \bibitem{Beisert} 
   N.~Beisert,
  ``The complete one loop dilatation operator of N=4 superYang-Mills theory,''
  Nucl.\ Phys.\ B {\bf 676} (2004) 3
  [hep-th/0307015]. $\bullet$
  N.~Beisert,
  ``The Dilatation operator of N=4 super Yang-Mills theory and integrability,''
  Phys.\ Rept.\  {\bf 405}, 1 (2005)
  [hep-th/0407277].
  


\bibitem{Belitsky:2005qn} 
  A.~V.~Belitsky and A.~V.~Radyushkin,
  ``Unraveling hadron structure with generalized parton distributions,''
  Phys.\ Rept.\  {\bf 418}, 1 (2005)
  [hep-ph/0504030].
  

 \bibitem{Elvang:2009wd}
  H.~Elvang, D.~Z.~Freedman and M.~Kiermaier,
  ``Solution to the Ward Identities for Superamplitudes,''
  JHEP {\bf 1010} (2010) 103
  [arXiv:0911.3169 [hep-th]].


  \bibitem{Beisert:2003ys}
  N.~Beisert,
  ``The su(2|3) dynamic spin chain,''
  Nucl.\ Phys.\ B {\bf 682} (2004) 487
  [hep-th/0310252].
  

\bibitem{addpapers1}  
  J.~M.~Drummond, J.~Henn, G.~P.~Korchemsky and E.~Sokatchev,
  ``Generalized unitarity for N=4 super-amplitudes,''
  arXiv:0808.0491 [hep-th]. $\bullet$ 
  A.~Brandhuber, P.~Heslop and G.~Travaglini,
  ``One-Loop Amplitudes in N=4 Super Yang-Mills and Anomalous Dual Conformal
  Symmetry,''
  JHEP {\bf 0908} (2009) 095
  [arXiv:0905.4377 [hep-th]].
$\bullet$ 
  H.~Elvang, D.~Z.~Freedman and M.~Kiermaier,
  ``Dual conformal symmetry of 1-loop NMHV amplitudes in N=4 SYM theory,''
  JHEP {\bf 1003} (2010) 075
  [arXiv:0905.4379 [hep-th]]. $\bullet$   N.~Arkani-Hamed, J.~L.~Bourjaily, F.~Cachazo and J.~Trnka,
  ``Local Integrals for Planar Scattering Amplitudes,''
  arXiv:1012.6032 [hep-th].

   

\bibitem{Beisert:2010jr}
  N.~Beisert, C.~Ahn, L.~F.~Alday, Z.~Bajnok, J.~M.~Drummond, L.~Freyhult, N.~Gromov and R.~A.~Janik {\it et al.},
  ``Review of AdS/CFT Integrability: An Overview,''
  Lett.\ Math.\ Phys.\  {\bf 99} (2012) 3
  [arXiv:1012.3982 [hep-th]].


\bibitem{AMapril}
  L.~F.~Alday and J.~Maldacena,
  ``Null polygonal Wilson loops and minimal surfaces in Anti-de-Sitter space,''
  JHEP {\bf 0911}, 082 (2009)
  [arXiv:0904.0663 [hep-th]].




\bibitem{BKSZ1} 
  N.~Beisert, V.~A.~Kazakov, K.~Sakai and K.~Zarembo,
  ``Complete spectrum of long operators in N=4 SYM at one loop,''
  JHEP {\bf 0507}, 030 (2005)
  [hep-th/0503200].
  


\bibitem{BKSZ2} 
  N.~Beisert, V.~A.~Kazakov, K.~Sakai and K.~Zarembo,
  ``The Algebraic curve of classical superstrings on AdS(5) x S**5,''
  Commun.\ Math.\ Phys.\  {\bf 263}, 659 (2006)
  [hep-th/0502226].
  


\bibitem{FaddeevReview}
  L.~D.~Faddeev,
  ``How algebraic Bethe ansatz works for integrable model,''
  hep-th/9605187.



\bibitem{KMMZ} 
  V.~A.~Kazakov, A.~Marshakov, J.~A.~Minahan and K.~Zarembo,
  ``Classical/quantum integrability in AdS/CFT,''
  JHEP {\bf 0405}, 024 (2004)
  [hep-th/0402207].
  


\bibitem{KZ} 
  V.~A.~Kazakov and K.~Zarembo,
 ``Classical / quantum integrability in non-compact sector of AdS/CFT,''
  JHEP {\bf 0410}, 060 (2004)
  [hep-th/0410105].
 


\bibitem{someReviewChapter}
  S.~Schafer-Nameki,
  ``Review of AdS/CFT Integrability, Chapter II.4: The Spectral Curve,''
  Lett.\ Math.\ Phys.\  {\bf 99} (2012) 169
  [arXiv:1012.3989 [hep-th]].


\bibitem{GKP}
  S.~S.~Gubser, I.~R.~Klebanov and A.~M.~Polyakov,
  ``A Semiclassical limit of the gauge / string correspondence,''
  Nucl.\ Phys.\ B {\bf 636} (2002) 99
  [hep-th/0204051].


\bibitem{Tseytlinetal}
R.~Roiban and A.~A.~Tseytlin,
  ``Spinning superstrings at two loops: Strong-coupling corrections to dimensions of large-twist SYM operators,''
  Phys.\ Rev.\ D {\bf 77} (2008) 066006
  [arXiv:0712.2479 [hep-th]].
  
  \bibitem{Freyhult:2007pz}
  L.~Freyhult, A.~Rej and M.~Staudacher,
  ``A Generalized Scaling Function for AdS/CFT,''
  J.\ Stat.\ Mech.\  {\bf 0807} (2008) P07015
  [arXiv:0712.2743 [hep-th]].
  
  \bibitem{Gromov:2008en}
  N.~Gromov,
  ``Generalized Scaling Function at Strong Coupling,''
  JHEP {\bf 0811} (2008) 085
  [arXiv:0805.4615 [hep-th]].
  

\bibitem{Drukker}
  H.~Dorn, N.~Drukker, G.~Jorjadze and C.~Kalousios,
  ``Space-like minimal surfaces in AdS x S,''
  JHEP {\bf 1004} (2010) 004
  [arXiv:0912.3829 [hep-th]].



\bibitem{Janik:2006dc} 
  R.~A.~Janik,
  ``The AdS(5) x S**5 superstring worldsheet S-matrix and crossing symmetry,''
  Phys.\ Rev.\ D {\bf 73}, 086006 (2006)
  [hep-th/0603038].
  

\bibitem{Basso:2011rc} 
  B.~Basso and A.~V.~Belitsky,
  ``Luescher formula for GKP string,''
  Nucl.\ Phys.\ B {\bf 860}, 1 (2012)
  [arXiv:1108.0999 [hep-th]].




\bibitem{Correa:2012hh} 
  D.~Correa, J.~Maldacena and A.~Sever,
  ``The quark anti-quark potential and the cusp anomalous dimension from a TBA equation,''
  arXiv:1203.1913 [hep-th].
  $\bullet$
  N.~Drukker,
  ``Integrable Wilson loops,''
  arXiv:1203.1617 [hep-th].
  


\bibitem{Gromov:2012eu} 
  N.~Gromov and A.~Sever,
  ``Analytic Solution of Bremsstrahlung TBA,''
  arXiv:1207.5489 [hep-th].
  


\bibitem{Alday:2005nd} 
  L.~F.~Alday, J.~R.~David, E.~Gava and K.~S.~Narain,
  ``Structure constants of planar N = 4 Yang Mills at one loop,''
  JHEP {\bf 0509}, 070 (2005)
  [hep-th/0502186].
  


\bibitem{Plefka:2012rd} 
  J.~Plefka and K.~Wiegandt,
  ``Three-Point Functions of Twist-Two Operators in N=4 SYM at One Loop,''
  arXiv:1207.4784 [hep-th].
  


\bibitem{Grass}
N.~Arkani-Hamed, F.~Cachazo, C.~Cheung and J.~Kaplan,
  ``A Duality For The S Matrix,''
  JHEP {\bf 1003} (2010) 020
  [arXiv:0907.5418 [hep-th]]. $\bullet$  N.~Arkani-Hamed, J.~L.~Bourjaily, F.~Cachazo, S.~Caron-Huot and J.~Trnka,
  ``The All-Loop Integrand For Scattering Amplitudes in Planar N=4 SYM,''
  JHEP {\bf 1101} (2011) 041
  [arXiv:1008.2958 [hep-th]].



\bibitem{Arutyunov:2004vx} 
  G.~Arutyunov, S.~Frolov and M.~Staudacher,
  ``Bethe ansatz for quantum strings,''
  JHEP {\bf 0410}, 016 (2004)
  [hep-th/0406256].
$\bullet$
  N.~Beisert and M.~Staudacher,
  ``Long-range psu(2,2|4) Bethe Ansatze for gauge theory and strings,''
  Nucl.\ Phys.\ B {\bf 727}, 1 (2005)
  [hep-th/0504190].
  $\bullet$
    N.~Beisert, B.~Eden and M.~Staudacher,
  ``Transcendentality and Crossing,''
  J.\ Stat.\ Mech.\  {\bf 0701}, P01021 (2007)
  [hep-th/0610251].
  
  


\bibitem{Belitsky:2006en}
  A.~V.~Belitsky, A.~S.~Gorsky and G.~P.~Korchemsky,
  ``Logarithmic scaling in gauge/string correspondence,''
  Nucl.\ Phys.\ B {\bf 748} (2006) 24
  [hep-th/0601112]. $\bullet$   L.~Freyhult, A.~Rej and M.~Staudacher,
  ``A Generalized Scaling Function for AdS/CFT,''
  J.\ Stat.\ Mech.\  {\bf 0807}, P07015 (2008)
  [arXiv:0712.2743 [hep-th]].

  
\end{thebibliography}
\end{document}